\documentclass[a4paper,11pt]{article}
\pdfoutput=1

\usepackage{xcolor}
\usepackage{jheppub}
\usepackage{mathtools}
\usepackage{dsfont}
\usepackage{braket}
\usepackage{appendix}
\newcommand\numberthis{\addtocounter{equation}{1}\tag{\theequation}}
\usepackage[linewidth=1pt]{mdframed}

\newcommand{\DD}{\mathrm{d}}
\renewcommand{\.}{\cdot}
\newcommand{\defined}{\coloneqq}
\newcommand{\e}{\epsilon}
\renewcommand{\d}{\partial}
\renewcommand{\t}[1]{\text{#1}}
\newcommand{\at}[1]{\bigg\vert_{#1}}

\renewcommand{\d}{\partial}

\usepackage{tikz}
\usetikzlibrary{decorations.pathmorphing} 
\usetikzlibrary{decorations.markings}

\title{Analytic properties of infrared-finite amplitudes in theories with long-range forces}

\author[a]{Luke Lippstreu}

\affiliation[a]{Higgs Centre, School of Physics \& Astronomy \\
		University of Edinburgh, EH9 3FD, United Kingdom}
\emailAdd{llippstr@ed.ac.uk}

\abstract{
Infrared divergences obscure important analytic properties of scattering amplitudes, indicating gaps in our understanding of unitarity, causality, and crossing symmetry in theories with long-range forces. Using the exactly solvable model of a charged scalar particle in a fixed Coulomb background, we demonstrate that novel analytic properties arise and can be systematically studied when long-range interactions are properly incorporated. We first canonically quantize a scalar particle in a Coulomb potential, confirming that basic conditions for unitarity and causality hold. We then examine the necessary modifications to the LSZ reduction formula, the general optical theorem, and the treatment of the disconnected components of scattering amplitudes. Next, we show that the Coulomb phase divergence is analytically related to real radiative divergences via crossing symmetry, implying that a well-defined treatment of the Coulomb phase divergence provides constraints on the real radiative divergence. In contrast to the Faddeev--Kulish approach, we propose that an effective way to eliminate infrared divergences and study these analytic properties is to fully solve the quantum theory associated with the asymptotic Hamiltonian.}

\begin{document}

\maketitle

\section{Introduction}
Infrared (IR) divergences occur in interacting theories with massless particles in four or fewer spacetime dimensions, making them relevant to a wide range of research programs, from Standard Model predictions at colliders to black-hole physics to the study of gravitational waves. Despite their prevalence and long history, these divergences are not yet fully under control. This is evident in two ways. First, there are very few explicitly computed IR-finite amplitudes.  Secondly, although important progress has been made, fundamental challenges remain in proving theorems on how causality, unitarity, and crossing symmetry constrain scattering amplitudes in theories with long-range forces—especially in comparison to theories with a mass gap, where such constraints are far more developed.
\subsection{Summary of main results}
In this paper we study a charged scalar in a fixed Coulomb background—an exactly solvable model that approximates $2\rightarrow 2$ scattering in scalar QED (SQED) where one of the charges is infinitely massive—to investigate novel properties that arise in quantum field theories with long-range interactions. For this setup, we show in section \ref{sect:unitarity} that the inner product of the in-state with the out-state lacks the usual disconnected ``no scattering'' \(\delta\)-component, meaning the usual decomposition,
\begin{gather}  
\braket{\t{out}|\t{in}}=\delta_{\t{in,out}}+ \delta^{(4)}(p_{\t{in}}-p_{\t{out}})\mathcal{M}_{\t{in}\rightarrow \t{out}},\label{decomp}  
\end{gather}  
does not accurately reflect the connectedness structure of the scattering amplitude\footnote{The modified connectedness structure of amplitudes in theories with long-range forces has been briefly mentioned in \cite{Eden:1966dnq} (pp. 191–192).}. This suggests that a similar feature may persist in full SQED, and more generally in theories with long-range forces. The general optical theorem, 
\begin{gather}
    \mathcal{M}_{\t{in}\rightarrow \t{out}}+\mathcal{M}^{\star}_{\t{in}\rightarrow \t{out}}=-\int\DD\alpha \, \delta^{(4)}(p_{\t{in}}-p_{\alpha})\mathcal{M}_{\t{in}\rightarrow \alpha}\mathcal{M}^{\star}_{\alpha\rightarrow \t{out}},
\end{gather}
where $\alpha$ denotes a complete set of states, relies on the decomposition (\ref{decomp}) and hence does not yield useful information in this setup. In section \ref{sect: modified elastic}, we derive a modification to the general optical theorem that is useful for relating different orders of perturbation theory and illustrate its use with an example.

Many theorems on the analytic properties of scattering amplitudes rely on the LSZ reduction formula, which is known not to apply in theories with long-range forces \cite{Kibble:1968oug, Abrikosov1954, Bogoliubov1956, Soloviev1963, Hagen1963, NuovoCimento1963, Fradkin1965}. In section \ref{sect:LSZ} we present a modified LSZ reduction formula that, in this model, enables a direct transition from correlation functions to IR-finite amplitudes.  This approach avoids any ambiguous scales or ill-defined integrals at intermediate steps.

In SQED, there are two types of infrared divergences: the Coulomb phase divergence and the real radiative divergence. Our setup omits soft photon production, thereby neglecting the more challenging real radiation divergence associated with radiating an infinite number of zero-energy photons. However, as we review in section~\ref{sect:crossing}, these two divergences are not independent. They are connected by analytic properties of the amplitude, including crossing symmetry and the absence of pseudothreshold singularities on the physical sheet. For general complex momenta, both divergences correspond to a single analytic function. This implies that if the IR-finite S-matrix satisfies crossing symmetry or the absence of pseudothresholds, then an unambiguous treatment of the Coulomb phase divergence necessarily constrains the real radiative divergence.

The Faddeev-Kulish \cite{Kulish:1970ut} framework (reviewed below) provides a method for handling IR divergences at the level of scattering amplitudes. While valid, this approach is challenging to work with, as it leads to infinitely ambiguous amplitudes which make it difficult to determine the fundamental analytic properties of IR-finite amplitudes, in addition to issues with Poincaré invariance. The FK method constructs asymptotic states by acting on free-particle states with asymptotic evolution operators. Here, we take a different approach: rather than using free states, we advocate for using states that are eigenstates of the asymptotic Hamiltonian. In our setup, we demonstrate in section \ref{sect:canonical} how to diagonalize the asymptotic Hamiltonian of a charged particle in the field of an infinitely massive source—an approach that we aim to extend to systems with generic masses.

An important analytic property of scattering amplitudes is their factorization on bound-state poles. In section \ref{sect:scatter to bound}, we compute the residue of the relativistic Coulomb amplitude and demonstrate that it factorizes into a product of scattering-to-bound-state transition amplitudes. At the bound-state energies, the Coulomb phase ceases to be a phase and contributes to the residue. Conventional approaches introduce an IR regulator to control the Coulomb phase divergence, causing the factorization at the bound-state poles to depend explicitly on the regulator scale. In contrast, our approach yields a consistent factorization relation with no dependence on arbitrary regularization scales.

\subsection{Motivation}
Infrared divergences commonly appear in scattering amplitude calculations in theories involving long-range forces, such as QED and gravity. These divergences occur because standard perturbation theory incorrectly assumes that in- and out-states behave as free-particle states at asymptotic times. The failure of this assumption is evident even at the classical level. Two charged particles interacting via their mutual Coulomb force do not follow free trajectories at asymptotic times but instead exhibit a slight logarithmic deviation from straight-line motion \cite{Sahoo_2019,PhysRevD.7.1082}. In the Feynman rules, this assumption of free-particle motion is reflected in the use of free plane waves meeting at the interaction vertices. To alleviate the so-called Coulomb-phase infrared divergence, which arises from this incorrect assumption of free motion, it is natural to instead use wavefunctions
\begin{gather}
e^{-iEt+i\vec{p}\cdot\vec{x}+i\gamma\log(|\vec{p}||\vec{x}|-\vec{p}\cdot\vec{x})}, \label{distorted wavefunctions}
\end{gather}
that include a slight logarithmic distortion factor that accounts for the deviation from straight-line motion, where $\gamma = \frac{e_1 e_2}{4\pi} \frac{E}{|\vec{p}|}$. See, for example, \cite{rowe1985hamilton} for how the wavefunctions (\ref{distorted wavefunctions}) encode the asymptotic trajectories. This approach has been successfully applied in non-relativistic quantum mechanics \cite{mulherin1970coulomb,barrachina1989scattering,Kadyrov_2005,KADYROV20091516,CROTHERS1992287,reinhold1987distorted}, although some of these methods \cite{mulherin1970coulomb,barrachina1989scattering} have been criticized for producing Møller operators that do not form an isometry \cite{dollardcrit}. It is difficult to extend these methods to relativistic quantum field theory, primarily because no completeness relations are known for the distorted plane-wave wavefunctions (\ref{distorted wavefunctions}). Without such relations, it is not possible to justify expanding quantum fields in this basis. However, to formulate a perturbation theory that computes the inner product of the in-state with the out-state, it is sufficient to use wavefunctions whose asymptotic behavior matches that of the actual in and out states. There exists a complete orthonormal basis of wavefunctions whose asymptotic behavior matches that of in and out states in which charged particle pairs continue to experience their mutual Coulomb interaction at late times. These are the Coulomb wavefunctions:
\begin{gather}
f_{\text{in}}(x,p) = \Gamma(1+i\gamma)  e^{-\frac{\pi\gamma}{2}}  e^{-i Et+i\vec{p}\cdot\vec{x}}  {}_1F_1\Big(-i\gamma,1,i(|\vec{p}||\vec{x}|-\vec{p}\cdot\vec{x})\Big), \label{Coulomb motivate} \\
\gamma(p) \defined \frac{e_1 e_2 E}{4\pi |\vec{p}|}, \qquad E = \sqrt{|\vec{p}|^2 + m^2}. \label{def:gamma motivate}
\end{gather}
These wavefunctions have the same asymptotic form as (\ref{distorted wavefunctions}). In particular, the large-$|\vec{x}|$ expansion of (\ref{Coulomb motivate}) reproduces (\ref{distorted wavefunctions}); see, for example, \cite{messiah1999quantum}. They therefore encode the same slight deviation from straight-line motion. Using these wavefunctions, the so-called Coulomb phase infrared divergence is completely eliminated in non-relativistic quantum mechanics \cite{Landau:1991wop}. In this paper, we demonstrate that the corresponding perturbative expansion for \(2 \rightarrow 2\) relativistic scattering in scalar QED, with one infinitely massive particle, is likewise free of the Coulomb phase infrared divergence. We also mention how this method could be extended to generic-mass scattering in future work.
This provides a well-defined framework for studying some of the novel analytic properties of amplitudes in theories with long-range forces, which are otherwise obscured by infrared divergences. Despite the seeming increase in complexity in going from distorted plane waves (\ref{distorted wavefunctions}) to Coulomb wavefunctions (\ref{Coulomb motivate}), the resultant integrals are often simpler for Coulomb wavefunctions \cite{gravielle1992some,colavecchia1997hypergeometric}. Moreover, it is reasonable to expect that Coulomb wavefunctions encode more information about the full scattering amplitude than distorted plane waves.

In abelian gauge theories, there are two distinct types of infrared divergences: the Coulomb phase divergence discussed above and the real radiative divergence. The physics underlying the Coulomb phase divergence is relatively simple—it stems from the incorrect assumption in standard perturbation theory that charged particles follow straight-line trajectories at asymptotic times. By contrast, the real radiative divergence is more subtle, as it is associated with the emission and absorption of an infinite number of zero-energy photons during every scattering event.

The perturbation theory described above, which employs Coulomb wavefunctions, addresses only the Coulomb phase divergence. However, these two divergences are not entirely independent. They are linked by analytic properties of scattering amplitudes, such as crossing symmetry. Consequently, completely eliminating the Coulomb phase divergence may provide insight into the real radiative divergence by exploiting this analytic connection. In section~\ref{sect:crossing}, we provide a pedagogical discussion of this relationship and highlight how resolving the Coulomb phase divergence could inform future approaches to the real radiative divergence.

\subsection{Historical overview}
\subsubsection*{IR-finite inclusive cross sections}
In 1937, Bloch and Nordsieck (BN) \cite{Bloch:1937pw} introduced a practical way to handle IR divergences. They showed that while individual cross sections can be IR-divergent, the physically relevant \textit{inclusive} cross sections—where one sums over all possible emissions of soft radiation below a detector threshold \(\Lambda_{\text{IR}}\)—remain IR-finite. This approach is still one of the most widely used methods for making reliable predictions in the presence of IR divergences. The proof of the BN cancellation came later in the works of \cite{yennie,Weinberg:1965nx,Grammer:1973db}.

This article does not consider charged massless particles and therefore does not address collinear divergences. Furthermore, we focus exclusively on theories with genuinely long-range forces, such as electromagnetism and gravity in four spacetime dimensions. Consequently, our results are not applicable to QCD, which has no long-range forces due to its mass gap and whose asymptotic dynamics are fundamentally different from QED and gravity.

However, for completeness, we briefly mention that for theories with massless charged particles, the Bloch-Nordsieck approach fails, as summing only over soft radiation is insufficient to cancel infrared divergences~\cite{Doria:1980ak}. In such cases, one turns to the KLN theorem~\cite{Kinoshita:1962ur,Lee:1964is}, which states that sufficiently inclusive sums over degenerate initial and final states yield IR-finite probabilities. Practical implementations—including recent refinements~\cite{Frye:2018xjj,Gonzo:2023cnv}—can sometimes achieve IR finiteness, although challenges remain~\cite{Lavelle_2006}. In QCD, whose asymptotic dynamics are non-perturbative, one considers infrared safe observables that factorize into a perturbatively calculable part and a non-perturbative component extracted from experiment~\cite{Collins:2011zzd,Agarwal:2021ais}.

\subsubsection*{IR-finite amplitudes: Faddeev-Kulish framework}
To study unitarity, causality, and crossing symmetry, it is preferable to address infrared divergences at the level of amplitudes rather than inclusive cross sections. Many foundational results on the analytic properties of the S-matrix assume the absence of long-range forces, making them inapplicable to QED and gravity \cite{Eden:1966dnq}. For an example of unitarity, the Froissart-Martin bound, central to the modern S-matrix bootstrap program \cite{deRham:2022hpx, Kruczenski:2022lot}, requires a mass gap. For an example of causality, the Steinmann relations, which play a key role in the $\mathcal{N}=4$ SYM bootstrap \cite{caronhuot2020steinmann}, also assume the absence of massless particles. Similarly, the proofs of crossing symmetry for $2\rightarrow 2$ and $2\rightarrow 3$ scattering \cite{Bros:1964iho, Bros:1965kbd,BROS1986325} do not apply to theories with long-range forces, leaving it unproven for QED and gravity—though recent work \cite{Mizera:2021fap} has established it in planar massless theories.

Despite significant progress in extending these theorems to gapless theories with long-range forces—see \cite{Bourjaily:2020wvq, Fevola:2023fzn, Berghoff:2022mqu, Hannesdottir:2022xki} for work on the Steinmann relations and \cite{caronhuot2023measuredasymptotically,caronhuot2023crossingscatteringamplitudes,Mizera:2021ujs} for advances on crossing symmetry—no framework exists that matches the unambiguous and systematic approach available for gapped theories in deriving such theorems.

During the late 1960s and early 1970s, significant efforts were made to develop a framework for obtaining IR-finite quantities directly at the level of amplitudes. The work of Chung \cite{Chung:1965zza}, Kibble \cite{Kibble:1968oug,Kibble:1968sfb}, and Dollard \cite{dollard1971quantum,Dollard1964AsymptoticCA} culminated in what became known as the Faddeev–Kulish (FK) approach \cite{Kulish:1970ut}. In theories with long-range forces, such as QED and gravity, the standard Dyson $S$-matrix fails to fully capture the asymptotic dynamics because the in- and out-states never asymptote to free-particle dynamics due to persistent long-range interactions. This is in contrast to theories with short-range forces, where asymptotic states approach free evolution at asymptotic times. The FK approach modifies the $S$-matrix by incorporating additional evolution operators that compensate for these residual interactions, effectively canceling the late-time asymptotic evolution. Specifically, the FK $S$-matrix takes the form
\begin{gather}
    S_{\text{FK}} = \lim_{t_{\pm} \to \pm\infty} e^{-R(t_+)} e^{-i\Phi(t_+)} \mathcal{S} e^{-i\Phi(t_-)} e^{R(t_-)},\label{both operators}
\end{gather}
where the two additional factors encode two distinct physical effects. The first, and the main focus of this article, is Dollard's evolution operator $\Phi$ \cite{Dollard1964AsymptoticCA,dollard1971quantum}, which accounts for the long-range Coulomb interactions between charged particles
\begin{gather}
    \Phi(t) = \frac{\alpha}{2} \int \frac{d^3p}{(2\pi)^3} \frac{d^3q}{(2\pi)^3} : \rho(p) \rho(q) :
\frac{p \cdot q}{\sqrt{(p \cdot q)^2 - m^4}} \ln |t/t_0|,\label{dollard operator}
\end{gather}
where $\rho$ is the charge density operator, measuring the charge of the states it acts upon
\begin{gather}
    \rho(p) = \sum_s \left( a_p^{s\dagger} a_p^s - b_p^s b_p^{s\dagger} \right),\label{rho}
\end{gather}
where $s$ sums over spin. Physically, it reflects that charged particles continue to interact through their mutual Coulomb potential, causing deviations from straight-line motion even at asymptotic times. The arbitrary reference time \( t_0 \) in the logarithm of Eq.~(\ref{dollard operator}) is often omitted, obscuring the fact that Dollard’s evolution operator retains an undetermined scale, leading to ambiguities in the resulting amplitudes. Different treatments of this scale yield amplitudes with distinct analytic properties. Another issue is that the Dollard evolution operator does not satisfy the group property of time translations, 
$U(t_1) U(t_2) \neq U(t_1 + t_2)$, which complicates the definition of time-covariant asymptotic dynamics. This issue has been discussed in \cite{Morchio:2014rfa,Morchio:2014zga}, where an approach to address it was also proposed.
 The only perturbative implementation of Dollard's evolution operator we are aware of is \cite{PAPANICOLAOU1976229}, where it becomes evident that the resulting scattering amplitudes depend explicitly on the choice of the reference time $t_0$\footnote{A more indirect perturbative implementation of the Dollard evolution operator appears in \cite{Hannesdottir:2019opa}, where Wilson line operators acting on free-particle states are shown to reproduce the effect of both operators in (\ref{both operators}). The resulting amplitudes also depend on the choice of the reference IR scale.}.

The real radiative evolution operator in the FK evolution operator (\ref{both operators}) is  
\begin{gather}
    R(t) = e \sum_{j=1}^{2} \int \frac{d^3p}{(2\pi)^3}\int^{\Lambda_{\t{IR}}} \frac{d^3k}{(2\pi)^3 \sqrt{2\omega_k}}
\left[ f_{j}(p,k) a_k^{j\dagger} e^{i \frac{p \cdot k}{\omega_p} t}
- f^{\star}_{j}(p,k) a_k^j e^{-i \frac{p \cdot k}{\omega_p} t} \right] \rho(\vec{p}).\label{R operator}
\end{gather}
This term describes how charged particles continue to radiate soft photons at late times, where $a_k^{j\dagger},a_k^{j}$ are the creation and annihilation operators for photons. The upper bound $\Lambda_{\t{IR}}$ on the soft photon momenta is often omitted, obscuring the fact that the resulting FK amplitudes depend on this arbitrary scale. In order to cancel infrared divergences, it is sufficient for the dressing functions $f_{j}(p,k)$ to satisfy
\begin{gather}
    \lim_{|\vec{k}|\rightarrow 0}f_{j}(p,k)=\Big(\frac{p^{\mu} }{p \cdot k}-c^{\mu}\Big)\epsilon_{\mu,j}^*(k)\,\, ,
\end{gather}
where $c^{\mu}$ satisfies $c^2=0,c\.k=1$.
As a result, there exists an infinite family of possible soft dressings $f_{j}(p,k)$, leading to an inherent ambiguity in the amplitudes. This ambiguity is difficult to interpret: depending on the choice of soft dressing, certain analytic properties—such as the Steinmann relations—may hold for one dressing but fail for another, as explicitly demonstrated in \cite{Hannesdottir:2019opa}. Moreover, the Faddeev-Kulish states do not have a well-defined action under Poincar\'e transformations unless one restricts to in- and out-states with zero net electric charge \cite{Prabhu:2022zcr}.

It is well established that the FK approach yields infrared-finite amplitudes, but explicit expressions that include the finite part are scarce. In the 50 years since the original FK paper, few works appear to have computed explicit amplitudes within this framework \cite{Hannesdottir:2019opa,Hannesdottir:2019umk,Forde:2003jt}. A major challenge is that FK amplitudes inherit an infinite ambiguity, making it difficult to address questions about the analytic properties of the S-matrix. While the FK approach is correct in principle, its practicality for explicit amplitude computations and the study of the analytic properties of the S-matrix has been limited.

\subsubsection*{Recent developments: the infrared triangle}
Over the past decade, significant progress has been made in understanding the infrared structure of gauge theories and gravity, leading to what is now called the infrared triangle—a relationship between soft theorems, memory effects, and asymptotic symmetries (see \cite{Strominger:2017zoo} for a review). These developments have provided new insights into the role of FK states.

This progress has clarified that FK states are eigenstates of soft charges associated with the asymptotic symmetry group in abelian gauge theories \cite{Kapec:2017tkm}. In this picture, each scattering event is accompanied by a vacuum transition necessary for the conservation of asymptotic symmetry charges, with the dressings of the states encoding these transitions. A consequence of this viewpoint is that it singles out certain FK dressings as preferable, potentially providing a criterion for selecting dressings in amplitude computations. The dressing procedure extends standard scattering theory to incorporate the memory effect as a quantum observable. However, this approach to incorporating the memory effect fails in QED with massless charged particles, Yang–Mills theory, and quantum gravity—primarily due to collinear divergences and, in the latter two cases, the difficulty of constructing states with vanishing large gauge charge at spatial infinity \cite{Prabhu:2022zcr}. Motivated by these issues, \cite{Prabhu:2024lmg,Prabhu:2022zcr} advocate for an algebraic approach to scattering theory that does not rely on constructing FK-like Hilbert spaces.

\subsection{Overview}
\subsubsection*{Diagonalizing the asymptotic Hamiltonian}
The FK approach acts on free-particle states with asymptotic evolution operators (\ref{both operators}). Instead of using free-particle states, we work with states that diagonalize the asymptotic Hamiltonian. But what is the asymptotic Hamiltonian? In the case of $2\rightarrow 2$ scattering in SQED with one of the particles infinitely massive—while ignoring zero-energy photon emission and absorption—an appropriate asymptotic Hamiltonian would be that the lighter particle moves in the heavy particle’s Coulomb potential. This can be stated as the distributional equality
\begin{gather}
    \lim_{t\rightarrow -\infty}e^{-i\hat{H}t}\ket{\psi_{\t{in}}(p_H,p_L)}=\lim_{t\rightarrow -\infty}e^{-i\hat{H}_{\t{as}}t}\ket{\Phi_{\t{in}}(p_H,p_L)},\label{start}
\end{gather}
where $\hat{H}$ is the full SQED Hamiltonian (with no background field), and $\hat{H}_{\t{as}}$ is the Hamiltonian for a light charged scalar moving in the Coulomb background generated by the heavy particle, namely the Hamiltonian (\ref{hamil}), with the gauge field given as
\begin{gather}
    \hat{A}^{\mu}(x)=\int\DD^3\vec{q}\frac{q^{\mu}}{\sqrt{(q\.x)^2-q^2x^2}}\, \hat{\rho}_H(q),
\end{gather}
where $\hat{\rho}_H$ is the charge density operator (\ref{rho}) which measures the charge of the heavy particle. 

In (\ref{start}) the states $\ket{\psi_{\t{in}}(p_H,p_L)}$ and $ \ket{\Phi_{\t{in}}(p_H,p_L)}$ are eigenstates of $\hat{H}$ and $\hat{H}_{\t{as}}$, respectively, with the standard properties that they are eigenstates of the total momentum operator with total momentum $(p_H+p_L)^{\mu}$ and the in-state boundary condition that at early times they look as much as possible like two freely moving particles. Standard perturbation theory uses a similar relation to (\ref{start}), except the right-hand side involves free states and the free Hamiltonian (see chapter 3 of \cite{Weinberg:1995mt}), which is not valid as a distributional equality, leading to infrared divergences.  In section \ref{sect:canonical}, we construct the Hilbert space of states that diagonalizes $\hat{H}_{\t{as}}$, study the resulting IR-finite amplitudes and correlation functions, and show that they are consistent with causality. Once the RHS of (\ref{start}) is fully understood, this equality could serve as the starting point for a perturbative expansion in SQED with an infinitely massive particle, free from the Coulomb phase divergences. However, we do not develop such a perturbation theory in this paper. Instead, our focus is on fully solving the quantum theory on the RHS, just as solving the free theory is a prerequisite for computing observables in an interacting quantum field theory.

Diagonalizing even this asymptotic Hamiltonian is already nontrivial. Having established that it can be done, a natural next step would be to relax the assumption of an infinitely massive particle. This could be addressed using the two-particle relativistic quantum mechanics developed by Sazdjian \cite{sazdjian1986scalar}, which we leave for future work. This would also address the shortcoming that (\ref{start}) does not account for the recoil of the heavy particle. Similarly, squeezed states may provide a way to diagonalize late-time soft-photon emission and absorption (\ref{R operator}).

\subsubsection*{An important lesson from non-relativistic quantum mechanics}
The absence of arbitrary IR scales in this article follows from an important lesson in non-relativistic quantum mechanics: when solving the non-relativistic Coulomb scattering problem non-perturbatively, no ambiguous scales appear in the scattering amplitudes.

In non-relativistic quantum mechanics, the exact in and out wavefunctions $\phi_{\text{in/out}}(\vec{x},\vec{p})$ for two-particle Coulomb scattering can be obtained by solving the Schrödinger equation exactly \cite{Landau:1991wop}. These wavefunctions correspond to the non-relativistic limit of (\ref{Coulomb motivate}). The scattering amplitude is given by the inner product of the in-state with the out-state (see sections 2.1-2.2 of \cite{Lippstreu:2023vvg} for a summary):
\begin{align}
    A^{\text{NRQM}}(\vec{p}_1,\vec{p}_2) &=\int \mathrm{d}^3 x\, \phi^{\star}_{\text{out}}(\vec{x},\vec{p}_2)\phi_{\text{in}}(\vec{x},\vec{p}_1) \\
    &=-\lim_{\e_{+}\rightarrow 0}\frac{i\pi}{2}\frac{e_1e_2}{|\vec{p}_1|^2}\delta(E_1-E_2)\frac{\Gamma(1+i\gamma_1)}{\Gamma(1-i\gamma_1)}\Bigg(\frac{4|\vec{p}_1|^2}{|\vec{p}_1-\vec{p_2}|^2}\Bigg)^{1+i\gamma_1-\e_{+}}\, ,\label{NRQM amplitude}
\end{align}
where $\gamma_1=\frac{e_1e_2}{4\pi}\frac{m}{E_1}$. Note the presence of the $|\vec{p}_1|^2$ term in the numerator, which arises from computing the inner product of the in-state with the out-state. Note that (\ref{NRQM amplitude}) is an unambiguous expression, whereas approaches involving the Dollard operator (\ref{dollard operator}), a photon mass, or dimensional regularization would typically introduce an arbitrary scale $\Lambda_{\t{IR}}$ in this numerator.

An important fact about (\ref{NRQM amplitude}) is that no theorems from scattering theory or perturbation theory are used to derive the amplitude (\ref{NRQM amplitude})—such theorems would not apply in this case (see section 2.2 of \cite{Lippstreu:2023vvg} for further discussion on this point). The only assumption is the fundamental axiom of quantum mechanics: probability amplitudes are computed via inner products. By contrast, attempting to apply standard perturbation theory in this setup leads to regulator-dependent results, and standard perturbation theory does not resum to the correct amplitude, as evidenced by the fact that the perturbation series retains explicit dependence on the regulator \cite{kang1962higher,Dalitz:1951ah,kacser1959higher}.

This conclusion about no arbitrary scales being present in the amplitude extends to relativistic Coulomb scattering amplitudes as we explain in section \ref{sect:canonical}. In essence, the non-relativistic $\DD^3\vec{x}$ inner product is upgraded to the SQED inner product, and the non-relativistic wavefunctions are upgraded to the relativistic wavefunctions.

The $\epsilon_{+}>0$ in (\ref{NRQM amplitude}) is a positive infinitesimal quantity that does not introduce any ambiguity in the amplitude; rather, it represents the unique distributional interpretation of the amplitude consistent with unitarity as we review in section \ref{sect:unitarity}.

One might argue that the difference between an arbitrary scale in (\ref{NRQM amplitude}) and the natural scale set by the exact wavefunctions, $|\vec{p}|$, affects the scattering amplitude only by a pure phase and is therefore physically irrelevant, and could be arbitrarily altered by choosing to rephase the states. However, one could make the same claim about the ratio of Gamma functions \( \frac{\Gamma(1+i\gamma)}{\Gamma(1-i\gamma)} = e^{i\delta} \) in (\ref{NRQM amplitude}). Yet, in the \textit{complex} $|\vec{p}|$ plane, this ratio exhibits poles corresponding to the bound states of hydrogen, consistent with standard results on the analytic properties of the S-matrix \cite{Mizera_2024}. Likewise, different choices for the numerator in (\ref{NRQM amplitude}) alter the analytic properties of the amplitude in the complex kinematic plane, for example the residue of the amplitude at the poles of the Gamma functions. We discuss this further in section \ref{sect:scatter to bound}.

\subsubsection*{Outline} 
In section \ref{sect:norms and completeness relations}, we set up relativistic scattering in a Coulomb background. We review the scalar QED inner product, introduce relativistic Coulomb wavefunctions, and establish their completeness and orthogonality relations. Additional identities, proved in appendix \ref{app: further identities}, are provided for later use.

In section \ref{sect:canonical}, we canonically quantize a charged scalar field in a background Coulomb field, diagonalize the Hamiltonian, and define the Hilbert space of states. We compute scattering amplitudes and the Feynman Green’s function, and verify that correlation functions respect causality. Additionally, we develop time-independent and time-dependent perturbation theories to account for short-distance interactions.

In section \ref{sect:LSZ}, we review why LSZ reduction fails in theories with long-range forces and derive a modified version inspired by the strong background field theory approach. We then explicitly demonstrate, through examples, how this modified LSZ reduction can be applied to transition from the correlation functions in section \ref{sect:canonical} to the IR-finite scattering amplitudes unambiguously.

In section \ref{sect:unitarity}, we show that the scattering amplitude lacks the usual disconnected piece and explain why this remains consistent with unitarity. Since the general optical theorem relies on the disconnected contribution, it no longer applies in this case. Therefore, in section \ref{sect: modified elastic}, we derive an alternative version of the general optical theorem and demonstrate, through an example, that this modified optical theorem is useful for relating different orders of perturbation theory.

In section \ref{sect:crossing}, we review that the real radiative divergence and the Coulomb phase divergence are analytically related through crossing symmetry and an additional analytic condition—the absence of a pseudothreshold on the physical Riemann sheet. This result is significant because it suggests that a well-defined treatment of the Coulomb phase divergence also implies an unambiguous treatment of the real radiative divergence, provided the IR-finite S-matrix satisfies crossing symmetry.

In section \ref{sect:scatter to bound}, we show that the amplitude factorizes at its bound-state poles. Specifically, we demonstrate that the residue at a bound-state pole factorizes into a product of residues corresponding to scattering-to-bound amplitudes. This factorization depends on the IR scale chosen by the Coulomb wavefunctions.

Appendix \ref{app: completeness relations and identities} proves the completeness relations and other key identities satisfied by the Coulomb wavefunctions and used throughout the text. Appendix \ref{app:greens} verifies that the Green's function derived in the text satisfies the relevant Green's function equation. Appendix \ref{app:latetimedynamics} examines the late-time dynamics of plane-wave creation and annihilation operators in a background Coulomb field, demonstrating that the asymptotic dynamics are not free.

\textbf{Conventions:} We work in $(+,-,-,-)$ metric signature.

\section{Relativistic Coulomb wavefunctions: norms and completeness relations}\label{sect:norms and completeness relations}
This section summarizes the norms, completeness relations, and key identities of the relativistic Coulomb wavefunctions used throughout the paper. In section \ref{sect:SQED inner product}, we review the properties of the SQED inner product. The following sections summarize the completeness relations of the Coulomb wavefunctions in the spherical basis (section \ref{sect:spherical basis}) and the scattering basis (section \ref{sect: scattering basis}). These latter two sections can be skimmed, with the key takeaway being that we have a complete orthonormal basis of wavefunctions for charged scalars on a Coulomb background.
\subsection{Scalar QED inner product}\label{sect:SQED inner product}
Consider the equation of motion (EQM) for a scalar field $\psi$ on a fixed background gauge field $A^{\mu}$,
\begin{equation}
    (\Box +m^2+2ie_1A^{\mu}\d_{\mu})\psi= (e_1^2A^{\mu}A_{\mu})\psi.\label{eq:intmotion}
\end{equation}
We can introduce an ``inner product''\footnote{Despite this being common terminology, this is not strictly an inner product as it does not satisfy the positive-definiteness condition for negative-energy solutions.} on the space of solutions to this equation,
\begin{align}
\braket{\psi_1|\psi_2}_{\text{SQED}}&=i\int_{\Sigma}\DD^3\Sigma_{\mu}J^{\mu}\\
    &=i\int_{\Sigma}\DD^3\Sigma_{\mu}\Big(\psi_1^{\star}(x)\d^{\mu}\psi_2(x)-\psi_2(x)\d^{\mu}\psi^{\star}_1(x)+2ie_1 \psi_1^{\star}\psi_2 A^{\mu}\Big),\label{eq:defJ}
\end{align}
where $\Sigma$ is an arbitrary Cauchy surface and $\DD^3\Sigma_{\mu}$ is the volume element which is normal to the surface. We add a subscript SQED to the inner product brackets to differentiate this from the inner product on the Hilbert space of states in the quantum theory. The SQED inner product is independent of the Cauchy surface\footnote{So long as the current falls off sufficiently fast at the boundaries of the Cauchy surface.} because the current is divergence-free, $\d_{\mu}J^{\mu}=0$, on the support of the equations of motion (\ref{eq:intmotion}) when working in Lorenz gauge $\d_{\mu}A^{\mu}=0$. Furthermore, the inner product is Lorentz invariant because $J^{\mu}$ is a Lorentz vector. The $i$ prefactor is necessary to ensure that the inner product is Hermitian. Let us note that the same inner product can be used for the classical solution space of
\begin{equation}
    (\Box +m^2+2ie_1A^{\mu}\d_{\mu})\phi= 0,\label{eq:freemotion}
\end{equation}
as the reader can verify that the same $J^{\mu}$ in (\ref{eq:defJ}) is divergence-less on the support of (\ref{eq:freemotion}). \\
In this paper we will fix the background gauge field to be the field produced by a massive charged particle moving with constant four-velocity
\begin{equation}
    A^{\mu}(x)=\frac{e_s}{4\pi}\frac{u_s^{\mu}}{\sqrt{(u_s\.x)^2-x^2}},\label{eq:gauge field}
\end{equation}
where $u^{\mu}_s=\frac{p^{\mu}_s}{m_s}$ is the four-velocity of the source particle, and the gauge field (\ref{eq:gauge field}) satisfies the Lorenz gauge condition $\d_{\mu}A^{\mu}=0$. Although the exact scattering solutions to the full EQM \eqref{eq:intmotion} in the relativistic Coulomb background \eqref{eq:gauge field} are known (see Appendix A of \cite{Lippstreu:2023vvg} and references therein), we will instead utilize the solutions to the simpler equations of motion \eqref{eq:freemotion}, where the \( e_1^2 A^{\mu} A_{\mu} \psi \) interaction term is omitted. 

The reasons for this choice are twofold: (1) we are primarily concerned with understanding how infrared divergences impact the analytic properties of scattering amplitudes, and the \( A^2\psi \) term is a short-range contribution\footnote{For instance, in the rest frame of the source this term is proportional to $\frac{1}{r^2}$, which falls off sufficiently rapidly not to affect the in-state at asymptotically early times; see appendix \ref{app:latetimedynamics}, in particular (\ref{eq:latetimeshortpotential}). See also Appendix A of \cite{Lippstreu:2023vvg} for a non-perturbative verification.} that does not affect the infrared-divergent structure of the theory; (2) the exact in- and out-wavefunctions for the full EQM are only known in terms of their partial-wave expansions. While the results in this paper could be reformulated using the full solutions, they would have to be expressed in a form where a partial-wave sum remains unresolved, making the final conclusions less transparent.\\
Often, our expressions take their most concise form in the rest frame of the source particle, where \( u_s^{\mu} = \delta^{\mu}_0 \). Consequently, we introduce the shorthand \( u_s^r \) to denote the rest frame of the source. For example, the EQM \eqref{eq:freemotion} in this frame takes the form:

\begin{gather}
    \Bigg( \Box + m^2 + 2i\frac{\alpha}{r} \partial_t \Bigg) \phi(x, u_s^r) = 0,\label{EQMrestframe}\\
    \alpha = \frac{e_1 e_s}{4\pi}.
\end{gather}

\subsection{Spherically symmetric wavefunctions}\label{sect:spherical basis}
\subsubsection*{Unbound wavefunctions}
The unbound (continuum) spherically symmetric wavefunctions which solve the EQM in the rest frame of the source \eqref{EQMrestframe} are (to avoid notational clutter, we often omit the dependence of wavefunctions on the source four-velocity):
    \begin{gather}
f^{\rho}_{lm}(x,|\vec{p}|)=e^{-i\rho E t}Y_{lm}(\theta,\phi)R_{l}(|\vec{p}|r,\rho\gamma)\label{eq:fullcont}\\
    R_{l}(|\vec{p}|r,\rho\gamma)=e^{-\frac{\pi\rho\gamma}{2}}\frac{|\Gamma(1+l+i\rho\gamma)|}{(2l+1)!}(2|\vec{p}|r)^le^{-i|\vec{p}|r}{}_1F_1\Big(1+l-i\rho\gamma,2l+2,2i|\vec{p}|r\Big)\label{eq:unboundradialsphericallysymetric}\\
    \gamma=\alpha \frac{E}{|\vec{p}|},\quad E=\sqrt{|\vec{p}|^2+m^2},\quad \rho=\pm.\label{energy}
\end{gather}
Here, \( \rho = \pm 1 \) distinguishes between positive- and negative-energy (particle/antiparticle) wavefunctions, respectively, while \( E \) is always positive, with \( \rho \) encoding the energy sign. The wavefunctions are normalized according to the SQED inner product:
\begin{gather}
    \Big\langle f_{l'm'}^{\rho'}(|\vec{p}\,'|)\Big\vert f_{lm}^{\rho}(|\vec{p}|)\Big\rangle_{\t{SQED}}=\rho\,\delta_{\rho\rho'}\delta_{mm'}\delta_{ll'}\frac{\pi}{|\vec{p}|}\delta(E-E').\label{contsymmetrinorm}
\end{gather}
The computation of the inner product (\ref{contsymmetrinorm}) can be done using the method in Appendix B of \cite{Lippstreu:2023vvg}.
\subsubsection*{Bound-state wavefunctions}
The spherically symmetric bound-state solutions in the rest frame of the source (\ref{EQMrestframe}) are
\begin{gather}
    B_{nlm}(x,u_s^r)=e^{i\, \t{sign}[\alpha] E_nt}\,Y_{lm}(\theta,\phi)R_{nl}(r),\label{eq:fullbound}
\end{gather}
where the $\t{sign}[\alpha]$ indicates that for repulsive potentials $\alpha>0$ only antiparticle states have bound-state wavefunctions, and vice versa for attractive potentials $\alpha<0$. The radial wavefunctions read:
\begin{gather}
     R_{nl}(r)=N(n,l)(2\eta_n r)^le^{-\eta_n r}{}_1F_1(1+l+n,2l+2,2\eta_n r)\label{eq:boundstateradial}\\
    \eta_n=-\frac{|\alpha| }{n}E_{n},\qquad E_n= \frac{m}{\sqrt{1+\frac{\alpha^2}{n^2}}},\qquad N(n,l)=\frac{|\alpha|^{\frac{3}{2}}}{m }\frac{E_n^2}{n^2}\sqrt{2\frac{(l+n)!}{(n-l-1)!}}\frac{1}{(2l+1)!}\label{eq:etas}
    \\ n=l+1,l+2,\dots\qquad\quad  l=0,1,2,\dots\label{eq:normbound}
\end{gather}
The normalization is chosen so that 
\begin{gather}
    \braket{B_{nlm}|B_{n'l'm'}}_{\text{SQED}}=-\t{sign}[\alpha]\,\delta_{nn'}\delta_{ll'}\delta_{mm'}.\label{boundortho}
\end{gather}
The bound-state wavefunctions are also orthogonal to the unbound wavefunctions
\begin{gather}
\braket{B_{nlm}|f_{l'm'}^{\rho}(|\vec{p}|)}_{\t{SQED}}=0\label{boundorthounbound}.
\end{gather}
\subsubsection*{Completeness relation for the spherically symmetric solutions}
In Appendix \ref{app:spherecomplete}, we prove the equal-time completeness relations for the spherically symmetric relativistic Coulomb wavefunctions in the rest frame of the source
   \begin{align*}
   &\delta(x_0-x_0')\sum_{l=0}^{\infty}\sum_{m=-l}^l\Bigg(  \sum_{n=l+1}^{\infty} E_n B_{nlm}(x)B^{\star}_{nlm}(x')+\frac{1}{\pi}\sum_{\rho=\pm}\int_{0}^{\infty}\DD |\vec{p}|\,\,|\vec{p}|^2 f^{\rho}_{lm}(x,|\vec{p}|)f^{\star \rho}_{lm}(x',|\vec{p}|)\Bigg)\\
   &=\delta^4(x-x').\numberthis\label{Completespherical}
\end{align*}
An important distinction between the relativistic and non-relativistic cases is the necessity of antiparticles in relativistic theories to obtain a completeness relation. Consequently, bound states always contribute to the completeness relation in relativistic theories, as either the antiparticles or the particles will bind to the potential. In contrast, in the non-relativistic case, bound states contribute to the completeness relation only for attractive potentials.\\
Similar completeness relations are known for the wavefunctions of the full SQED equations of motion (\ref{eq:intmotion}), see for instance \cite{hostler1964coulomb}.
\subsection{Scattering wavefunctions}\label{sect: scattering basis}
The in/out scattering wavefunctions which solve the EQM in the rest frame of the source (\ref{EQMrestframe}) are
\begin{align}
f^{\rho}_{\text{in}}\Big(x,p,u_s^{r}\Big)&=\Gamma(1+i\rho\gamma)e^{-\frac{\pi\rho\gamma}{2}}e^{-i\rho Et+i\vec{p}\.\vec{x}}{}_1F_1\Big(-i\rho\gamma,1,i(|\vec{p}||\vec{x}|-\vec{p}\.\vec{x})\Big)\label{restframeinstate}\\
    f^{\rho}_{\text{out}}\Big(x,p,u_s^{r}\Big)&=\Gamma(1-i\rho\gamma)e^{-\frac{\pi\rho \gamma}{2}}e^{-i\rho Et+i\vec{p}\.\vec{x}}{}_1F_1\Big(i\rho\gamma,1,-i|\vec{p}||\vec{x}|-i\vec{p}\.\vec{x}\Big)\label{restframeoutstate}
    \end{align}
    \begin{equation}
        \gamma(p,u_s^r)\defined \frac{e_1e_2 E}{4\pi |\vec{p}|},\qquad E=\sqrt{|\vec{p}|^2+m^2},\label{def:gamma}
    \end{equation}
    and the covariant expression for these wavefunctions in an arbitrary reference frame can be found in  \cite{Lippstreu:2023vvg}.
    We can also express these in terms of the spherical wavefunctions
    \begin{gather}
    f^{\rho}_{\t{in}}(x,p,u_s^r)=4\pi\sum_{l,m}i^l\sqrt{\frac{\Gamma(1+l+i\rho\gamma)}{\Gamma(1+l-i\rho\gamma)}}f^{\rho}_{lm}(x,|\vec{p}|)Y^{\star}_{lm}(\hat{p})\label{partialwavescatter}\\
    f^{\rho}_{\t{out}}(x,p,u_s^r)=4\pi\sum_{l,m}i^l\sqrt{\frac{\Gamma(1+l-i\rho\gamma)}{\Gamma(1+l+i\rho\gamma)}}f^{\rho}_{lm}(x,|\vec{p}|)Y^{\star}_{lm}(\hat{p}),\label{partialwaveout}
\end{gather}
see, e.g., section 10 of chapter 11 of \cite{messiah1999quantum} for proofs of (\ref{partialwavescatter},\ref{partialwaveout}).
These are normalized to (see Appendix B of \cite{Lippstreu:2023vvg} for the computation)
\begin{align}
  \Big\langle f^{\rho}_{\t{in}}(\vec{p}\,')\Big\vert f^{\rho'}_{\t{in}}(\vec{p})\Big\rangle_{\t{SQED}}= \Big\langle f^{\rho}_{\t{out}}(\vec{p}\,')\Big\vert f^{\rho'}_{\t{out}}(\vec{p})\Big\rangle_{\t{SQED}}
  &=\rho\delta_{\rho\rho'}(2\pi)^3 2E\delta^3(\vec{p}-\vec{p}').\label{normofinout}
\end{align}
The scattering wavefunctions are also orthogonal to the bound-state wavefunctions
\begin{gather}
\big\langle B_{nlm}\,\big|\,f^{\rho}_{\t{in/out}}(\vec{p})\big\rangle_{\t{SQED}}=0\label{boundscatterortho}.
\end{gather}
The inner product of the in-state with the out-state in the rest frame of the source reads
\begin{align}
    \Big\langle f^{\rho}_{\t{out}}(\vec{p}_2,u_s^r)\Big\vert f^{\rho}_{\t{in}}(\vec{p}_1,u_s^r)\Big\rangle_{\t{SQED}}&=  \mathcal{A}^{\rho}_0\Big(p_1\rightarrow p_2;u^{r}_s\Big)\\
    &=-i\rho\gamma\frac{4\pi^2}{|\vec{p}_1|} \delta(E_1-E_2) \frac{\Gamma(1+i\rho\gamma)}{\Gamma(1-i\rho\gamma)}\Bigg(\frac{2|\vec{p}_1|}{|\vec{p}_1-\vec{p}_2|}\Bigg)^{2+2i\rho\gamma},\label{treerestfframe}
\end{align}
where the first equality is the definition of $\mathcal{A}^{\rho}_0\Big(p_1\rightarrow p_2;u^{r}_s\Big)$. Note the important fact that the inner product (\ref{treerestfframe}) does not contain an IR divergence, an arbitrary cutoff scale, or an arbitrary dimensional-regularization scale $\mu$. The covariant expression for the inner product in an arbitrary reference frame is
\begin{align*}
    &\mathcal{A}^{\rho}_0\Big(p_1\rightarrow p_2;u_s\Big)\\
    &=i\pi\, \rho \,e e_s\delta\Big(u_s\.(p_1-p_2)\Big)\frac{\Gamma(1+i\rho\gamma)}{\Gamma(1-\rho i\gamma)}\frac{p_1\.u_s}{m^2-(p_1\.u_s)^2}\Bigg(\frac{4\big(m^2-(p_1\.u_s)^2\big)}{(p_1-p_2)^2}\Bigg)^{1+i\rho\gamma},\numberthis\label{eq:LOcovamp}
\end{align*}
where $\gamma$ in an arbitrary reference frame reads
\begin{gather}
    \gamma(p, u_s) = \frac{e_p e_s}{4\pi} \frac{p \cdot u_s}{\sqrt{(p \cdot u_s)^2 - p^2}}.\label{gamma}
\end{gather}
\subsubsection*{Completeness relations for scattering solutions}
In Appendix \ref{app:scattercomplete}, we prove the equal-time completeness relation of the in/out scattering solutions and bound-state wavefunctions in the rest frame of the source,
    \begin{align*}
   &\delta(x_0-x_0')\Bigg( \sum_{l=0}^{\infty}\sum_{m=-l}^l \sum_{n=l+1}^{\infty} E_n B_{nlm}(x)B^{\star}_{nlm}(x')+\frac{1}{2}\sum_{\rho=\pm}\int \frac{\DD^3\vec{p}}{(2\pi)^3}\, f_{\t{in}}^{\rho}(x,\vec{p})f_{\t{in}}^{\star\rho}(x',\vec{p})\Bigg)\\
   &=\delta^4(x-x').\numberthis\label{Completescattermain}
\end{align*}
Note that we can take either the in-states or the out-states in the completeness relation (\ref{Completescattermain}) as both sets of wavefunctions form a complete basis when combined with the bound states.

\section{Canonical quantization in a Coulomb background}\label{sect:canonical}
In standard QFT with only short-range interactions, one first solves the free theory completely before using a relation like (\ref{start}) to develop perturbation theory for the interacting case. However, in quantum field theories with long-range interactions, the asymptotic Hamiltonian is not the free Hamiltonian. Infrared divergences arise precisely because standard perturbation theory incorrectly assumes that it is.

A natural approach to handling IR divergences then is to fully solve the dynamics of the asymptotic Hamiltonian. Once this is done, a relation like (\ref{start}) could be used to develop a perturbation theory for the fully interacting theory without IR divergences. However, we do not develop such a perturbation theory in this paper. Instead, we focus on the first step: solving the quantum theory associated with the asymptotic Hamiltonian.

In SQED, the asymptotic Hamiltonian consists of two components: (1) the continual absorption and emission of low-energy photons and (2) the Coulomb interaction between particle pairs. In this section, we focus exclusively on the late-time Coulomb interaction and ignore the dynamics of soft radiation. To further simplify the problem, we assume that one of the charged particles is infinitely massive. We then fully solve the quantum theory governing the asymptotic dynamics of such particle pairs in SQED. 

It would be interesting to move beyond the infinite mass limit using the two-particle relativistic quantum mechanics developed by Sazdjian \cite{sazdjian1986scalar}. In section \ref{sect:crossing}, we show that the Coulomb phase divergence is analytically related to the real radiative divergence through crossing symmetry. This suggests that insights into the Coulomb phase divergence could be leveraged to better understand the real radiative divergence.

\subsection{Diagonalizing the Hamiltonian}
The Hamiltonian density for a charged scalar field in a fixed background gauge field is
\begin{gather}
    \mathcal{H}=\pi^{\dagger}\pi+ieA^0(\pi^{\dagger}\phi^{\dagger}-\phi\pi)+\phi^{\dagger}(-\vec{D}^2+m^2)\phi+e^2A^{\mu}A_{\mu}\phi^{\dagger}\phi,\label{hamil}
\end{gather}
where the spatial covariant derivative is defined by
\begin{gather}
    \vec{D}^2\phi=(\vec{\d}-ie\vec{A})\.(\vec{\d}-ie\vec{A})\phi.
\end{gather}
The term \( e^2A^{\mu}A_{\mu}\phi^{\dagger}\phi \) has been included in (\ref{hamil}) to remove the quartic interaction from this Hamiltonian, as will become clear shortly. The quartic term will be reintroduced later as a perturbation.
The momentum conjugates \(\pi, \pi^{\dagger}\) are obtained from Hamilton's equations,  
\(\dot{\phi} = \frac{\delta \mathcal{H}}{\delta \pi}\), \(\dot{\phi}^{\dagger} = \frac{\delta \mathcal{H}}{\delta \pi^{\dagger}}\),  
which yield  
\begin{alignat}{2}  
    \pi^{\dagger} &= (\partial_0 + ie A^0) \phi &&= D_0 \phi, \label{eq:pi} \\  
    \pi &= (\partial_0 - ie A^0) \phi^{\dagger} &&= (D_0 \phi)^{\dagger}. \label{eq:pidagger}  
\end{alignat}  
Applying Hamilton's second equation, \(\dot{\pi}^{\dagger} = -\frac{\delta \mathcal{H}}{\delta \phi^{\dagger}}\), and fixing the background field to Lorenz gauge, \(\partial_{\mu} A^{\mu} = 0\), we obtain the SQED equation of motion with the quartic interaction omitted:
\begin{gather}  
    \Big(\Box + m^2 + 2ieA^{\mu} \partial_{\mu} \Big) \phi = 0. \label{semi free}  
\end{gather}  
We now specialize to the Coulomb field by setting the background gauge field to  
\begin{equation}
    A^{\mu}(x,u_s)=\frac{e_s}{4\pi}\frac{u_s^{\mu}}{\sqrt{(u_s\.x)^2-x^2}}.\label{eq:Coulomb field}
\end{equation}
The scattering wavefunctions \( f_{\text{in/out}, \pm} \) are given in (\ref{restframeinstate}, \ref{restframeoutstate}), while the bound-state wavefunctions \( B_{nlm}(x) \) are given in (\ref{eq:fullbound}). Together, they form a complete basis for functions in \( L^2(\mathbb{R}^3) \), as shown in (\ref{Completescatter}). Therefore, any solution to the semi-free equation of motion (\ref{semi free}) can be expanded as
\begin{gather}
    \hat{\phi}(\vec{x},t)=\int\frac{\DD^3\vec{p}}{(2\pi)^3 2E_p}\Bigg(\hat{a}_{\t{in}}(\vec{p})f_{\t{in},+}(x,p)+\hat{b}^{\dagger}_{\t{in}}(\vec{p})f_{\t{in},-}(x,p)\Bigg)+\sum_{n,l,m}\hat{c}_{nlm}B_{nlm}(x).\label{expansion}
\end{gather}
Since both the in-wavefunctions with bound states and the out-wavefunctions with bound states form a complete basis, we are free to expand the field using either set. We will demonstrate and exploit this flexibility throughout.
The placement of the dagger on \(\hat{c}_{nlm}\) depends on the sign of \(\alpha = \frac{ee_s}{4\pi}\):  
\begin{itemize}  
    \item If \(\alpha > 0\), bound states correspond to antiparticles, so \(\hat{c}_{nlm}\) carries a dagger.
    \item If \(\alpha < 0\), bound states correspond to particles, so \(\hat{c}_{nlm}\) remains undaggered.  
\end{itemize}  

This convention is consistent with our choice of daggers on \(\hat{a}_{\text{in}}\) and \(\hat{b}_{\text{in}}^{\dagger}\), ensuring that \(\hat{\phi}\) annihilates particles and creates antiparticles when acting on the vacuum \(\ket{\Omega}\). For simplicity, we will assume \(\alpha < 0\) throughout, so the undaggered \(\hat{c}_{nlm}\) in (\ref{expansion}) is correct.

We now specialize to the rest frame of the source particle, where the gauge field takes the form  
\begin{gather}  
    A^{\mu}(u_s^r) = \frac{e_s}{4\pi} \frac{\delta^{\mu}_0}{r}. \label{eq:Coulomb field rest}  
\end{gather}  
We work in this frame throughout, as the Hamiltonian is only diagonal in this frame. Covariantly, this means that \( u_s \cdot p \) is the only conserved component of the probe particle's momentum.

The quantum field in (\ref{expansion}) is in the Heisenberg picture, meaning all time dependence is carried by the operators. However, the wavefunctions \( f_{\text{in},\pm} \) and \( B_{nlm} \) in (\ref{expansion}) already include their \( e^{-i\rho E t} \) time dependence, while the creation and annihilation operators remain time-independent. This may seem inconsistent, but we will soon verify that the Hamiltonian is diagonal in this basis. As a result, the Heisenberg operators evolve as  
\begin{gather}  
    \hat{a}(\vec{p}, t) = e^{-iE_p t} \hat{a}(\vec{p}, 0),  
\end{gather}  
with a similar relation for the other operators. In (\ref{expansion}), this time dependence has been absorbed into the wavefunctions, ensuring a consistent Heisenberg picture.

We now verify that the mode expansion (\ref{expansion}) diagonalizes the Hamiltonian. Substituting the expressions for the canonical momentum conjugates \(\pi, \pi^{\dagger}\) from (\ref{eq:pi}, \ref{eq:pidagger}) and using the equation of motion (\ref{semi free}), we find that the Hamiltonian density in the rest frame of the source, \( u_s^r \), takes the form  
\begin{align}  
    \mathcal{H}(u_s^r)  
    &=\,\,: \dot{\phi} \dot{\phi}^{\dagger} - \phi^{\dagger} \ddot{\phi} - 2ie \phi^{\dagger} A^0 \dot{\phi} :\,\, , \label{looks sqed}  
\end{align}  
where \( :\mathcal{O}(x): \) denotes normal ordering when quantizing. It is useful to recognize that (\ref{looks sqed}) resembles the SQED inner product, with an extra time derivative that introduces an additional factor of energy. Substituting the mode expansion (\ref{expansion}), the Hamiltonian produces six types of terms: \( a^{\dagger}a, b^{\dagger}b, c^{\dagger}c \), and the cross terms \( ab, ac, bc \). The coefficients of the cross terms vanish, while the diagonal terms remain.  For example, defining \( \mathcal{H}_{aa} \) as the contribution from \( a^{\dagger}a \) terms in \( \mathcal{H} \), we find that (\ref{looks sqed}) gives  
\begin{align}
   \int\DD^3\vec{x} \,\, \mathcal{H}_{aa}(u_s^r)&=\int\frac{\DD^3\vec{p}\,\,\DD^3\vec{q}}{(2\pi)^6 4E_pE_q}\hat{a}^{\dagger}_{\t{in}}(\vec{p})\hat{a}_{\t{in}}(\vec{q})\,E_q\,\braket{f_{\t{in},+}(p)|f_{\t{in},+}(q)}_{\t{SQED}}\\
    &=\frac{1}{2}\int\frac{\DD^3\vec{p}}{(2\pi)^3 }\hat{a}^{\dagger}_{\t{in}}(\vec{p})\hat{a}_{\t{in}}(\vec{p}).\label{aa terms}
\end{align}
Here, we used \( \dot{f}_{\text{in},+}(x, p) = -iE_p f_{\text{in},+}(x, p) \) and the orthonormality of these wavefunctions with respect to the SQED inner product (\ref{normofinout}).  The presence of the SQED inner product now makes it evident why the coefficients of the \( ab, ac, bc \) terms vanish—they do so because the corresponding inner products vanish. Applying the same calculation as in (\ref{aa terms}) to the \(\mathcal{H}_{bb}\) and \(\mathcal{H}_{cc}\) terms, we obtain  
\begin{align}
    \hat{H}_0(u_s^r)&=\int\DD^3\vec{x} \,\, \mathcal{H}(u_s^r)\\
    &=\frac{1}{2}\int\frac{\DD^3\vec{p}}{(2\pi)^3 }\Big(\hat{a}^{\dagger}_{\t{in}}(\vec{p})\hat{a}_{\t{in}}(\vec{p})+\hat{b}^{\dagger}_{\t{in}}(\vec{p})\hat{b}_{\t{in}}(\vec{p})\Big)+\sum_{nlm}E_n\hat{c}^{\dagger}_{nlm}\hat{c}_{nlm},\label{diagonal H}
\end{align}
which confirms that the Hamiltonian is diagonal, as expected.  We emphasize that one could just as well expand half the fields in the Hamiltonian (\ref{looks sqed}) in the in-basis and the other half in the out-basis, or use the out-basis for both. In all cases, the result remains the same as (\ref{diagonal H}). However, explicitly demonstrating this requires the Bogoliubov transformation coefficients, which we discuss in the next section.  
\subsubsection*{Canonical commutation relations}
To construct the Hilbert space of states and verify that (\ref{diagonal H}) acts as expected, we require the canonical commutation relations of the mode coefficients. The commutation relations for the in- and out-operators are  
\begin{alignat}{3}
     &[\hat{a}_{\text{in}}(p_1),\hat{a}^{\dagger}_{\text{in}}(p_2)]&&=[\hat{b}_{\text{in}}(p_1),\hat{b}^{\dagger}_{\text{in}}(p_2)]&&=(2\pi)^32E_p\delta^3(\vec{p}_1-\vec{p}_2)\label{canonical for a}\\
     &[\hat{a}_{\text{out}}(p_1),\hat{a}^{\dagger}_{\text{out}}(p_2)]&&=[\hat{b}_{\text{out}}(p_1),\hat{b}^{\dagger}_{\text{out}}(p_2)]&&=(2\pi)^32E_p\delta^3(\vec{p}_1-\vec{p}_2),
\end{alignat}
and the bound-state commutation relations read
\begin{gather}
    [\hat{c}_{nlm},\hat{c}_{nlm}^{\dagger}]=\delta_{nn'}\delta_{ll'}\delta_{mm'}.
\end{gather}
The following commutation relations vanish:
\begin{gather}
  [\hat{a}_{\text{in}}(p_1),\hat{a}_{\text{in}}(p_2)]= [\hat{b}_{\text{in}}(p_1),\hat{b}_{\text{in}}(p_2)] = [\hat{a}_{\text{in}}(p_1),\hat{b}_{\text{in}}(p_2)]= [\hat{a}_{\text{in}}(p_1),\hat{b}^{\dagger}_{\text{in}}(p_2)]=0\\
  [\hat{a}_{\text{in}}(p),\hat{c}_{nlm}]=[\hat{b}_{\text{in}}(p),\hat{c}_{nlm}]=0,\label{zeros}
\end{gather}
The same holds for the Hermitian conjugates of these equations and when every $\t{in}$ is replaced by $\t{out}$.
\subsubsection*{Hilbert space of Coulomb states}
We now define the in- and out-vacua as the states annihilated by the corresponding annihilation operators:  
\begin{gather}
    \hat{a}_{\t{in}}(p)\ket{\Omega_{\t{in}}}=\hat{b}_{\t{in}}(p)\ket{\Omega_{\t{in}}}=\hat{c}_{nlm}\ket{\Omega_{\t{in}}}=0\\
    \hat{a}_{\t{out}}(p)\ket{\Omega_{\t{out}}}=\hat{b}_{\t{out}}(p)\ket{\Omega_{\t{out}}}=\hat{c}_{nlm}\ket{\Omega_{\t{out}}}=0,
\end{gather}
for all \(\vec{p}\). In the next section, we use the Bogoliubov coefficients to show that the in- and out-vacua are related by a phase. We construct the one-particle in-states as  
\begin{gather}
    \ket{\Phi^{+}_{\t{in}}(p)}=\hat{a}_{\t{in}}^{\dagger}(\vec{p})\ket{\Omega_{\t{in}}}\\
    \ket{\Phi^{-}_{\t{in}}(p)}=\hat{b}_{\t{in}}^{\dagger}(\vec{p})\ket{\Omega_{\t{in}}},
\end{gather}
where the \( + \) and \( - \) notation denotes particles and antiparticles, respectively. One could construct the full Fock space by taking direct products of these states, but we will not need to go beyond single-particle states. Similarly, the one-particle out-states are given by
\begin{gather}
      \ket{\Phi^{+}_{\t{out}}(p)}=\hat{a}_{\t{out}}^{\dagger}(\vec{p})\ket{\Omega_{\t{out}}}\\
      \ket{\Phi^{-}_{\t{out}}(p)}=\hat{b}_{\t{out}}^{\dagger}(\vec{p})\ket{\Omega_{\t{out}}}.
\end{gather}
It is now straightforward to verify that the Hamiltonian (\ref{diagonal H}) acts as expected on these states using the canonical commutation relations (\ref{canonical for a}). For example, acting on a one-particle in-state gives  
\begin{gather}
    \hat{H}_0  \ket{\Phi^{+}_{\t{in}}(p)}=E_p  \ket{\Phi^{+}_{\t{in}}(p)},
\end{gather}
as expected.
\subsection{Bogoliubov coefficients and scattering amplitudes}\label{sect:bogo}  
To analyze the stability of the vacuum and determine the commutator 
\begin{gather}
       [\hat{a}_{\text{out}}(p_1),\hat{a}^{\dagger}_{\text{in}}(p_2)],
\end{gather}
we need to express the out-operators in terms of the in-operators via the Bogoliubov transformation. Starting from the expansion of \(\hat{\phi}(x,t)\) in terms of the out-basis, (\ref{out expansion}), we obtain
\begin{gather}
    \hat{a}_{\t{out}}(p)=\braket{f_{\t{out},+}(p)|\hat{\phi}(x,t)}_{\t{SQED}}\label{bogo coefficient},
\end{gather}
which follows from the orthonormality of the out-wavefunctions (\ref{normofinout}) and (\ref{boundscatterortho}). Expressing \(\hat{\phi}(x,t)\) in (\ref{bogo coefficient}) instead in the in-basis (\ref{expansion}) leads to the Bogoliubov transformation
\begin{align}
    \hat{a}_{\t{out}}(p)&=\int\frac{\DD^3\vec{q}}{(2\pi)^3 2E_q}\braket{f_{\t{out},+}(p)|f_{\t{in},+}(q)}_{\t{SQED}}\hat{a}_{\t{in}}(q)\\
    &=\int\frac{\DD^3\vec{q}}{(2\pi)^3 2E_q}\mathcal{A}_0^{+}(q\rightarrow p)\hat{a}_{\t{in}}(q),\label{out as in}
\end{align}
where \(\mathcal{A}_0^{+}(q\rightarrow p)\), given in (\ref{treerestfframe}), is defined as the SQED inner product of the in-states with the out-states. This confirms the well-known result that the Bogoliubov coefficient corresponds to the scattering amplitude.  Similarly, for antiparticles, we obtain
\begin{align}
    \hat{b}_{\text{out}}(p)&=-\braket{\hat{\phi}^{\dagger}(x,t)|f_{\t{out},-}(p_1)}_{\t{SQED}}\\
    &=-\int\frac{\DD^3\vec{q}}{(2\pi)^3 2E_q}\braket{f_{\t{in},-}(q)|f_{\t{out},-}(p)}_{\t{SQED}}\hat{b}_{\t{in}}(\vec{q})\\
    &=-\int\frac{\DD^3\vec{q}}{(2\pi)^3 2E_q}\mathcal{A}^{-,\star}_0(q\rightarrow p)\,\hat{b}_{\t{in}}(\vec{q})\label{bb}.
\end{align}
These transformations lead to the commutation relations
\begin{align}
     [\hat{a}_{\text{out}}(p_1),\hat{a}^{\dagger}_{\text{in}}(p_2)]&=\mathcal{A}^{+}_0(p_2\rightarrow p_1)\label{commutation a out with a in}\\
     [\hat{b}_{\text{out}}(p_1),\hat{b}^{\dagger}_{\text{in}}(p_2)]&=-\mathcal{A}^{-,\star}_0(p_2\rightarrow p_1).
\end{align}
These follow directly by substituting (\ref{out as in}) into the commutators and using the canonical relations (\ref{canonical for a}).
\subsubsection*{Vacuum stability}
The Bogoliubov coefficients also determine how the in-vacuum relates to the out-vacuum. Computing the expectation value of the out-number operator in the in-vacuum, we find
\begin{gather}
\braket{\Omega_{\t{in}}|\hat{a}^{\dagger}_{\t{out}}\hat{a}_{\t{out}}+\hat{b}^{\dagger}_{\t{out}}\hat{b}_{\t{out}}|\Omega_{\t{in}}}=0.
\end{gather}
Since the expectation value is zero, the in-vacuum contains no out-modes and is annihilated by the out-annihilation operators, as can be seen from (\ref{out as in}, \ref{bb}). This implies that the in-vacuum can be identified with the out-vacuum up to a phase:
\begin{gather}
   | \braket{\Omega_{\text{out}}|\Omega_{\text{in}}}|=1\label{vacuum transtion}\\
   \braket{\Omega_{\text{out}}|\Omega_{\text{in}}}=c_v.
\end{gather}
We often omit the vacuum transition phase factor from amplitudes, as it is common to all amplitudes and therefore not physically measurable.
In full SQED, where charged particles can radiate soft photons, the in-to-out vacuum transition amplitude is generally not a pure phase. Instead, as shown in \cite{Kapec:2017tkm}, this transition is necessary to account for large gauge charge conservation.  We also note that (\ref{vacuum transtion}) ceases to be a pure phase when the quartic interaction term \(|\phi|^2 A^2\) is reinstated, and we enter the strong coupling regime \(\alpha > \frac{1}{2}\). In this regime, the Coulomb field becomes supercritical, meaning it is strong enough to destabilize the vacuum and induce spontaneous particle-antiparticle pair production. See \cite{GreinerMuellerRafelski1985} for a discussion of pair production in strong Coulomb fields.  

\subsubsection*{Scattering amplitudes}
We now compute the scattering amplitudes. The explicit expression for \(\mathcal{A}_{0}^{+}(p \to q)\) is given in (\ref{treerestfframe}), and the antiparticle amplitude follows similarly, allowing both amplitudes to be written as
\begin{align}
    \braket{\Phi^{\rho}_{\t{out}}(q)|\Phi^{\rho}_{\t{in}}(p)}c_{v}^{-1}
    &=\mathcal{A}_{0}^{\rho}(p\rightarrow q).\label{the scattering amplitude}
\end{align}
A key feature of (\ref{the scattering amplitude}) is that it is free of arbitrary scales: it does not depend on a soft cutoff \(\Lambda_{\text{IR}}\) or an arbitrary dimensional regularization scale \(\mu\). Instead, using Coulomb wavefunctions naturally sets the infrared scale to \((u_s \cdot p) - m^2\).

\subsubsection*{Expanding in the in- or out-basis}
It was mentioned below (\ref{diagonal H}) that we could expand each \(\hat{\phi}\) in either the in- or out-basis while still obtaining the same Hamiltonian (\ref{diagonal H}). To see that this flexibility extends to any composite operator or correlation function, we note that if we expand \(\hat{\phi}(x,t)\) in terms of the out-basis instead, we obtain
\begin{align*}
    \hat{\phi}(\vec{x},t)&=\int\frac{\DD^3\vec{p}}{(2\pi)^3 2E_p}\Bigg(\hat{a}_{\t{out}}(\vec{p})f_{\t{out},+}(x,p)+\hat{b}^{\dagger}_{\t{out}}(\vec{p})f_{\t{out},-}(x,p)\Bigg)+\sum_{n,l,m}\hat{c}_{nlm}B_{nlm}(x)\numberthis\label{out expansion}\\
    &=\int\frac{\DD^3\vec{p}\DD^3\vec{q}}{(2\pi)^6 4E_pE_q}\Bigg(\mathcal{A}^{+}_0(q\rightarrow p)\hat{a}_{\t{in}}(\vec{q})f_{\t{out},+}(x,p)\label{halfways}\\
    &\hspace{3.3cm}-\mathcal{A}^{-}_0(q\rightarrow p)\hat{b}^{\dagger}_{\t{in}}(\vec{q})f_{\t{out},-}(x,p)\Bigg)+\sum_{n,l,m}\hat{c}_{nlm}B_{nlm}(x)\numberthis\\
    &=\int\frac{\DD^3\vec{q}}{(2\pi)^3 2E_q}\Bigg(\hat{a}_{\t{in}}(\vec{q})f_{\t{in},+}(x,q)+\hat{b}^{\dagger}_{\t{in}}(\vec{q})f_{\t{in},-}(x,q)\Bigg)+\sum_{n,l,m}\hat{c}_{nlm}B_{nlm}(x),\numberthis\label{expand in out}
\end{align*}
where in (\ref{halfways}), we used the Bogoliubov transformations (\ref{out as in}, \ref{bb}), and in the final step, we applied the relation  
\begin{gather}
     f^{\rho}_{\t{out}}(p,x)=\rho\int\frac{\DD^3 \vec{q}}{(2\pi)^32E_q}\mathcal{A}^{\star,\rho}_0(q\rightarrow p)f^{\rho}_{\t{in}}(q,x).\label{relation}
\end{gather}
The equality (\ref{expand in out}) confirms that all results remain unchanged regardless of whether the field is expanded in the in- or out-basis. The relation (\ref{relation}) follows by assuming that \( f^{\rho}_{\text{out}}(p,x) \) can be expressed in terms of in-scattering wavefunctions \( f^{\rho}_{\text{in}}(q,x) \) and bound states, then solving for the coefficients using the orthogonality relations of the wavefunctions. Thus, from (\ref{expand in out}), we see that the choice of in- or out-basis for expanding the field is arbitrary. 
\subsection{Equal-time commutation relations}
As a consistency check, we verify that the canonical commutation relations of the creation and annihilation operators (\ref{canonical for a}-\ref{zeros}) correctly imply the equal-time commutation relation:
\begin{gather}
    [\hat{\phi}(t,\vec{x}),\hat{\pi}(t,\vec{y})]=i\delta^{(3)}(\vec{x}-\vec{y}).\label{equal time commutation}
\end{gather}
Using the definition of the canonical momentum (\ref{eq:pidagger}), we write
\begin{align*}
    &\pi(y)=(\d_0-ieA_0)\phi^{\dagger}(y)\numberthis\\
    &=i\int\frac{\DD^3\vec{p}}{(2\pi)^3 2}\Bigg(\hat{a}^{\dagger}_{\t{in}}(\vec{p})f^{\star}_{\t{in},+}(y,p)-\hat{b}_{\t{in}}(\vec{p})f^{\star}_{\t{in},-}(y,p)\Bigg)+i\sum_{n,l,m}E_n\hat{c}^{\dagger}_{nlm}B^{\star}_{nlm}(y)\\
    &-ieA_0\Bigg(\int\frac{\DD^3\vec{p}}{(2\pi)^3 2E_p}\Bigg(\hat{a}^{\dagger}_{\t{in}}(\vec{p})f^{\star}_{\t{in},+}(y,p)+\hat{b}_{\t{in}}(\vec{p})f^{\star}_{\t{in},-}(y,p)\Bigg)+\sum_{n,l,m}\hat{c}^{\dagger}_{nlm}B^{\star}_{nlm}(y)\Bigg),\numberthis\label{2 terms}
\end{align*}
where in the second line we used
\begin{gather}
    \d_0f_{\t{in},\rho}=-i\rho E f_{\t{in},\rho}\\
    \d_0B_{nlm}=i\,\t{sign}[\alpha]E_nB_{nlm},
\end{gather}
These relations hold only in the rest frame of the source. We also took $\alpha<0$, as discussed below (\ref{expansion}). When we take the commutator (\ref{equal time commutation}), there are two terms:
\begin{align}
      [\hat{\phi}(t,\vec{x}),\hat{\pi}(t,\vec{y})]=I_1+I_2  ,
\end{align}
where \( I_1 \) corresponds to the first term in (\ref{2 terms}), and \( I_2 \) corresponds to the second line with the \(-ie A_0\) prefactor in (\ref{2 terms}). For $I_2$, we find
\begin{align*}
    I_2&=-ieA_0\Bigg(\int\frac{\DD^3\vec{p}}{(2\pi)^3 2E_p}\Bigg(f_{\t{in},+}(x,p)f^{\star}_{\t{in},+}(y,p)-f_{\t{in},-}(x,p)f^{\star}_{\t{in},-}(y,p)\Bigg)\\
    &\hspace{2cm}+\sum_{n,l,m}B_{nlm}(x)B^{\star}_{nlm}(y)\Bigg)\numberthis\\
    &=0,\numberthis\label{zerohero}
\end{align*}
where in the first line we used the canonical commutation relations (\ref{canonical for a}-\ref{zeros}). In appendix \ref{app: further identities}, we demonstrate that the second line indeed vanishes. Next, we compute $I_1$:
\begin{align}
    I_1&=\frac{i}{2}\int\frac{\DD^3\vec{p}}{(2\pi)^3 }\Bigg(f_{\t{in},+}(x,p)f^{\star}_{\t{in},+}(y,p)+f_{\t{in},-}(x,p)f^{\star}_{\t{in},-}(y,p)\Bigg)+i\sum_{n,l,m}E_nB_{nlm}(x)B^{\star}_{nlm}(y)\\
    &=i\delta^{(3)}(\vec{x}-\vec{y}),
\end{align}
This is precisely the completeness relation (\ref{Completescattermain}), proved in appendix \ref{app:scattercomplete}. We have therefore verified the equal-time commutation relation (\ref{equal time commutation}).
\subsection{Feynman Green's function}
We now compute the time-ordered Green’s function, also known as the Green’s function with Feynman boundary conditions,
\begin{align}
  G_F(x,y)&=  \braket{\Omega|T(\phi^{\dagger}(x)\phi(y))|\Omega}\\
  &=\theta(x^0-y^0)\braket{\Omega|\phi^{\dagger}(x)\phi(y)|\Omega}+\theta(y^0-x^0)\braket{\Omega|\phi(y)\phi^{\dagger}(x)|\Omega},
\end{align}
where we omit the $\ket{\Omega_{\t{in/out}}}$ labels on the vacuum, as they affect only the overall phase of the Green’s function. Using the expansion of the fields in terms of the in-operators (\ref{expansion}) and the in-state commutation relations, we find
\begin{align*}
    G_F(x,y)
    &=\int\frac{\DD^3\vec{p}}{(2\pi)^3 2E_p}\Bigg(\theta(x^0-y^0)f^{\star}_{\t{in},-}(x,p)f_{\t{in},-}(y,p)+\theta(y^0-x^0)f^{\star}_{\t{in},+}(x,p)f_{\t{in},+}(y,p)\Bigg)\\
    &+\theta(y^0-x^0)\sum_{nlm}B_{nlm}(y)B^{\star}_{nlm}(x).\numberthis\label{greens in main text}
\end{align*}
In appendix \ref{app:greens}, we verify that (\ref{greens in main text}) satisfies the Green's function equation:
\begin{gather}
       \left( \Box + m^2 + 2i\frac{\alpha}{r} \partial_t \right) G_F(x, y, u_s^r) = -i \delta^{(4)}(x - y). \label{deltagreen}
\end{gather}
Note that (\ref{greens in main text}) admits the usual interpretation as in free field QFT---for $x^0>y^0$ we have antiparticles propagating backwards in time and for $x^0<y^0$ we have particles traveling from $x$ to $y$ forwards in time. The \(\theta\)-function multiplying the bound-state term also follows this interpretation. Since we assumed \(\alpha < 0\), bound states correspond to particles, not antiparticles. As particles only propagate for \( y^0 > x^0 \), the bound-state contribution appears for this time ordering.

In section \ref{sect:exampleLSZ}, we will demonstrate how a modified form of LSZ reduction connects the correlation function (\ref{greens in main text}) to the IR-finite scattering amplitude (\ref{the scattering amplitude}).

It would be interesting to investigate the relativistic counterpart of Schwinger’s \cite{Schwinger:1964zzb} one-parameter integral representation of the non-relativistic Coulomb Green’s function, which exploits the hidden Runge-Lenz symmetry.
\subsection{Causality}
Causality in QFT requires that field commutators vanish outside the light cone. To verify this, we examine the causal Green’s function
\begin{gather}
   G_C(x,y,u_s^{r})=  \braket{\Omega|[\phi^{\dagger}(x),\phi(y)]|\Omega}.\label{causal commutator} 
\end{gather}
Expanding the fields in terms of in-wavefunctions and using the canonical commutation relations for the in-operators, we obtain
\begin{align*}
    G_C(x,y,u_s^r) =\int\frac{\DD^3\vec{p}}{(2\pi)^3 2E_p}\Big(f^{\star}_{\t{in},-}(x,p)f_{\t{in},-}(y,p)&-f^{\star}_{\t{in},+}(x,p)f_{\t{in},+}(y,p)\Big)\\
   & -\sum_{nlm}B_{nlm}^{\star}(x)B_{nlm}(y).\numberthis
\end{align*}
In appendix \ref{app: further identities}, and specifically in (\ref{equal time vanishing covariant}), we prove that this quantity vanishes for spacelike separations, thus proving that the commutator satisfies the causality condition. The proof assumes that, for the classical equations of motion (\ref{semi free}), the initial data \(\phi\) and \(n \cdot \partial \phi\) can be independently specified on a spacelike hypersurface with normal vector \(n^\mu\).
\subsection{Perturbation theory}\label{sect:perturb}
In this section, we develop both time-dependent and time-independent perturbation theory to incorporate the quartic vertex. The goal is not to derive highly nontrivial results, but rather to highlight key analogies to standard (non-background) perturbation theory. In particular, we pinpoint where common incorrect assumptions—responsible for IR divergences—are typically introduced, and what the correct modification is in these circumstances.

While the modifications here are relatively simple, they serve as a useful toy example and a warm-up for the fully dynamical theory of SQED. In that context, we expect similar modifications to perturbation theory to play a role in obtaining IR-finite scattering amplitudes.
\subsubsection{Time-dependent perturbation theory}
We now develop a perturbation theory for the scattering eigenstates of the full Hamiltonian, which includes the quartic interaction,
\begin{gather}
    H=H_{0}+V\\
    V=-e^2\int\DD^3\vec{x}\,\, A^{\mu}A_{\mu}|\phi|^2,\label{potential}
\end{gather}
where $H_0$ is the Coulomb Hamiltonian in (\ref{diagonal H}). We work in the rest frame of the source, as energy conservation for the probe particle holds only in this frame.\footnote{More covariantly, one could consider eigenstates of \( u_s^{\mu}  \hat{P}_\mu \), since this is the conserved component of the probe particle’s momentum.} In this frame, the in- and out-states are defined as eigenstates of the full Hamiltonian
\begin{gather}
    \hat{H}\ket{\psi^{\rho}_{\t{in/out}}(p^{\mu})}=p^0\ket{\psi^{\rho}_{\t{in/out}}(p^{\mu})}.\label{eigenfull}
\end{gather}
The fundamental quantity of interest is the transition amplitude between an in-state and an out-state, given by their inner product
\begin{gather}
   \mathcal{A}^{\rho}(p\rightarrow q)= \braket{\psi^{\rho}_{\t{out}}(q^{\mu})|\psi^{\rho}_{\t{in}}(p^{\mu})},
\end{gather}
Here, $\mathcal{A}$ has no $0$ subscript; we reserve it for amplitudes without the quartic vertex.
We presume that in-states $\ket{\psi^{\rho}_{\t{in}}(p^{\mu})}$ resemble Coulomb in-states $\ket{\Phi^{\rho}_{\t{in}}(p^{\mu})}$ at asymptotically early times, while out-states resemble Coulomb out-states at asymptotically late times. The precise form of this statement is the assumption that
\begin{gather}
  \lim_{t\rightarrow-/+\infty}  e^{-i\hat{H}t}\ket{\psi^{\rho}_{\t{in/out}}(p^{\mu})}=\lim_{t\rightarrow-/+\infty}e^{-i\hat{H}_{0}t}\ket{\Phi^{\rho}_{\t{in/out}}(p^{\mu})},\label{assume asymptotics}
\end{gather}
where the equality holds in the sense of distributions and is meaningful for smooth wave-packet superpositions. This assumption is precisely where standard perturbation theory introduces an incorrect step, leading to IR divergences (see, e.g., Chapter 14 of \cite{Taylor:1972pty} for a discussion). Consider, for simplicity, scattering in a Coulomb background in non-relativistic quantum mechanics or relativistic quantum field theory. Standard perturbation theory mistakenly replaces the right-hand side of (\ref{assume asymptotics}) with the free Hamiltonian and free-particle states. This is incorrect, as the exact in-state wavefunctions, which are known, do not behave as plane waves at early times. Consequently, (\ref{assume asymptotics}) does not hold, and the resulting perturbation theory does not correctly compute the inner product of the in- and out-states. The same issue arises in standard SQED without a background field, where an analogous incorrect assumption of the form (\ref{assume asymptotics}) is made.

However, here the right-hand side of (\ref{assume asymptotics}) is taken to include the Coulomb Hamiltonian without the quartic vertex (\ref{hamil}) and Coulomb states rather than free-particle states. Since the quartic vertex only affects short-distance physics, the long-distance asymptotics of the full in-states should match those of the Coulomb states. Consequently, the perturbation theory based on (\ref{assume asymptotics}) is expected to be IR-finite.

Following the standard textbook steps (e.g., chapter 3.5 of \cite{Weinberg:1995mt}), the assumption (\ref{assume asymptotics}) allows us to derive a perturbative expression for the amplitude,
\begin{gather}
     \braket{\psi^{\rho}_{\t{out}}(q^{\mu})|\psi^{\rho}_{\t{in}}(p^{\mu})}= \braket{\Phi^{\rho}_{\t{out}}(q^{\mu})|\hat{S}|\Phi^{\rho}_{\t{in}}(p^{\mu})},
\end{gather}
where $\hat{S}$ is the time-ordered exponential of the interaction, given by
\begin{gather}
    \hat{S}=T \,\, \exp\Big(-i\int\DD^4x \, V(t) \Big).\label{exponential}
\end{gather}
The operator \( V(t) \) can be viewed as evolving in a modified interaction picture, where \( V(t) = e^{iH_0 t} V(0) e^{-iH_0 t} \) with \( H_0 \) being the Coulomb Hamiltonian.\footnote{This is not the Furry representation, where \( V(t) \) would evolve under the full background field Hamiltonian.}  Expanding the exponential in (\ref{exponential}), the $\mathds{1}$ term gives the leading-order amplitude,
\begin{align}
   \braket{\psi^{\rho}_{\t{out}}(q^{\mu})|\psi^{\rho}_{\t{in}}(p^{\mu})}_{\t{LO}}  &= \braket{\Phi^{\rho}_{\t{out}}(q^{\mu})|\Phi^{\rho}_{\t{in}}(p^{\mu})}\\
   &=\mathcal{A}_0^{\rho}(p\rightarrow q).\label{LO}
\end{align}
Note that \(\mathcal{A}_0^{\rho}(p \to q)\) is IR-finite (see (\ref{treerestfframe})) and does not depend on any arbitrary reference scales.
\subsubsection*{Next-to-leading order correction}  
The next-to-leading order (NLO) term in the expansion of (\ref{exponential}) is  
\begin{gather}
    \braket{\psi^{\rho}_{\t{out}}(q^{\mu})|\psi^{\rho}_{\t{in}}(p^{\mu})}_{\t{NLO}}= ie^2\int\DD^4 xA^{\mu}A_{\mu}\braket{\Phi^{\rho}_{\t{out}}(q^{\mu})|:\hat{\phi}(x)\hat{\phi}^{\dagger}(x):|\Phi^{\rho}_{\t{in}}(p^{\mu})}.\label{perturb expand}
\end{gather}
For illustration, we consider the case of particle scattering (\(\rho = +\)). In this case, a convenient (though not necessary) choice is to expand \(\hat{\phi}\) in terms of in-wavefunctions and operators, while expanding \(\hat{\phi}^{\dagger}\) in terms of out-wavefunctions and operators. Using the commutators of the creation and annihilation operators, this leads to
\begin{align}
     &\braket{\psi^{+}_{\t{out}}(q^{\mu})|\psi^{+}_{\t{in}}(p^{\mu})}=ie^2\int\DD^4 xA^{\mu}A_{\mu}f_{\t{out}}^{+,\star}(q,x)f^{+}_{\t{in}}(p,x)\\
     &=2\pi i \, e^2e^2_s\delta(E_p-E_q)\int\DD^3\vec{x}\frac{1}{|\vec{x}|^2}f_{\t{out}}^{+,\star}(q,x)f^{+}_{\t{in}}(p,x),\label{correct NLO integrand}\\
     &=\alpha^2\frac{4\pi^2\delta\big(E_1-E_2\big)}{|\vec{p}_1|}\sum_{l}\frac{\Gamma(1 + l + i \gamma)}{\Gamma(1 + l - i \gamma)}
\left( i \pi + H_{l - i \gamma} - H_{l + i \gamma} \right)P_l(\hat{p}_1\.\hat{p}_2)\label{finalNLO}
\end{align}
where the \( \DD x^0 \) integral is trivial in the rest frame of the source, as the time dependence of \( f^{+}_{\text{in/out}}(x, p, u_s^r) \) is simply an exponential \( e^{-iE_p x^0} \). The integral in (\ref{correct NLO integrand}) was shown in \cite{Lippstreu:2023vvg} to yield (\ref{finalNLO}), which is indeed the $\alpha^2$ term of the exact non-perturbative amplitude (\ref{exactampinapp}).

It is interesting to note that, if one had instead expanded the fields in (\ref{perturb expand}) purely in terms of out-wavefunctions and out-operators, one would obtain
\begin{align*}
   &\braket{\psi^{+}_{\t{out}}(q^{\mu})|\psi^{+}_{\t{in}}(p^{\mu})}_{\t{NLO}}\\
   &= -ie^2e^2_s\delta(E_p-E_q)\frac{|\vec{p}|^2}{(2\pi)^32E_p}\int\DD^2\Omega_{k}\mathcal{A}^{+}_{0}(p\rightarrow k)\int\DD^3\vec{x}\frac{1}{|\vec{x}|^2}f_{\t{out}}^{+,\star}(q,x)f^{+}_{\t{out}}(k,x).\numberthis
\end{align*}
This representation of the NLO amplitude consists of an initial on-shell Coulomb scattering process with the characteristic \( t^{-1 + i\gamma} \) factor, followed by a correction to the out-state. It factors the usually troublesome long-distance physics encoded in $\mathcal{A}_{0}^{+}$ from the short-distance physics---the scattering of the out-state on the $\frac{1}{|\vec{x}|^2}$ potential. Such factorization is generic in this perturbation theory because one can always trade an $f_{\t{in}}$ for an $f_{\t{out}}$ using (\ref{relation}).

\subsubsection{Time-independent perturbation theory}
We begin with the eigenvalue equation for the full in/out states,  
\begin{gather}
    (\hat{H}_0+\hat{V})\ket{\psi^{\rho}_{\t{in/out}}(p)}=E_p\ket{\psi^{\rho}_{\t{in/out}}(p)}.
\end{gather}
Rearranging this gives
\begin{gather}
   (\hat{H}_0-E_p) \ket{\psi^{\rho}_{\t{in/out}}(p)}=-\hat{V}\ket{\psi^{\rho}_{\t{in/out}}(p)}.
\end{gather}
Multiplying both sides by the inverse operator \((\hat{H}_0-E_p)^{-1}\), and assuming the correct element of the kernel is selected, we obtain the Lippmann-Schwinger equation for the energy eigenstates:  
\begin{gather}
    \ket{\psi^{\rho}_{\t{in/out}}(p)}=\ket{\Phi^{\rho}_{\t{in/out}}(p)}+\underset{\beta}{\int}\!\!\!\!\mathclap{\sum} \,\,\,\,\,\frac{\ket{\beta}\braket{\beta|\hat{V}|\psi^{\rho}_{\t{in}}(p)}}{E_{p}-E_{\beta}\pm i\e},\label{LS eqn}
\end{gather}
where the \(+i\e\) prescription applies to in-states and \(-i\e\) to out-states. The sum over \(\beta\) runs over a complete basis of bound and continuum states, as indicated by the integral-sum notation. We choose this basis to consist of Coulomb states \(\ket{\Phi^{\rho}_{\t{in/out}}(p^{\mu})}\) (using either the in- or out-states, depending on convenience) along with the bound states.

A key assumption in (\ref{LS eqn}) is that the selected element of the kernel of \((\hat{H}_0-E_p)\) corresponds to the Coulomb state \(\ket{\Phi^{\rho}_{\t{in/out}}(p)}\). This is precisely where standard time-independent perturbation theory in QFT introduces an incorrect assumption.

To illustrate, consider scattering in a Coulomb background, either in non-relativistic quantum mechanics or relativistic quantum field theory. In standard perturbation theory, \(\hat{H}_0\) is taken to be the free Hamiltonian. When inverting \((\hat{H}_0-E_p)\), standard perturbation theory assumes that the element of the kernel corresponds to a single plane wave state. This assumption is incorrect and leads to IR divergences.  

If one insists on using the free Hamiltonian, the correct element of the kernel should instead be a superposition of plane waves that matches the Fourier transform of the in-Coulomb wavefunction at early times.

In (\ref{LS eqn}), the kernel element is chosen to be the Coulomb state. We are therefore perturbing the exactly solved theory of scattering on a $1/r$ potential by a $1/r^2$ potential, using the Coulomb Hamiltonian and Coulomb-state basis. Because all long-distance physics is already accounted for exactly, the resulting perturbation theory is expected to be IR-finite.

The leading-order (LO) term in the expansion of (\ref{LS eqn}) for computing the inner product of the in-state with the out-state is straightforward
\begin{align}
   \braket{\psi^{\rho}_{\t{out}}(q^{\mu})|\psi^{\rho}_{\t{in}}(p^{\mu})}_{\t{LO}}  &= \braket{\Phi^{\rho}_{\t{out}}(q^{\mu})|\Phi^{\rho}_{\t{in}}(p^{\mu})}\\
   &=\mathcal{A}_0^{\rho}(p\rightarrow q),
\end{align}
exactly as in the time-dependent approach (\ref{LO}).
\subsubsection*{Next-to-leading order correction}  
For the next-to-leading order (NLO) contribution, there is one perturbative correction to the in-state and one to the out-state
\begin{gather}
   \braket{\psi_{\t{out}}(q)|\psi_{\t{in}}(p)}_{\t{NLO}} =\braket{\Phi_{\t{out}}(q)|\psi_{\t{in}}(p)}_{\t{NLO}}+\braket{\psi_{\t{out}}(q)|\Phi_{\t{in}}(p)}_{\t{NLO}}.
\end{gather}
The first term on the right-hand side evaluates to
\begin{align}
    \braket{\Phi_{\t{out}}(q)|\psi_{\t{in}}(p)}_{\t{NLO}}=\frac{\braket{\Phi_{\t{out}}(q)|\hat{V}|\Phi_{\t{in}}(p)}}{E_p-E_q+i\e}.\label{NLO}
\end{align}
A convenient way to compute this is to choose the complete basis \(\ket{\beta}\bra{\beta}\) to consist of the out-states \(\ket{\Phi_{\t{out}}(k)}\bra{\Phi_{\t{out}}(k)}\) along with bound states. This choice simplifies the sum over \(\beta\) due to the orthogonality relations of the out-states, though the final result remains independent of this choice.  

The second term is 
\begin{gather}
    \braket{\psi_{\t{out}}(q)|\Phi_{\t{in}}(p)}_{\t{NLO}}=\frac{\braket{\Phi_{\t{out}}(q)|\hat{V}|\Phi_{\t{in}}(p)}}{E_q-E_p+i\e}.
\end{gather}
Here, the \(i\e\) arises because of the complex conjugation of the out-state wavefunction. Using the identity
\begin{gather}
    \frac{1}{E_q-E_p+i\e}-\frac{1}{E_q-E_p-i\e}=2\pi i\,  \delta(E_p-E_q),
\end{gather}
we obtain
\begin{gather}
    \braket{\psi_{\t{out}}(q)|\psi_{\t{in}}(p)}_{\t{NLO}}=   2\pi i\,  \delta(E_p-E_q)\braket{\Phi_{\t{out}}(q)|\hat{V}|\Phi_{\t{in}}(p)}.\label{correct integrand again}
\end{gather}
This reproduces the correct NLO correction, as in (\ref{correct NLO integrand}).

\subsection{Path integral}\label{sect:path integral}
Consider the path integral generating functional for a charged scalar field in a fixed background Coulomb gauge field $A^{\mu}$:
\begin{gather}
    Z[J,J^{\star}]=\int \mathcal{D}\phi\mathcal{D}\phi^{\star} \,\, e^{i\int \DD^4 x \mathcal{L}[\phi,A(u_s)]+J\phi+J^{\star}\phi^{\star}}\\
    \mathcal{L}=|D_{\mu}\phi|^2-m^2|\phi|^2,\qquad D_{\mu}\phi=\d_{\mu}\phi+ieA_{\mu}\phi.
\end{gather}
We treat the quartic interaction term, $A^{\mu}A_{\mu}|\phi|^2$, as a perturbation, since it is a short-range interaction and our focus is on solving the long-range dynamics exactly.  The path integral is Gaussian and is therefore straightforward to evaluate. At leading order—i.e., omitting the quartic vertex—the two-point function is
\begin{align}
  \braket{\Omega|T(\phi^{\dagger}(x)\phi(y))|\Omega}&=\frac{1}{Z}\Big[\frac{\delta}{\delta J(y)}\frac{\delta}{\delta J^{\star}(x)}Z[J,J^{\star}]\Big]\at{J=J^{\star}=0}\\
  &=G_F(x,y;u_s)+\mathcal{O}\Big(e^2A^{\mu}A_{\mu}|\phi|^2\Big),
\end{align}
where $G_F(x,y;u_s)$ is the Green's function satisfying
\begin{gather}
\left( \Box + m^2 + 2i e A^\mu \partial_\mu \right) G_F(x,y;u_s) = i\delta^{(4)}(x - y),
\end{gather}
which is given explicitly in (\ref{greens in main text}). We remind the reader that \( u_s^\mu \) in $G_F$ denotes the four-velocity of the heavy particle generating the background Coulomb field (\ref{eq:gauge field}).  

More generally, a \( 2n \)-point function involving \( n \) insertions of \( \phi \) and \( n \) insertions of \( \phi^\star \), at leading order (i.e., omitting the quartic vertex), is given by the sum over all possible Wick contractions of the \( \phi \)'s with the \( \phi^\star \)'s, where each contraction corresponds to the Feynman propagator \( G_F(x,y;u_s) \).

Next, we compute the first-order correction to the two-point function from the quartic interaction. At next-to-leading order (NLO), the correction is given by
\begin{align*}
     \braket{\Omega|T(\phi^{\dagger}(x)\phi(y))|\Omega}_{\t{NLO}}&=ie^2\braket{\Omega|T\Big(\phi^{\dagger}(x)\phi(y)\textstyle\int\DD^4zA^{\mu}(z)A_{\mu}(z)\phi(z)\phi^{\dagger}(z)\Big)|\Omega}  \\
   &=ie^2\int\DD^4z A^{\mu}(z)A_{\mu}(z)G_F(x,z)G_F(z,y).\label{NLO from path integral}\numberthis
\end{align*}
The first line follows from expanding the interaction term \( \exp(ie^2\textstyle \int \DD^4z A^2 |\phi|^2) \), while the second line follows from Wick contracting the fields.  

We will use this result (\ref{NLO from path integral}) in section \ref{sect:exampleLSZ} to perform a modified LSZ reduction, relating (\ref{NLO from path integral}) to the IR-finite NLO amplitude (\ref{correct NLO integrand}).  

\section{LSZ revisited}\label{sect:LSZ}
The Kinoshita-Poggio-Quinn (KPQ) theorem \cite{Poggio:1976qr,Sterman:1976jh,Kinoshita:1977dd} establishes that the Fourier transforms of off-shell position-space correlation functions are free from infrared divergences. Consequently, infrared divergences only begin to arise when transitioning from off-shell Green's functions to on-shell S-matrix elements via the LSZ reduction formula.

One of the key assumptions in the original LSZ paper \cite{Lehmann:1954rq}---see \cite{Klein1961DispersionRA} for the English translation, and \cite{Collins:2019ozc} for a modern approach that addresses several subtleties overlooked in the original derivation---is that quantum fields behave like free fields at asymptotically early and late times, as will be made precise below. However, in theories with long-range forces, such as QED and quantum gravity, this assumption fails, and the LSZ reduction formula does not hold. This is especially important since much of our understanding of the analytic properties of the S-matrix relies on the assumption of the LSZ prescription \cite{Kruczenski:2022lot}. 

This failure of the LSZ prescription is evident in the Fourier transform of position-space Green’s functions in QED, which do not exhibit a simple $(p^2 - m^2)^{-1}$ pole. Instead, they display a branch point singularity near the mass shell of the form $(p^2 - m^2)^{-1+\beta+i\gamma}$ \cite{Kibble:1968oug, Abrikosov1954, Bogoliubov1956, Soloviev1963, Hagen1963, NuovoCimento1963, Fradkin1965}, where $i\gamma$ is purely imaginary and associated with the Coulomb phase, and $\beta$ is a real function capturing the effects of soft photon emission. This observation led Zwanziger \cite{ZwanzLSZiswrong,Zwanzigereduction} to propose a modification of the LSZ formula: instead of extracting the coefficient of the $(p^2 - m^2)^{-1}$ pole in the Fourier transform of position-space correlation functions, one should isolate the coefficient of the branch point singularity $(p^2 - m^2)^{-1+\beta + i\gamma}$.
However, both Dollard’s and Zwanziger’s approaches have been criticized for breaking spacetime covariance due to the failure of their modified asymptotic dynamics to form a proper unitary group \cite{Morchio:2014zga, Morchio:2014rfa}\footnote{The issue raised is that the asymptotic Hamiltonian contains logarithmic time-dependent terms, such as \( e^{i\alpha \log(t)} \), which fail to satisfy the group property of time translations, \( U(t_1 + t_2) = U(t_1) U(t_2) \).}. These approaches also exhibit an inherent ambiguity, which we discuss in detail below (\ref{reftime}).

Rather than pursuing these approaches, we return to the original starting point of the LSZ paper: the precise mathematical condition for asymptotic free dynamics, expressed through the existence of the limit in (\ref{KGassumption}). In section \ref{sec:LSZreview}, we analyze this condition from multiple perspectives and explain why it fails in the presence of long-range interactions. In section \ref{sect:strongfield LSZ}, this analysis will lead to a natural modification of the LSZ reduction formula, inspired by approaches used in the strong-background-field literature \cite{Fradkin:1991zq}. In section \ref{sect:exampleLSZ}, we demonstrate that the modified LSZ procedure applies successfully to SQED in a Coulomb background, yielding IR-finite scattering amplitudes directly from position-space correlation functions, whereas standard LSZ reduction results in IR-divergent expressions. Unlike the Dollard and Zwanziger approaches, our modified prescription avoids inherent ambiguities and our asymptotic evolution operator, $e^{-iH_{\t{as}}t}$ with $H_{\t{as}}$ given by (\ref{hamil}), satisfies the group property of time translations.

Throughout, we use QFT in a Coulomb background as our primary example, corresponding to the limit of SQED where one particle becomes infinitely massive. We expect that relaxing this limit will require a similar modification of the LSZ formula in standard SQED, a topic we leave for future work and briefly discuss in section \ref{eq:tothefuture}.

\subsection{Breakdown of LSZ in the presence of long-range forces}\label{sec:LSZreview}
The starting point of the LSZ paper is the assumption that, at asymptotically early times, the Klein-Gordon (KG) inner product of a quantum field with a positive-frequency on-shell plane wave yields the desired in-state. For simplicity, we consider a real scalar field:
\begin{align*}
    &\braket{p_1,\dots,\t{out}|\hat{\phi}(x_2)\dots|q_1,\dots\t{in}}\\
    &=-i\lim_{x_1^0\rightarrow -\infty} Z^{-\frac{1}{2}}\int_{x_1^0}\DD^3\vec{x}_1\,e^{-iq_1\.x_1}\overset{\leftrightarrow}{\d}_{x_1^0}    \braket{p_1,\dots,\t{out}|\hat{\phi}(x_1)\hat{\phi}(x_2)\dots|q_2,\dots\t{in}}.\numberthis\label{KGassumption}
\end{align*}
This equation is understood as a distributional equality between wave packets. Here, $Z$ denotes the field-strength renormalization constant. A similar expression applies for constructing out-states. The existence of this asymptotic-time limit precisely captures the LSZ assumption that quantum fields behave as free fields at asymptotic times. Starting from the assumption (\ref{KGassumption}), LSZ derive their well-known formula relating S-matrix elements to the residues of mass-shell poles in momentum-space Green’s functions. See \cite{Jain:2023fxc,Kim:2023qbl} for recent works that employ this more primitive form of LSZ reduction, (\ref{KGassumption}). It is this starting formula (\ref{KGassumption}) that we will modify to account for IR divergences.

Let us first review why (\ref{KGassumption}) fails in theories without a mass gap. To do this, we introduce shorthand notation for the time-dependent operator in (\ref{KGassumption}) and make explicit use of wave packets:
\begin{gather}
    \hat{\phi}^f(t)=i\int_t\DD^3\vec{x}\,f^{\star}(x)\overset{\leftrightarrow}{\d_{0}}\hat{\phi}(x)\label{wavepacketoperator}\\
    f(x)=\int \frac{\DD^3k}{(2\pi)^32E_k}F(\vec{k})e^{ik\.x}\\
    (\Box+m^2)f(x)=0,
\end{gather}
where $F(\vec{k})$ defines a wave packet. The key question is whether \( \lim_{t \to -\infty} \hat{\phi}^f(t) \) acts as a creation operator for one-particle in-states. We investigate this by examining its overlap with a multiparticle state of definite momentum, \( \hat{P}^{\mu} \ket{n} = P^{\mu}_n \ket{n} \), where \( \hat{P}^\mu \) is the momentum operator. Using translational invariance, one finds (the explicit steps are detailed in \cite{Coleman:2011xi}):
\begin{gather}
    \braket{n|\hat{\phi}^f(t)|\Omega}=\frac{\omega_{\vec{P}_n}+P^0_n}{2\omega_{\vec{P}_n}}F(\vec{P}_n)\braket{n|\hat{\phi}(0)|\Omega}e^{-i(\omega_{\vec{P}_n}-P^0_n)t}\label{overlap}\\
    \omega_{\vec{P}_n}=\sqrt{m^2+|\vec{p}_1+..+\vec{p}_n|^2},\qquad P^0_n=\sum_{i=1}^n\sqrt{m_i^2+|\vec{p}_i|^2},
\end{gather}
where $\ket{\Omega}$ is the vacuum state. The key point is that in theories with a mass gap, we always have \( P^0_n > \omega_{\vec{P}_n} \). As a result, the overlap (\ref{overlap}) tends to zero as \( t \to \pm \infty \) when \( \ket{n} \) is smeared with a wave packet. This follows from the rapidly oscillating phase, which vanishes under integration according to the Riemann-Lebesgue lemma. We therefore conclude that \( \lim_{t \to -\infty} \hat{\phi}^f(t) \) only has overlap with single-particle states.
However, in the absence of a mass gap, we can consider the overlap of the operator (\ref{wavepacketoperator}) with a state containing a single scalar particle of momentum \( p^{\mu} \) and a photon of momentum \( k^{\mu} \). In this case, the relevant phase is  
\begin{align}
    \omega_{\vec{P}_n}-P^0_n&=\sqrt{m^2+|\vec{p}+\vec{k}|^2}-\sqrt{m^2+|\vec{p}|^2}-|\vec{k}|\\
    &=-\frac{p\.k}{E_p}+\mathcal{O}\left(\frac{|\vec{k}|^2}{E_p}\right),\label{massless photons}
\end{align}
where, in the second line, we have taken the limit of small photon momenta. From (\ref{massless photons}), we see that at late times \( t \), photons with momenta satisfying\footnote{Here we see the well-known observation that late-time dynamics are dominated by photons with momenta transverse to the source particle’s four-momentum, i.e., satisfying \( p \cdot k \approx 0 \). For example, the Coulomb field (\ref{eq:gauge field}) is generated by off-shell photons with \( p \cdot k = 0 \), with both the Weinberg soft-photon factor $\frac{p\.\e}{p\.k}$ and the eikonal propagator \( \frac{\alpha}{p \cdot k} \) also exhibiting this dominance.}

\begin{equation}
    \frac{p \cdot k}{E_p} \sim \frac{1}{t}, \label{photon_momentum_condition}
\end{equation}  
are still created by the operator (\ref{wavepacketoperator}). Hence, (\ref{wavepacketoperator}) cannot be interpreted as a creation operator for single-particle in-states.

Not only can (\ref{wavepacketoperator}) not be interpreted as a creation operator for a single-particle in-state, but the limit (\ref{KGassumption}) itself does not exist. This provides a precise formulation of the statement that quantum fields do not behave as free fields at asymptotic times. While the failure of the late-time limit can be verified explicitly in a fixed Coulomb background using exact correlation functions (e.g., (\ref{greensinLSZ}) or (\ref{1pt})), this relies on having exact solutions. To develop a more general method—applicable even to standard SQED without a background—we focus instead on deriving the late-time evolution equation for the on-shell plane-wave expansion coefficients of correlation functions. We derive the late-time evolution equations explicitly in appendix \ref{app:latetimedynamics}. The key results can be summarized as follows. For correlation functions in a \( \frac{\alpha}{|\vec{x}|} \) potential, the late-time evolution of a plane-wave mode with on-shell momentum \( \vec{p} \) scales as
\begin{gather}
\lim_{|t| \to \infty} \dot{a}_{\vec{p}} \sim i \frac{\alpha}{|\vec{x}|} a_{\vec{p}} \sim i \alpha \frac{E_p}{|\vec{p}| t} a_{\vec{p}},
\end{gather}
where, in the second relation, we have used the fact that at late times, wave packets localize along classical trajectories, \( \vec{x} = \frac{\vec{p}}{E_p} t \). This leads to the solution (with $\gamma=\frac{E_p}{|\vec{p}|}\alpha$):
\begin{gather}
a_{\vec{p}}(t) = a_{\vec{p}}(t_0) \left( \frac{t}{t_0} \right)^{i \gamma}, \label{reftime}
\end{gather}  
showing that the mode coefficients do not asymptote to constant values at large times. The late-time logarithmic behavior is precisely what the Dollard--Faddeev--Kulish (DFK) approach \cite{Dollard1964AsymptoticCA,dollard1971quantum} factors out. However, this approach introduces an inherent ambiguity: the solution (\ref{reftime}) depends on an arbitrary choice of reference time \( t_0 \), making the resulting amplitudes explicitly dependent on this choice. The extent to which such ambiguities affect the analytic properties of scattering amplitudes remains an open question. For example, in \cite{Hannesdottir:2019opa}, different treatments of infrared divergences using the DFK approach lead to IR-finite \( S \)-matrix elements with different analytic properties, such as whether the Steinmann relations are satisfied. In that context, ambiguities also manifest through dependence on the arbitrary dimensional regularization scale \( \mu \). The method we outline in the next section avoids these ambiguities, albeit in a much simpler setting than \cite{Hannesdottir:2019opa}, as we focus solely on scattering in a fixed Coulomb background.

In contrast to the long-range Coulomb potential \( \frac{\alpha}{|\vec{x}|} \), where the mode coefficients do not asymptote to constants, for a short-range potential \( \frac{\alpha}{|\vec{x}|^2} \), we find  
\begin{gather}
\lim_{|t| \to \infty} \dot{a}_{\vec{p}}(t) \sim i\frac{\delta}{t^2} a_{\vec{p}}(t),
\end{gather}
where $\delta$ is a function of the charge and momenta. This integrates to
\begin{gather}
a_{\vec{p}}(t) = C \, e^{-i \frac{\delta}{t}},
\end{gather}
where \( C \) is a constant of integration. In this case, \( a_{\vec{p}}(t) \) approaches a constant as \( t \to \infty \), indicating that plane waves form an appropriate basis for the Green's function when only short-range interactions are present. Consequently, the limit in (\ref{KGassumption}) exists, and the standard LSZ reduction procedure applies without modification in such cases.

\subsection{IR-finite LSZ reduction}\label{sect:strongfield LSZ}

To identify an appropriate modification of (\ref{KGassumption}) in theories with long-range forces, we begin with the expectation that position-space correlation functions
\begin{gather}
    G(x_1, \dots, x_n) = \braket{\Omega | \mathrm{T} \hat{\phi}(x_1) \dots \hat{\phi}(x_n) | \Omega},
\end{gather}
are well defined and IR finite at fixed external separations when all charged particles are massive. This expectation is supported by the KPQ theorem for off-shell Fourier-transformed Green's functions and by the coordinate-space Landau analysis of massive lines in \cite{Erdogan:2013bga}, although neither provides a general proof directly in position space. Without making any assumptions, we can expand the \( x_1 \)-dependence of \( G \) in terms of on-shell plane waves with time-dependent coefficients \( a_{\vec{p}}(x_1^0) \), since on-shell plane waves form a complete basis for \( L^2(\mathbb{R}^3) \). The Klein-Gordon inner product allows these coefficients and their time derivatives to be extracted at any finite time. The key question, and the core assumption underlying (\ref{KGassumption}), is whether these coefficients asymptote to constants at early and late times. It is precisely this assumption that fails in theories with long-range interactions, where these coefficients do not asymptote to constants, as seen in (\ref{reftime}). This can also be restated as the Green’s function failing to satisfy the Klein-Gordon equation at asymptotic times.

A natural modification of the LSZ prescription is to replace the plane-wave basis with another complete basis whose late-time behavior matches that of the Green’s functions. More precisely, we should choose a complete basis that closely resembles plane waves at large distances, with the property that the expansion coefficients of the Green’s function asymptote to constants at early and late times. Relativistic scattering in a Coulomb background provides a setting where this can be implemented directly, as we demonstrate in the next section \ref{sect:exampleLSZ}. In this case, we expand correlation functions in terms of Coulomb wavefunctions, which we established in appendix \ref{app:scattercomplete} form a complete basis. We find that the corresponding coefficients are constant at asymptotic times, even in the presence of short-range perturbations, and we can transition between correlation functions and IR-finite amplitudes straightforwardly. In summary, we will promote the KG inner product to the SQED inner product and replace the plane-wave basis with the Coulomb wavefunction basis in the original LSZ reduction formula (\ref{KGassumption}).

The approach presented here can be viewed as a specific implementation of the reduction formulas used in the strong-background-field literature, as discussed in \cite{Fradkin:1991zq} (though the Coulomb case is explicitly omitted there). The key insight is that the LSZ prescription developed for background fields naturally resolves the Coulomb phase infrared divergence—an aspect that, to our knowledge, has not been previously recognized. This connection seems natural in hindsight: the Coulomb phase divergence arises precisely because standard LSZ reduction assumes that interactions vanish at asymptotic times, whereas the background-field framework is designed to handle interactions that persist over long distances and times. While the background-field literature typically employs wavefunctions that solve the full equations of motion, including all interaction terms, we find that this is not necessary. It suffices to use wavefunctions that incorporate only the long-range interaction terms, such as the \( A^{\mu} \partial_{\mu} \phi \) coupling in SQED. This is because it is precisely these long-range interactions that cause the mismatch between the asymptotic behavior of Green’s functions and plane waves.

The perturbiner approach, also known as the Arefeva-Faddeev-Slavnov (AFS) procedure, is a lesser-known alternative to LSZ reduction \cite{Arefeva:1974jv,Jevicki,ABBOTT1983372,Schrauaner}. It computes S-matrix elements directly from the path integral without relying on Green’s functions. In \cite{Adamo:2023cfp}, this method was shown to yield IR-finite scattering amplitudes for $1\rightarrow 1$ scattering on a Coulomb background. It would be interesting to investigate whether the AFS/perturbiner approach can be extended to compute IR-finite amplitudes for generic-mass $2\to 2$ scattering.

\subsection{Example: IR-finite LSZ reduction in a Coulomb background}\label{sect:exampleLSZ}

In this section, we demonstrate that replacing the Klein-Gordon inner product and on-shell plane waves in (\ref{KGassumption}) with the SQED inner product and Coulomb wavefunctions allows for the extraction of IR-finite scattering amplitudes from position-space correlation functions in a Coulomb background.

We begin by showing that this approach works in the absence of the \( A^2|\phi|^2 \) interaction. Subsequently, we incorporate this term as a first-order perturbation and demonstrate that the modified LSZ prescription is still useful for extracting IR-finite amplitudes even when short-range perturbations are included. Specifically, we treat the \( A^2|\phi|^2 \) term as a first-order perturbation and show that the background-field LSZ prescription matches the corresponding term in the perturbative expansion of the known exact IR-finite result.

The modified LSZ prescription relies only on the knowledge of a complete orthonormal basis of wavefunctions that resemble on-shell plane waves at large distances and whose expansion coefficients for the position-space correlation function approach constant values at asymptotic times.

\subsubsection*{Leading order}

Consider the time-ordered two-point function of a free scalar field in a background Coulomb field:
\begin{gather}
  G_F(x,y;u_s)=  \braket{\Omega|T(\phi(x)\phi^{\dagger}(y))|\Omega}. \label{Greens}
\end{gather}
Here, $u_s^{\mu}$ is the four-velocity of the source particle generating the Coulomb background.
In this subsection, we omit the \( A^2|\phi|^2 \) interaction term, in which case (\ref{Greens}) satisfies the following Green's function equation (specializing to the rest frame of the source for simplicity):
\begin{gather}
       \left( \Box + m^2 + 2i\frac{\alpha}{r} \partial_t \right) G_F(x, y, u_s^r) = -i \delta^{(4)}(x - y), \label{deltagreens}
\end{gather}
where the subscript \( F \) indicates that this Green’s function satisfies Feynman (time-ordered) boundary conditions. The explicit expression for this Green’s function in the scattering basis was derived in (\ref{greens in main text}). As verified in appendix \ref{app:greens}, it satisfies (\ref{deltagreens}) and is given by
        \begin{align*}
   G_F(x,y; u_s^r)=\Bigg(& \sum_{l=0}^{\infty}\sum_{m=-l}^l \sum_{n=l+1}^{\infty} \theta\Big(-\t{sign}[\alpha](x^0-y^0)\Big) B_{nlm}(x)B^{\star}_{nlm}(y)\\
   &+\frac{1}{2}\sum_{\rho=\pm}\int \frac{\DD^3\vec{p}}{(2\pi)^3E_p}\theta(\rho(x^0-y^0)) f_{\t{in}}^{\rho}(x,\vec{p})f_{\t{in}}^{\star\rho}(y,\vec{p})\Bigg).\numberthis\label{greensinLSZ}
\end{align*}
To compute the scattering amplitude for particle scattering \( \braket{\Phi^{+}_{\text{out}}(p_2) | \Phi^{+}_{\text{in}}(p_1)} \), we first project onto the in-state using the SQED inner product and the in-state Coulomb wavefunction\footnote{We use the shorthand \( a^{\star} \overset{\leftrightarrow}{D}_{\mu} b \equiv a^{\star} D_{\mu} b - (D_{\mu} a)^{\star} b \), where \( D_{\mu} = \partial_{\mu} + i e_1 A_{\mu}(x) \) is the covariant derivative. This provides a compact expression for the SQED inner product.}
\begin{gather}
     \braket{\Omega|\phi(x)|\Phi^{+}_{\text{in}}(p_1)} = -i \int\limits_{y^0 = -\infty} \!\!\!\!\mathrm{d}^{3} \vec{y} \,\,\, f^{+}_{\mathrm{in}}(p_1, y) \overset{\leftrightarrow}{D}_{y^0} \braket{\Omega|T(\phi(x)\phi^{\dagger}(y))|\Omega}. \label{proectin}
\end{gather}
Recalling the normalization and orthogonality relations of the Coulomb wavefunctions (\ref{normofinout}, \ref{boundorthounbound}):
\begin{gather}
  \Big\langle f^{\rho}_{\t{in}}(\vec{p}\,')\Big\vert f^{\rho'}_{\t{in}}(\vec{p})\Big\rangle_{\t{SQED}}= \Big\langle f^{\rho}_{\t{out}}(\vec{p}\,')\Big\vert f^{\rho'}_{\t{out}}(\vec{p})\Big\rangle_{\t{SQED}}
  =\rho\delta_{\rho\rho'}(2\pi)^3 2E\delta^{(3)}(\vec{p}-\vec{p}').\label{orthonormalagain}\\
  \braket{B_{nlm}|f_{l'm'}^{\rho}(|\vec{p}|)}_{\t{SQED}}=0,
\end{gather}
the integral in (\ref{proectin}) evaluates to:\footnote{When the time derivative $\d_{y^0}$ in the SQED inner product acts on $\theta(x^0-y^0)$, one obtains the vanishing equal-time relation (\ref{equal time vanishing}). However, we can ignore the time-ordering $\theta$-functions altogether because we set $x^{0}=+\infty$ and $y^{0}=-\infty$.}
\begin{gather}
     \braket{\Omega|\phi(x)|\Phi^{+}_{\text{in}}(p_1)} = f_{\t{in}}^{+}(x,\vec{p}_1).\label{1pt}
\end{gather}
To project onto the out-state, we apply the SQED inner product with the outgoing Coulomb wavefunction:
\begin{gather}
 \braket{\Phi^{+}_{\text{out}}(p_2) | \Phi^{+}_{\text{in}}(p_1)} = i\int\limits_{x^0 = +\infty} \!\!\!\!\mathrm{d}^{3} \vec{x} \,\,\,   f^{+,\star}_{\t{out}}(x,p_2)\overset{\leftrightarrow}{D}_{x^0}\braket{\Omega|\phi(x)|\Phi^{+}_{\text{in}}(p_1)}.\label{projectout}
\end{gather}
Combining these results (\ref{1pt},\ref{projectout}), we obtain the IR-finite scattering amplitude:
\begin{align}
    &\braket{\Phi^{+}_{\text{out}}(p_2) | \Phi^{+}_{\text{in}}(p_1)} =\braket{f^{+}_{\t{out}}(p_2)|f^{+}_{\t{in}}(p_1)}_{\t{SQED}}\\
    &=i\pi\, e_1e_s\delta\Big(u_s\.(p_1-p_2)\Big)\frac{\Gamma(1+i\gamma)}{\Gamma(1-i\gamma)}\frac{p_1\.u_s}{m^2-(p_1\.u_s)^2}\Bigg(\frac{4\big(m^2-(p_1\.u_s)^2\big)}{(p_1-p_2)^2}\Bigg)^{1+i\gamma},\label{exactamplitude}
\end{align}
In the last equality, we have given the covariant expression for the amplitude, valid in any reference frame rather than only the rest frame of the source particle. Here, \( \gamma \) corresponds to the same quantity defined in the scattering wavefunctions (\ref{def:gamma}), \( \gamma = \alpha \frac{E_p}{|\vec{p}|} \), now also expressed in covariant form:
\begin{gather}
    \gamma(p, u_s) = \frac{e_p e_s}{4\pi} \frac{p \cdot u_s}{\sqrt{(p \cdot u_s)^2 - p^2}}.
\end{gather}
We have thus successfully established a transition from the position-space correlation function to the IR-finite amplitudes computed in section \ref{sect:bogo}. Notably, unlike the DFK approach, there is no ambiguous scale—such as a dimensional regularization $\mu$ scale—present in (\ref{exactamplitude}). Instead, the Coulomb wavefunctions naturally select the kinematic scale \( m^2 - (p_1 \cdot u_s)^2 \).

\subsubsection*{Next-to-leading order}

If we treat the \( A^2|\phi|^2 \) term as a perturbation, the first-order correction to the two-point function (next-to-leading order, NLO) is given by (see section \ref{sect:path integral} for details on the computation):
\begin{align}
     G^{\mathrm{NLO}}\big(x, y \,|\, A(u_s)\big) &= i e^2 \int \mathrm{d}^4 z \, G_F(x, z) A^{\mu}(z) A_{\mu}(z) G_F(z, y).
\end{align}
Projecting onto the in- and out-states using the same procedure as in (\ref{proectin}) and (\ref{projectout}) gives
\begin{align*}
 &\braket{\Phi^{+}_{\text{out}}(p_2) | \Phi^{+}_{\text{in}}(p_1)} _{\mathrm{NLO}} \\
    &= \int\limits_{\substack{x^0 = +\infty \\ y^0 = -\infty}} \!\!\!\mathrm{d}^3 \vec{x} \, \mathrm{d}^3 \vec{y} \,\,\, f^{+,\star}_{\mathrm{out}}(x, p_2) \overset{\leftrightarrow}{D}_{x^0} f^{+}_{\mathrm{in}}(p_1, y) \overset{\leftrightarrow}{D}_{y^0} \, G^{\mathrm{NLO}}\big(x, y \,|\, A(u_s)\big). \numberthis
\end{align*}
The spatial integrals over \( \vec{x} \) and \( \vec{y} \) are straightforward, yielding momentum-conserving delta functions that localize the momentum integrals in the Feynman propagators \( G_F \) (see (\ref{greensinLSZ})). This follows directly from the orthonormality of the Coulomb wavefunctions with respect to the SQED inner product (\ref{orthonormalagain}). Working in the rest frame of the source simplifies the \(\mathrm{d} z^0\) integral, as the time dependence of the Coulomb wavefunctions reduces to simple factors of \(e^{-iE z^0}\). This leads to:
\begin{gather}
  \braket{\Phi^{+}_{\text{out}}(p_2) | \Phi^{+}_{\text{in}}(p_1)}_{\mathrm{NLO}}=i\alpha^2\delta(E_1-E_2)\int\DD^3\vec{z}\, f_{\t{out}}^{+,\star}(z,p_2)\frac{1}{|\vec{z}|^2}f_{\t{in}}^{+}(z,p_1).\label{NLOs}
\end{gather}
Equation (\ref{NLOs}) agrees with both perturbative calculations. The time-independent and time-dependent results appear in (\ref{correct NLO integrand}) and (\ref{correct integrand again}), respectively. This integral was evaluated in \cite{Lippstreu:2023vvg} and shown to precisely match the first-order term (\ref{finalNLO}) in the expansion of the exact non-perturbative scattering amplitude (\ref{exactampinapp}) in powers of the \( A^2|\phi|^2 \) coupling. Notably, this amplitude is IR-finite, as evident from the fact that the potential \( 1/|\vec{z}|^2 \) is short-ranged.

This demonstrates that the modified LSZ prescription remains useful for extracting the IR-finite scattering amplitudes from position-space correlation functions, even when short-range perturbations are included.
\subsection{Outlook and prospects for generalizing to full SQED}\label{eq:tothefuture}
We have demonstrated that promoting the KG inner product to the SQED inner product, and replacing plane waves with relativistic Coulomb wavefunctions in the starting assumption (\ref{KGassumption}) of the original LSZ paper, allows for the extraction of unambiguous, IR-finite scattering amplitudes directly from position-space correlation functions for QFT on a Coulomb background. Notably, this approach avoids the introduction of arbitrary scales, such as the dimensional regularization parameter \( \mu \) or an IR cutoff scale \( \Lambda_{\mathrm{IR}} \), which commonly appear in alternative frameworks like the Faddeev--Kulish formalism. The presence of these arbitrary scales can obscure important analytic properties of scattering amplitudes—for instance, different choices of scale can affect whether the Steinmann relations are preserved \cite{Hannesdottir:2019opa}.

An interesting feature of this modified LSZ prescription is that it provides yet another confirmation of the well-known intuition—also made explicit in the Faddeev--Kulish approach—that even an in-state containing two matter particles necessarily includes the accompanying photon field. This is evident from the modified LSZ prescription’s explicit dependence on the Coulomb field when constructing such an in-state.

In extending these results to fully dynamical SQED, two generalizations are necessary. First, we must move beyond the limit of an infinitely massive source to consider scattering between particles of generic mass. The natural generalization would be to use Coulomb wavefunctions to describe the relative coordinates and momenta of charged particle pairs. Here, Sazdjian's relativistic quantum mechanics \cite{sazdjian1986scalar} may be of use. Second, we must address IR divergences arising from soft-photon production, which is distinct from the Coulomb phase divergence discussed here. In this case, it is not immediately clear whether the amplitude should be extracted from a position-space correlation function of the matter fields alone or from one involving gauge-invariant combinations of matter fields with Wilson lines. To address this, it may be beneficial to first determine the correct IR-finite scattering amplitude, including radiative effects, before relating it to correlation functions via a modified LSZ prescription. As discussed in section \ref{sect:crossing}, leveraging our understanding of the Coulomb phase divergence and imposing crossing symmetry on the amplitude could offer a pathway to determining these amplitudes.

\section{Unitarity and the disconnected piece}\label{sect:unitarity}
For \textit{short-range} interactions, the inner product of the in-state with the out-state describing a scattering process naturally decomposes into two terms
\begin{gather}
    \braket{\t{out}|\t{in}}=\mathds{1}+iT\,\, ,\label{short range}
\end{gather}
where \(\mathds{1}\) represents the ``no scattering'' (or disconnected) part, while \(T\) accounts for the ``scattering'' (or connected) contribution.\footnote{Strictly speaking, \(T\) still contains some disconnected components, a fact recently used in \cite{caronhuot2023crossingscatteringamplitudes,caronhuot2023measuredasymptotically} to derive generalized crossing relations.} We emphasize that the presence of a disconnected part in (\ref{short range}) is not guaranteed for all potentials but arises specifically for short-range interactions \cite{Taylor:1972pty}. When the decomposition (\ref{short range}) holds, unitarity imposes the constraint
\begin{gather}
T - T^{\dagger} = i\, T\,T^{\dagger}, \label{generalized optical}
\end{gather}
commonly known as the general optical theorem. This relation plays a fundamental role across various areas of physics. It is a key constraint in the S-matrix bootstrap program (see, for instance, \cite{correia2020analyticaltoolkitsmatrixbootstrap}); it underpins the generalized unitarity approach used to construct higher-loop amplitudes from lower-loop ones (see \cite{Mizera_2024,brandhuber2023sagexreviewscatteringamplitudes,Britto_2025} for recent pedagogical discussions); and it is important for deriving the analytic properties of the S-matrix \cite{Hannesdottir_2022,Eden:1966dnq}.

There have been brief mentions in the literature that for long-range interactions, such as Coulomb and gravitational interactions, a disconnected part may not exist, meaning interactions always occur\footnote{One might object that, in the real world, all charges are shielded \cite{Taylor:1972pty}. However, this argument does not apply to gravity, where no such shielding is possible. Our approach is to explore whether a consistent formulation of SQED exists in which scattering amplitudes lack a disconnected component.}. This question has been mentioned in \cite{Eden:1966dnq} (pp. 191–192), \cite{Taylor:1972pty} (p. 42), and \cite{PhysRevD.7.1082} (section 4). The most thorough analysis, however, comes from Herbst \cite{herbst1974connectedness}, who demonstrated that nonrelativistic Coulomb scattering lacks a disconnected component. His argument is as follows: for nonforward \(2 \to 2\) scattering, the inner product of the in-state with the out-state can be explicitly computed using the known Coulomb wavefunctions. Herbst then shows that in the forward direction, there is a unique distributional completion consistent with unitarity (the \(\epsilon_+\) in (\ref{NRQMamplitudeeps})), and this completion does not contain a disconnected part.

Although the unitarity of the Coulomb amplitude has been verified in the partial-wave basis (see, for instance, \cite{herbst1974connectedness}), we have found no such verification in the momentum basis. We verify unitarity in the momentum basis in section \ref{sect:unitarity momentum basis} for two reasons. First, in the momentum basis, it becomes apparent that unitarity of Coulomb scattering is equivalent to the orthogonality of states in the unitary principal series representation of \( SO(1,3) \). Establishing this relation is useful because the integrals associated with the principal series representation of \( SO(1,3) \), as well as its unitarity properties, are much better understood \cite{gelfand1966generalized} than the unitarity of scattering on a Coulomb background. As this holds in both the relativistic and nonrelativistic cases, the \( SO(1,3) \) symmetry should be understood not as the Lorentz symmetry of spacetime but as a two-dimensional conformal symmetry associated with the hidden Runge-Lenz symmetry (see section 2.3 of \cite{Lippstreu:2023vvg} for a brief review, where references on the Runge-Lenz symmetry of scattering states and the Coulomb amplitude are collected). Second, modern approaches to unitarity-based amplitude computations primarily operate in the momentum basis rather than the partial-wave basis, making such a verification especially relevant.

Since the general optical theorem (\ref{generalized optical}) does not yield useful constraints in the absence of a disconnected component, we derive an alternative relation in section \ref{sect: modified elastic}. We then demonstrate this relation in a specific example, showing that it leads to nontrivial constraints connecting different orders of perturbation theory.

\subsection{Verifying unitarity in the momentum basis}\label{sect:unitarity momentum basis}
In this section, we verify that the relativistic amplitude for scattering on a Coulomb background satisfies the unitarity condition that the total probability sums to one. Notably, this verification demonstrates that unitarity permits the probability amplitude to lack the usual connectedness structure. Since the proof is identical in the nonrelativistic case, we illustrate it using the relativistic case.

The probability amplitude for relativistic $1\rightarrow 1$ scattering on a Coulomb background, evaluated in the rest frame of the source for simplicity, is given by (to avoid notational clutter, we write $\braket{p_{2,+}^{\t{out}}| p_{1,+}^{\t{in}}}$ instead of $\braket{\Phi^{+}_{\text{out}}(p_2) | \Phi^{+}_{\text{in}}(p_1)}$):
\begin{gather}
         \braket{p_{2,+}^{\t{out}}| p_{1,+}^{\t{in}}}=  -i\gamma\frac{4\pi^2}{|\vec{p}_1|} \delta(E_1-E_2) \frac{\Gamma(1+i\gamma)}{\Gamma(1-i\gamma)}\Bigg(\frac{2}{1-\hat{p}_1\.\hat{p}_2}\Bigg)^{1+i\gamma}\label{NRQMamplitude},
    \end{gather}
where $\hat{p}=\frac{\vec{p}}{|\vec{p}|}$ is a unit three-vector. Verifying unitarity amounts to checking that
    \begin{gather}
       \sum_{\rho=\pm} \int\frac{\DD^3\vec{l}}{(2\pi)^32E_l}\, \braket{l_{\rho}^{\t{out}}| p_1^{\t{in}}}\braket{l_{\rho}^{\t{out}}| p_2^{\t{in}}}^{\star}=(2\pi)^32E_1\delta^{(3)}(\vec{p}_1-\vec{p}_2)\quad ,\label{unitarity to verify}
    \end{gather}
where \(\rho = \pm\) labels particles and antiparticles. In principle, bound states should also be included in the intermediate state sum, but they do not contribute, as the transition amplitude between bound states and scattering states $\braket{B_{nlm}|p_{1,+}^{\t{in}}}$ is identically zero. Similarly, the sum over antiparticles is superfluous since the transition amplitude between a particle and an antiparticle also vanishes. Henceforth, we omit the $\rho$ subscripts, with the understanding that all particles considered are particles and not antiparticles.  

It is important to highlight that the amplitude (\ref{NRQMamplitude}) does not take the familiar form of a disconnected component,  
\(\delta^{(3)}(\vec{p}_1 - \vec{p}_2)\), plus a connected component, \(iT\), in which case verifying unitarity would instead require checking  
\(i(T - T^{\dagger}) = TT^{\dagger}\).  
We now evaluate (\ref{unitarity to verify}) using (\ref{NRQMamplitude}):
    \begin{align}
      &\int\frac{\DD^3\vec{l}}{(2\pi)^32E_l}\,\braket{l^{\t{out}}| p_1^{\t{in}}}\braket{l^{\t{out}}| p_2^{\t{in}}}^{\star}=\frac{\pi \gamma^2}{|\vec{p}|}\delta(E_1-E_2)I(\hat{p}_1,\hat{p}_2)\label{first eqn}\\
      &I(\hat{p}_1,\hat{p}_2)=\int\DD^2\Omega_l  \Bigg(\frac{2}{1-\hat{p}_1\.\hat{p}_l}\Bigg)^{1+i\gamma}\Bigg(\frac{2}{1-\hat{p}_2\.\hat{p}_l}\Bigg)^{1-i\gamma}\quad ,\label{2evaluate}
    \end{align}
where we used the fact that the ratio of Gamma factors in (\ref{NRQMamplitude}) is a pure phase and thus cancels out after being multiplied by its complex conjugate. To evaluate (\ref{2evaluate}), we take advantage of its \(2D\) conformal symmetry by parameterizing all momenta in stereographic coordinates  
\begin{gather}
    \hat{k} = \frac{1}{1 + |z|^2} 
    \left( z + \bar{z}, i(\bar{z} - z), 1 - |z|^2 \right) \\ 
    z = e^{i\phi} \tan \frac{\theta}{2}\, .
\end{gather}
The measure in these coordinates is given by
\begin{equation}
    \int d\theta \, d\phi \, \sin\theta =4 \int \frac{ \, d^2z}{(1 + |z|^2)^2}.
\end{equation}
Additionally, in these coordinates, we have
\begin{equation}
     \frac{2}{1 - \hat{p} \cdot \hat{l}} 
    =
    \frac{(1 + |z_p|^2)(1 + |z_l|^2)}{|z_p - z_l|^2}.
\end{equation}
This allows us to simplify the integral to
\begin{gather}
    I(\hat{p}_1,\hat{p}_2)=4(1+|z_1|^2)^{1+i\gamma}(1+|z_2|^2)^{1-i\gamma}\int\DD^2z_l \frac{1}{|z_l-z_1|^{2+2i\gamma}|z_l-z_2|^{2-2i\gamma}}.\label{def of i}
\end{gather}
This integral is conformally covariant and is a standard one from conformal field theory.\footnote{This conformal symmetry originates from the hidden Runge-Lenz symmetry, which combines with rotational symmetry to generate a unitary representation of \( SO(1,3) \), the 2D Euclidean conformal group. The conformal dimension of each scattering state lies on the principal series, \(\Delta = 1 + i\gamma\). For a broader discussion, see section 2.3 of \cite{Lippstreu:2023vvg}.}
For example, this integral is evaluated in Appendix A of \cite{dolan2012conformalpartialwavesmathematical} (see also \cite{Symanzik:1972wj} and Appendix A of \cite{guevara2021celestialopeblocks}):
\begin{gather}
    \lim_{\e_{+}\rightarrow 0}\int\DD^2z_l \frac{1}{|z_l-z_1|^{2+2i\gamma-\e_{+}}|z_l-z_2|^{2-2i\gamma-\e_{+}}}=\frac{\pi^2}{\gamma^2}\delta^{(2)}(z_1-z_2),\label{deltaeval}
\end{gather}
where the integral is evaluated via analytic continuation. An infinitesimal term, \(\epsilon_{+}\), is introduced in the exponents to ensure convergence, and the result is then analytically continued to \(\epsilon_{+} \to 0\). This is in line with the work of Herbst \cite{herbst1974connectedness}, who proved that the amplitude (\ref{NRQMamplitude}) possesses a unique unitary extension for all scattering angles. Specifically, the correct distributional interpretation of the amplitude is\footnote{It would be interesting to explore whether the \(\epsilon_{+}\) prescription could be related to the causal \(i\epsilon\) prescription of the S-matrix \cite{Hannesdottir_2022}.}
\begin{gather}
         \braket{p_2^{\t{out}}| p_1^{\t{in}}}= \lim_{\e_{+}\rightarrow 0} -i\gamma\frac{4\pi^2}{|\vec{p}_1|} \delta(E_1-E_2) \frac{\Gamma(1+i\gamma)}{\Gamma(1-i\gamma)}\Bigg(\frac{2}{1-\hat{p}_1\.\hat{p}_2}\Bigg)^{1+i\gamma-\e_{+}}.\label{NRQMamplitudeeps}
    \end{gather}
It is worth noting that if soft-photon production were included, the exponent would be damped with the correct sign, eliminating the need for the regulator \(\epsilon_{+}\) to ensure convergence; that is, the $R$ term in (\ref{eq:Weinbergexponential}) is always negative \cite{Weinberg:1995mt}. Nonetheless, our focus here is to verify that the Coulomb amplitude itself, in the absence of photon production, remains a well-defined and unitary amplitude.

Combining the results (\ref{first eqn}, \ref{def of i}, \ref{deltaeval}), we obtain
    \begin{align}
        \int\frac{\DD^3\vec{l}}{(2\pi)^32E_l}\,\braket{l^{\t{out}}| p_1^{\t{in}}}\braket{l^{\t{out}}| p_2^{\t{in}}}^{\star}&=\frac{4\pi^3}{|\vec{p}_1|}(1+|z_1|^2)^2\delta(E_1-E_2)\delta^{(2)}(z_1-z_2)\\
        &=(2\pi)^32E_1\delta^{(3)}(\vec{p}_1-\vec{p}_2),
    \end{align}
thus confirming that the amplitude (\ref{NRQMamplitudeeps}) satisfies unitarity\footnote{A minor subtlety arises if we start with the distribution (\ref{NRQMamplitudeeps}), in which case the integrand in (\ref{deltaeval}) includes an additional factor of \((1+|z_l|^2)^{\e_{+}}\). However, this factor does not introduce any new singularities in the integration domain and therefore does not affect the analytic continuation of the integrated result to \(\e_{+} \to 0\).}. The unitarity of the amplitude is unaffected by the scale in its numerator. Specifically, if \(\mu^{i\gamma}\) appeared in the numerator of the amplitude (\ref{NRQMamplitude}), the resulting amplitude would still satisfy the unitarity condition, as this phase cancels out when multiplied by its complex conjugate in (\ref{unitarity to verify}).

A notable aspect of this verification is that the probability amplitude does not manifestly contain a disconnected part, making the unitarity condition different from the standard case in theories with a mass gap, where unitarity instead constrains the connected amplitude via \( i(T - T^{\dagger}) = TT^{\dagger} \). Nevertheless, it is worth examining whether the distribution (\ref{NRQMamplitudeeps}) implicitly defines a disconnected component through the \(\epsilon_{+}\) prescription.

\subsection{Modified general optical theorem}   \label{sect: modified elastic}  

When the S-matrix admits the decomposition  
\begin{gather}  
    S = \mathds{1} + iT,  
\end{gather}  
the general optical theorem  
\begin{gather}  
    T - T^{\dagger} = iTT^{\dagger}, \label{elastic}  
\end{gather}  
can be used to relate amplitudes at different orders in a perturbative expansion. However, since the Coulomb amplitude (\ref{NRQMamplitude}) does not contain the usual disconnected component, the general optical theorem (\ref{elastic}) does not yield useful information in this case. Nevertheless, it is still possible to derive useful analogs of (\ref{elastic}) even in the absence of the usual disconnected component in the scattering amplitude.  

To construct such an example, let us consider \(1 \to 1\) scattering in a fixed Coulomb background, now including the \(|\phi|^2 A^2\) vertex. We decompose the full scattering amplitude \(P(p_1 \to p_2)\) into the Coulomb amplitude \(A(p_1 \to p_2)\) (which corresponds to the case without the \(|\phi|^2 A^2\) vertex) plus corrections denoted by \(T(p_1 \to p_2)\):  
\begin{gather}
    P(p_1\rightarrow p_2)=A(p_1\rightarrow p_2)+ T(p_1\rightarrow p_2).\label{def of T}
\end{gather}
Then, the unitarity relation
\begin{gather}
    \int\frac{\DD^3k}{(2\pi)^32E_k}P(p_1\rightarrow k)P^{\star}(p_2\rightarrow k)=(2\pi)^32E_1\delta^{(3)}(p_1-p_2),
\end{gather}
can be rewritten, given that the Coulomb amplitude $A(p_1\rightarrow p_2)$ already satisfies the unitarity relation (\ref{unitarity to verify}), as \footnote{A similar relation was derived in section 7 of \cite{PAPANICOLAOU1976229}, although there they further factored (\ref{def of T}) as $T(p_1\rightarrow p_2)= A(p_1\rightarrow p_2)\mathcal{T}(p_1\rightarrow p_2)$.}
\begin{gather}
    \int\DD^3k \Big(A(p_1\rightarrow k)T^{\star}(p_2\rightarrow k)+A^{\star}(p_2\rightarrow k)T(p_1\rightarrow k)+T(p_1\rightarrow k)T^{\star}(p_2\rightarrow k)\Big)=0.\label{modified elastic}
\end{gather}
This enables us to relate different orders of the perturbative expansion through unitarity, mirroring the structure of the unitarity relation (\ref{elastic}). We now verify this relation for the amplitude with the quartic vertex included. The amplitude is known in the partial-wave basis (see Appendix A of \cite{Lippstreu:2023vvg} for a derivation)\footnote{It is evident that this amplitude obeys unitarity, as the coefficient of \((2l+1) P_{l}(\hat{p}_1 \cdot \hat{p}_2)\) is a pure phase.} \footnote{As is typical for amplitudes with long-range interactions, the partial-wave expansion in (\ref{exactampinapp}) formally diverges when summed. However, it remains a well-defined distribution, meaning it has a finite and well-defined action on any sufficiently smooth wave packet \cite{Taylor}; see also Appendix B.3 of \cite{Lippstreu:2023vvg}.}:
\begin{align*}
    &P(p_1\rightarrow p_2)=\braket{\psi_{\text{out}}(p_2^{\mu},u^r_s)|\psi_{\text{in}}(p_1^{\mu},u^r_s)}_{\text{SQED}}\\
   &=\frac{4\pi^2\delta\big(E_1-E_2\big)}{|\vec{p}_1|}\sum_{l=0}^{\infty}(2l+1)\frac{\Gamma\Big(\sqrt{(l+\frac{1}{2})^2-\alpha^2}+\frac{1}{2}+i\gamma\Big)}{\Gamma\Big(\sqrt{(l+\frac{1}{2})^2-\alpha^2}+\frac{1}{2}-i\gamma\Big)}e^{i\pi\Big(l+\frac{1}{2}-\sqrt{(l+\frac{1}{2})^2-\alpha^2}\Big)}P_{l}(\hat{p}_1\.\hat{p}_2),\numberthis\label{exactampinapp}
\end{align*}
where $P_l$ denotes the Legendre polynomial of degree $l$. We can expand this in powers of the quartic vertex, which corresponds to powers of $\alpha^2=\big(\frac{e_1e_2}{4\pi}\big)^2$. Expanding the amplitude (\ref{exactampinapp}) at order $\alpha^0$ gives the Coulomb amplitude (\ref{NRQMamplitude}) in the partial-wave basis:
\begin{gather}
     A(p_1\rightarrow p_2)=\frac{4\pi^2\delta\big(E_1-E_2\big)}{|\vec{p}_1|}\sum_{l}(2l+1)\frac{\Gamma(1 + l + i \gamma)}{\Gamma(1 + l - i \gamma)}P_l(\hat{p}_1\.\hat{p}_2).
\end{gather}
The remaining terms—\(\mathcal{O}(\alpha^2)\) and higher—in the \(\alpha\) expansion of the exact amplitude (\ref{exactampinapp}) correspond to \(T(p_1\rightarrow p_2)\) by definition (\ref{def of T}). The \(\mathcal{O}(\alpha^2)\) term from expanding (\ref{exactampinapp}) reads  
\begin{gather}
    T^{(\alpha^2)}(p_1\rightarrow p_2)=\alpha^2\frac{4\pi^2\delta\big(E_1-E_2\big)}{|\vec{p}_1|}\sum_{l}\frac{\Gamma(1 + l + i \gamma)}{\Gamma(1 + l - i \gamma)}
\left( i \pi + H_{l - i \gamma} - H_{l + i \gamma} \right)P_l(\hat{p}_1\.\hat{p}_2),\label{alpha2}
\end{gather}
where \( H_l \) denotes the \( l^{\text{th}} \) harmonic number, and the superscript on \( T^{(\alpha^2)} \) indicates its order in the \(\alpha^n\) expansion of \( T \). The \(\mathcal{O}(\alpha^4)\) term reads  
\begin{align*}
T^{(\alpha^4)}(p_1\rightarrow p_2)&= \alpha^4\frac{4\pi^2\delta\big(E_1-E_2\big)}{|\vec{p}_1|} \frac{1}{2(2l+1)^2} \frac{\Gamma(1 + l + i \gamma)}{ \Gamma(1 + l - i \gamma)}J_l(\gamma),\label{alpha4}\numberthis
\end{align*}
where we have defined
\begin{align*}
    J_l(\gamma)&\defined  (2l + 1)\Big( H_{l - i \gamma} - H_{l + i \gamma} +i\pi \Big)^2\\
    &+2 i \pi  +2  
\Big( H_{l - i \gamma} - H_{l + i \gamma} \Big) + (2l + 1)\Big( \psi^{(1)}(1 + l + i \gamma)  -\psi^{(1)}(1 + l - i \gamma) \Big),\numberthis\label{JL}
\end{align*}
where \(\psi^{(a)}(z)\) is the polygamma function of order \(a\), and (\ref{JL}) has been organized so that the top line is purely real, while the bottom line is purely imaginary. We now verify the unitarity relation (\ref{modified elastic}) at order $\mathcal{O}(\alpha^4)$. Using (\ref{alpha2}), the \( TT \) term at this order is given by
\begin{align*}
    &\int\DD^3k \,T^{(\alpha^2)}(p_1\rightarrow k)T^{(\alpha^2)\,\star}(p_2\rightarrow k)\\
    &=\alpha^4\frac{64\pi^5\delta\big(E_1-E_2\big)}{|\vec{p}_1|^2}\sum_{l}\frac{1}{2l+1}\left| i \pi + H_{l - i \gamma} - H_{l + i \gamma} \right|^2P_{l}(\hat{p_1}\.\hat{p}_2)\numberthis
\end{align*}
where we used:
\begin{gather}
    \int\DD^2 \Omega\,\, P_{l}(\hat{k}_1\.\hat{x})P_{l'}(\hat{k}_2\.\hat{x})=\delta_{ll'}\frac{4\pi}{2l+1}P_{l}(\hat{k}_1\.\hat{k}_2).
\end{gather}
Similarly, we compute one of the \( A^{\star}T \) terms using (\ref{alpha4}):
\begin{gather}
    \int\DD^3k\, A^{\star}(p_2\rightarrow k)T^{(\alpha^4)}(p_1\rightarrow k)=\alpha^4\frac{64\pi^5\delta\big(E_1-E_2\big)}{|\vec{p}_1|^2} \sum_{l}\frac{1}{2(2l+1)^2}J_l(\gamma)P_{l}(\hat{p_1}\.\hat{p}_2),\label{to add conjugate}
\end{gather}
After adding the complex conjugate of (\ref{to add conjugate}), the unitarity relation (\ref{modified elastic}) requires that
\begin{gather}
    \sum_{l}\frac{1}{2(2l+1)^2}\Big(J_l(\gamma)+J^{\star}_{l}(\gamma)\Big)P_{l}(\hat{p_1}\.\hat{p}_2)=-\sum_{l}\frac{1}{2l+1}\left| i \pi + H_{l - i \gamma} - H_{l + i \gamma} \right|^2P_{l}(\hat{p_1}\.\hat{p}_2).
\end{gather}
However, since the Legendre polynomials \( P_l(x) \) form a complete basis on \( -1 < x < 1 \), this relation must hold for each angular momentum mode:
\begin{gather}
    \frac{1}{2(2l+1)}\Big(J_l(\gamma)+J^{\star}_{l}(\gamma)\Big)=-\left| i \pi + H_{l - i \gamma} - H_{l + i \gamma} \right|^2.\label{doesmodified hold}
\end{gather}
Examining the top line of (\ref{JL}) confirms that (\ref{doesmodified hold}) holds, thus verifying the unitarity relation (\ref{modified elastic}) at order \(\alpha^4\).

\section{Analytic constraints relating the Coulomb phase and real radiation}\label{sect:crossing}

Throughout this article, we have demonstrated that the Coulomb phase infrared divergence can be analyzed unambiguously. In particular, we obtain scattering amplitudes that do not introduce arbitrary scales. However, in theories with long-range forces, there is a second infrared divergence—associated with the emission of zero-energy massless particles, which we will refer to as the real radiative term. This divergence is more challenging to treat, as it involves the physics of an infinite number of zero-energy photons. Moreover, unlike the Coulomb phase, it affects the modulus of the amplitude, directly impacting cross sections. By contrast, the Coulomb phase is a pure phase (for real kinematics) and thus cancels in cross-section calculations. 

In this section, we show that the real radiative term and the Coulomb phase are not independent but instead arise from a single analytic function when treating scattering amplitudes as analytic functions of complexified kinematic variables. This function only decomposes into two distinct terms when restricted to real kinematics above threshold. We will demonstrate that only their combination satisfies key analytic properties, namely: (1) the absence of a pseudothreshold on the physical Riemann sheet and (2) crossing symmetry. This perspective is advantageous because it allows us to leverage our understanding of the Coulomb phase to gain insight into the real radiative term. If the full IR-finite S-matrix is required to satisfy either of these conditions—a nontrivial assumption, as neither crossing symmetry nor the absence of the pseudothreshold has been rigorously established in this context—then we can use our understanding of the Coulomb phase to bootstrap, or at the very least gain insight into, the IR-finite real radiative term.  

It is already known that the real radiative divergence and Coulomb phase divergence are related via crossing symmetry (see, for instance, Eq. 59 of \cite{Korchemsky1987}, Eq. 2.5 of \cite{Laenen_2015b}, and section 5.3 of \cite{DiVecchia:2021bdo}). However, we have been unable to find a detailed exposition of the analytic continuation required to relate the relevant scattering channels. We consider such an understanding necessary if one is to eventually bootstrap the crossing-symmetric IR-finite combination of the real radiative and Coulomb phase terms.

To be more precise about these two divergences, the abelian exponentiation theorem \cite{yennie,Weinberg:1965nx} states that the infrared divergences in an amplitude $\mathcal{A}$ for abelian gauge theories factorize and exponentiate. The exponent is given by the 1-loop soft singularities (see chapter 13 of \cite{Weinberg:1995mt} for a textbook discussion): 
\begin{gather}
    \mathcal{A}=\Bigg(\frac{\Lambda_{\text{IR}}}{\lambda_{\text{IR}}}\Bigg)^{R+I}\mathcal{A}_0(\Lambda_{\text{IR}}),
    \end{gather}
    where the real $R$ and imaginary $I$ exponents are given by
    \begin{gather}
    R=\frac{1}{16\pi^2}\sum_{i,j}\frac{\eta_i\eta_je_ie_j}{\beta_{ij}}\log\Bigg(\frac{1+\beta_{ij}}{1-\beta_{ij}}\Bigg)\quad,\quad I=\frac{1}{8\pi}\sum_{\eta_i\eta_j=+}-i\frac{e_ie_j}{\beta_{ij}} \numberthis\label{eq:Weinbergexponential},\\
    \beta_{ij}\defined \sqrt{1-\frac{m_i^2m_j^2}{(p_i\.p_j)^2}},\label{def of beta}
\end{gather}
Here, $\eta=\pm1$ distinguishes incoming from outgoing states. The scales $\lambda_{\text{IR}}$ and $\Lambda_{\text{IR}}$ are the lower and upper infrared cutoffs, respectively, for the energies of exchanged photons. The factor $\mathcal{A}_0(\Lambda_{\text{IR}})$ includes only photon exchanges with energies greater than $\Lambda_{\text{IR}}$. The real part $R$ in (\ref{eq:Weinbergexponential}) is associated with real photon emission, as it is this term which cancels against real photon production when computing inclusive cross sections that involve soft radiative processes. The second term $I$ in (\ref{eq:Weinbergexponential}) is the Coulomb phase divergence. It is purely imaginary and appears only for particle pairs that are both incoming or both outgoing (i.e., $\eta_i\eta_j = +1$). Physically, the Coulomb phase divergence reflects the logarithmic correction to free particle motion at asymptotic times. Classical particles interacting via a $\frac{1}{r}$ potential do not asymptote to free trajectories; instead, they follow hyperbolic trajectories with logarithmic corrections (see, e.g., \cite{PhysRevD.7.1082,Sahoo_2019} for the trajectories, and appendix H of \cite{Hirai:2021ddd} for the relation between the trajectories and this phase).

It is immediately apparent that the Coulomb phase term, $I$ in (\ref{eq:Weinbergexponential}), alone breaks crossing symmetry, as it appears only when both interacting particles are either incoming or outgoing, but not for mixed configurations of incoming and outgoing particles. In this section, we show that only the \textit{combination} of the real radiative term $R$ and the Coulomb phase $I$ satisfies crossing symmetry. For a pedagogical discussion of crossing symmetry, see \cite{Mizera_2024}. Rigorous proofs of crossing symmetry for \( 2 \to 2 \) and  \( 2 \to 3 \) scattering exist in theories with a mass gap \cite{Bros:1964iho,Bros:1965kbd,BROS1986325}, and in the planar limit even when a mass gap is absent \cite{Mizera:2021fap} for any particle multiplicity. Therefore, imposing crossing symmetry in our context is not \textit{a priori} justified, as infrared divergences complicate the question of whether the IR-finite S-matrix should obey it. 

Similarly, the absence of the $s = (m_1 - m_2)^2$  pseudothreshold on the physical Riemann sheet is not a guaranteed property of the IR-finite S-matrix, but it is physically plausible, as there is no expectation that a physical state in the theory exists with center-of-mass energy $s = (m_1 - m_2)^2$. Such singularities are known to contribute to the analytic properties of the S-matrix, appearing generically on higher Riemann sheets. Individual Feynman integrals have been bootstrapped using this constraint along with other conditions in \cite{hannesdottir2024landaubootstrap}.

\subsubsection*{Correlators of Wilson lines}  

Obtaining an analytic expression for a Feynman integral or scattering amplitude that faithfully captures its full analytic structure for complex kinematics is nontrivial. While commonly used expressions for the Coulomb phase and real radiative term, like (\ref{eq:Weinbergexponential}), correctly describe amplitudes for real kinematics above threshold, they do not necessarily extend properly to the full complex plane. A simple example is the presence of the implicit theta functions \( \theta(\eta_i\eta_j) \) in the Coulomb phase (\ref{eq:Weinbergexponential}), which restrict both particles to be either incoming or outgoing. These functions render the Coulomb phase non-analytic. To address this, we first derive an analytic representation of the Coulomb phase and real radiative term that remains valid throughout the complex kinematic plane.

The purpose of this subsection is to derive a less familiar, but analytically correct, representation of the Coulomb phase and real radiative exponent. However, readers interested only in the final result may skip this derivation and take (\ref{rightanalytic}) as the correct analytic expression before proceeding.

One way to obtain this representation is by recalling that infrared divergences in abelian gauge theory scattering amplitudes are identical to those in Wilson line correlators \cite{Polyakov1980, Arefeva1980, Dotsenko1980, Brandt1981, Korchemsky1986a, Korchemsky1986b, Korchemsky1987}. In abelian gauge theory, the exponent of the correlator of two Wilson lines for outgoing particles moving along the directions \( u_{1,2}^{\mu} = p_{1,2}^{\mu} / m_{1,2} \) is computed in dimensional regularization as  
(for a more detailed discussion, see e.g. \cite{Laenen_2015,Laenen_2015b}):  
\begin{gather}
    W=\alpha\mu^{2\e}\int_0^{\infty}\DD\tau_1\int_0^{\infty}\DD\tau_2\frac{u_1\.u_2}{\Big[-(\tau_1u_1-\tau_2 u_2)^2+i\e\Big]^{1-\e}},
\end{gather}
where the proper-time integration limits are chosen for outgoing particles. We adopt the all-outgoing convention, where outgoing particles have positive energy, and incoming particles have negative energy. In this convention, \( p_i \cdot p_j < 0 \) for an incoming-outgoing pair, while \( p_i \cdot p_j>0 \)  for pairs that are either both incoming or both outgoing. To isolate the overall divergence, we introduce the change of variables \((\tau_1,\tau_2)=(\lambda x,\lambda (1-x))\), leading to  
\begin{gather}
W=\alpha\mu^{2\e}\int_0^{\infty}\frac{\DD\lambda}{\lambda^{1-2\e}}\int_0^{1}\DD x\frac{u_1\.u_2}{\Big[-(xu_1-(1-x)u_2)^2+i\e\Big]^{1-\e}}.\label{scaleless}
\end{gather}
The $\lambda$ integral is scaleless and exhibits both ultraviolet and infrared divergences, rendering it formally zero in dimensional regularization.
To regularize, we introduce a cutoff on long-distance exchanges by adding the regulator \( e^{-\Lambda \lambda} \) (see \cite{Gardi:2014kpa,Falcioni:2014pka,Gardi:2011wa} for a regulator that preserves the rescaling symmetry of the Wilson line correlators)  
\begin{gather}
    \mu^{2\epsilon} \int_{0}^{\infty} \frac{\mathrm{d} \lambda \, e^{-\Lambda \lambda}}{\lambda^{1 - 2\epsilon}}
= \Gamma(2\epsilon) \left( \frac{\mu}{\Lambda} \right)^{2\epsilon}
= \frac{1}{2\epsilon} \left( \frac{\mu}{\Lambda} \right)^{2\epsilon} + \mathcal{O}(\epsilon^0).
\end{gather}
We now examine the kinematic dependence of (\ref{scaleless}), specifically its dependence on 
\( u_1 \cdot u_2 \). By analyzing the zeros of the denominator in (\ref{scaleless}), we see that for real 
\( u_1 \cdot u_2 \) with \( (u_1 \cdot u_2) < 1 \), the zeros lie outside the domain of integration, 
\( x \in [0,1] \). Since we are only interested in parameterizing the \( \frac{1}{\epsilon} \) coefficient of 
\( W \), we set \( 1 - \epsilon \to 1 \) in the exponent. Evaluating (\ref{scaleless}) in this regime, we obtain   
\begin{gather}
    W= \frac{i\alpha}{\epsilon} \left( \frac{\mu}{\Lambda} \right)^{2\epsilon} \frac{z}{\sqrt{1-z^2}}\log\Bigg(\frac{\sqrt{1-z}+i\sqrt{1+z}}{\sqrt{2}}\Bigg)+\mathcal{O}(\e^0)\quad \text{for}\quad |z|<1,\label{rightanalytic}
\end{gather}
where we have defined  
\begin{gather}
    z=u_1\.u_2=\frac{p_1\.p_2}{m_1m_2}.
\end{gather}
Equation (\ref{rightanalytic}) is the expression we will analytically continue to different values of 
\( z \) in order to obtain the Coulomb phase and real radiative term in different scattering channels.
\subsubsection*{Relation to the standard formulae}
The function (\ref{rightanalytic}) has potential branch points at $z=\pm 1$ from the square roots $\sqrt{1\pm z}$. Some simple algebra shows that these are the threshold and pseudothreshold singularities
\begin{alignat}{3}
    s&=(m_1+m_2)^2 \,\, &&\leftrightarrow\,\, &&z=1\label{threshold singularity}\\
    s&=(m_1-m_2)^2\,\, &&\leftrightarrow &&z=-1,\label{pseudothreshold}
\end{alignat}
where $s=(p_1+p_2)^2$.
For the scattering of two outgoing particles with physical kinematics, we must analytically continue \( z \) beyond the threshold, \( z > 1 \). To obtain the amplitude consistent with causality, \( z \) must be continued to the real axis while approaching the threshold branch cut (\ref{threshold singularity}) from above \cite{Hannesdottir_2022}, corresponding to \( \text{Im}[z] = +i\epsilon_+ \). This translates to resolving   
    $\sqrt{1 - z} = -i\sqrt{z - 1}$  
for real \( z > 1 \). Thus, analytically continuing to physical kinematics where \( \text{Re}[z] > 1 \) and \( \text{Im}[z] = +i\epsilon \) gives  
\begin{align}
     W&=\alpha \frac{1}{2\epsilon} \left( \frac{\mu}{\Lambda} \right)^{2\epsilon} 2i\frac{z}{-i\sqrt{z^2-1}}\log\Bigg(\frac{-i\sqrt{z-1}+i\sqrt{1+z}}{\sqrt{2}}\Bigg)\\
     &=\alpha \frac{1}{2\epsilon} \left( \frac{\mu}{\Lambda} \right)^{2\epsilon} \frac{z}{-\sqrt{z^2-1}}\Bigg[2\log\Bigg(\frac{-\sqrt{z-1}+\sqrt{1+z}}{\sqrt{2}}\Bigg)+2\log(i)\Bigg]\\
     &=\alpha \frac{1}{2\epsilon} \left( \frac{\mu}{\Lambda} \right)^{2\epsilon} \frac{z}{-\sqrt{z^2-1}}\Bigg[\log\Bigg(\frac{[-\sqrt{z-1}+\sqrt{1+z}]^2}{2}\Bigg)+i\pi\Bigg]\\
           &=\alpha \frac{1}{2\epsilon} \left( \frac{\mu}{\Lambda} \right)^{2\epsilon} \frac{1}{\beta}\Bigg[\frac{1}{2}\log\Bigg(\frac{1+\beta}{1-\beta}\Bigg)-i\pi\Bigg],\qquad \text{Im}[z]=+i\e\, , \text{Re}[z]>1,\label{chill}
\end{align}
where $\beta=\sqrt{1-\frac{1}{z^2}}$ is the same relative velocity variable as in (\ref{def of beta}). The expression (\ref{chill}) matches (\ref{eq:Weinbergexponential}) up to a regulator-dependent factor, and corresponds to the familiar form of the real radiative and Coulomb phase divergence commonly found in the literature and textbooks. The series of manipulations leading from (\ref{rightanalytic}) to (\ref{chill}) are valid only when evaluating the function at purely real values of \( z \) with \( z > 1 \), strictly on the first Riemann sheet, and only when approaching the real \( z \)-axis from above. However, steps such as \( \log(AB) = \log A + \log B \) and \( 2\log(\sqrt{A}) = \log A \) are not generally valid if we seek to preserve the full analytic structure across the entire \( z \)-plane. This is precisely why we did not start with the expression (\ref{chill}) when analyzing its analytic structure—it is not immediately clear how to extend it consistently to the full complex plane.
\subsection{Absence of the pseudothreshold on the physical Riemann sheet}  

We now show that the Wilson line correlator (\ref{rightanalytic}) does not have the pseudothreshold at \( z=-1 \) (\ref{pseudothreshold}) as a branch point on the first Riemann sheet, though it does appear on higher Riemann sheets. Since we are only interested in the kinematic dependence, we omit the non-kinematic prefactors from (\ref{rightanalytic}) and define the function  
\begin{gather}  
   f(z)= \frac{z}{\sqrt{1-z^2}}\log\Bigg(\frac{\sqrt{1-z}+i\sqrt{1+z}}{\sqrt{2}}\Bigg),\label{just the important bits}  
\end{gather}  
 We now ask whether \( f(z) \) has a nonzero monodromy around \( z=-1 \), i.e., whether \( z=-1 \) is a branch point.  

To determine this, we consider a closed path encircling \( z=-1 \) without enclosing \( z=+1 \). This analytic continuation flips the sign of the square root,  
\begin{gather}  
    \sqrt{1+z} \to -\sqrt{1+z}.  \label{signfliip}
\end{gather}  
To compute the monodromy \(\mathcal{M}_{z \circlearrowleft -1}\) around this branch point, we compare the function before continuation, \( f(z) \), with its analytic continuation after encircling \( z = -1 \), denoted \( f(z)_{z \circlearrowleft -1} \):
\begin{align}
    \mathcal{M}_{z \circlearrowleft  -1}\, f(z)&=f(z)-f(z)_{z \circlearrowleft  -1}\\
    &=\frac{z}{\sqrt{1-z^2}}\Bigg[\log\Bigg(\frac{\sqrt{1-z}+i\sqrt{1+z}}{\sqrt{2}}\Bigg)+\log\Bigg(\frac{\sqrt{1-z}-i\sqrt{1+z}}{\sqrt{2}}\Bigg)\Bigg]\label{plus sign why}\\
    &=\frac{z}{\sqrt{1-z^2}}\log\Bigg(\frac{[\sqrt{1-z}+i\sqrt{1+z}][\sqrt{1-z}-i\sqrt{1+z}]}{2}\Bigg)\\
   &=\frac{z}{\sqrt{1-z^2}}\log\Bigg(\frac{1-z+1+z}{2}\Bigg)\\
   &=\frac{z}{\sqrt{1-z^2}}\log(1)\\
   &=0,\label{zero monodromy}
\end{align}
where the plus sign between the two terms in (\ref{plus sign why}) arises because the prefactor \( \sqrt{1 - z^2} = \sqrt{1 + z} \sqrt{1 - z} \) also undergoes a sign flip (\ref{signfliip}). From (\ref{zero monodromy}), we conclude that \( f(z) \) does not have a branch point at the pseudothreshold—at least not on the first Riemann sheet (it does appear on higher Riemann sheets).  

Let us emphasize this key point: only the combination of the Coulomb phase and the real radiative factor ensures that the pseudothreshold does not appear on the physical Riemann sheet. This is significant because if we assume that this is a fundamental property of the IR-finite S-matrix, we can leverage our understanding of how to obtain an IR-finite Coulomb phase to constrain the IR-finite real radiative term.

We note that performing a similar calculation, but instead flipping \( \sqrt{1 - z} \to -\sqrt{1 - z} \), one can show that the threshold (\ref{threshold singularity}) is indeed a branch point of the function (\ref{just the important bits}). Moreover, the monodromy around this branch point is proportional to the Coulomb phase—a result that has been thoroughly explored from a physical perspective in \cite{Laenen_2015,Laenen_2015b}.  
\subsection{Crossing symmetry}  

Let us analytically continue $f(z)$ to the region corresponding to one incoming and one outgoing particle, meaning we extend to \( \text{Re}[z] < -1 \) and then approach the real axis to obtain real kinematics. Since we have verified that there is no branch point at \( z = -1 \), it does not matter whether we approach the real \( z \)-axis from above or below. In this region, \( 1+z \) becomes negative, requiring us to resolve the square root \( \sqrt{1+z} \). The choice of \( \pm i \) is inconsequential because no branch point exists at \( z=-1 \). Taking  
\begin{gather}  
    \sqrt{1+z} \to +i\sqrt{-1-z} , 
\end{gather}  
for example, gives  
\begin{align}
    f(z)&=\frac{z}{i\sqrt{z^2-1}}\log\Bigg(\frac{\sqrt{1-z}-\sqrt{-1-z}}{\sqrt{2}}\Bigg)\\
    &=\frac{z}{i\sqrt{z^2-1}}\frac{1}{2}\log\Bigg(\frac{[\sqrt{1-z}-\sqrt{-1-z}]^2}{2}\Bigg)\\
    &=\frac{1}{4i\beta}\log\Bigg(\frac{1-\beta}{1+\beta}\Bigg),\qquad \text{Im}[z]=0,\,\, \t{Re}[z]<-1.
\end{align}
So in this region we have
\begin{align}
     W
     &= \frac{\alpha}{4\epsilon} \left( \frac{\mu}{\Lambda} \right)^{2\epsilon} \frac{1}{\beta}\log\Bigg(\frac{1-\beta}{1+\beta}\Bigg),\qquad \text{Im}[z]=0,\,\, \t{Re}[z]<-1.\label{real radiative}
\end{align}
As expected, this result contains only the real radiative divergence and not the Coulomb phase divergence, consistent with the case of one incoming and one outgoing particle. Note the relative minus sign in the real radiative term here (\ref{real radiative}) compared to the case of two outgoing particles (\ref{chill}). This matches the expected \(\eta_i \eta_j\)-sign dependence of the real radiative term (\ref{eq:Weinbergexponential}).  

Thus, it is only the combination of the Coulomb phase and real radiative term that satisfies crossing symmetry, ensuring that we have a single analytic function (\ref{rightanalytic}) that relates different scattering channels. Once again, this is advantageous: an unambiguous treatment of the Coulomb phase divergence, whose physics is much easier to understand, namely the fact that the late-time trajectories are not straight-line trajectories, suggests that a similarly well-defined treatment of the real radiative divergence, whose physics is slightly harder to understand as it requires understanding an infinite number of zero-energy photons, should exist.

Let us note that currently there are two additional sources of crossing symmetry violation and the presence of a pseudothreshold on the physical Riemann sheet in the Coulomb amplitude,
\begin{align}
    &\braket{\Phi^{+}_{\text{out}}(p_2) | \Phi^{+}_{\text{in}}(p_1)} =\braket{f^{+}_{\t{out}}(p_2)|f^{+}_{\t{in}}(p_1)}_{\t{SQED}}\\
    &=i\pi\, e_1e_s\delta\Big(u_s\.(p_1-p_2)\Big)\frac{\Gamma(1+i\gamma)}{\Gamma(1-i\gamma)}\frac{p_1\.u_s}{m^2-(p_1\.u_s)^2}\Bigg(\frac{4\big(m^2-(p_1\.u_s)^2\big)}{(p_1-p_2)^2}\Bigg)^{1+i\gamma},\label{amp explaining}
\end{align}
where 
\begin{align}
      \gamma(p, u_s) &= \frac{e_p e_s}{4\pi} \frac{p \cdot u_s}{\sqrt{(p \cdot u_s)^2 - p^2}}\\
      &=\frac{e_p e_s}{4\pi} \frac{2p\.p_s}{\sqrt{[s-(m+m_s)^2][s-(m-m_s)^2]}},\label{gamma has pseudo}
\end{align}
where $s=(p+p_s)^2$. From (\ref{gamma has pseudo}) we note that \(\gamma\) has the pseudothreshold \(s = (m - m_s)^2\) as a branch point, which implies that the ratio of Gamma functions \(\frac{\Gamma(1+i\gamma)}{\Gamma(1-i\gamma)}\) also has a branch point there, as no other function in the amplitude cancels this branch cut. The same applies to the numerator \(m^2 - (p_1 \cdot u_s)^2\).

Furthermore, under crossing—where the incoming probe is replaced by an outgoing probe and the outgoing source by an incoming source—the analytically continued amplitude should contain a ratio of Gamma functions with poles at the source–anti-source and probe–anti-probe bound-state energy levels. However, the analytic continuation of (\ref{amp explaining}) to this scattering channel lacks these poles, violating crossing symmetry. Similarly, the numerator \(m^2 - (p_1 \cdot u_s)^2\) also fails to respect crossing symmetry.

In light of the previous section, the violation of these analytic properties is expected, as we have seen that only the combined Coulomb phase and real radiative terms ensure these properties hold. By contrast, the computation leading to (\ref{amp explaining}) ignored radiative effects, as well as the recoil of the heavy source particle. This can be viewed as advantageous, as these constraints could instead be used to bootstrap the amplitude which incorporates radiative effects.

\section{Factorization on the bound-state poles}\label{sect:scatter to bound}
We have emphasized that the scattering amplitudes (\ref{start amp}), obtained from the inner product of relativistic Coulomb wavefunctions, contain no arbitrary scales. This contrasts with approaches that introduce an IR regulator, such as a photon mass $\mu$, leading to amplitudes that depend on the arbitrary scale $\mu^{2i\gamma}$ (see, e.g., \cite{Kabat_1992}), rather than the natural scale $|\vec{p}|^{2i\gamma}$ that emerges from the inner product of Coulomb wavefunctions.

In this section, we present an example where the IR-finite scale set by the Coulomb wavefunction plays a key role in a specific analytic property of the scattering amplitude in the complex kinematic plane—its residue at the bound-state poles. In particular, we highlight that the scale set by the Coulomb wavefunctions, $|\vec{p}|$, acquires a definite value at these poles, unlike an arbitrary regulator scale $\mu$.

It is a general theorem of quantum mechanics that continuum radial wavefunctions are related to bound-state wavefunctions via the residue of the former (see, for instance, chapter 2 of \cite{Mizera_2024} for a pedagogical discussion, as well as chapters 12 and 15 of \cite{Newton:1982qc}, chapter 8 of \cite{gottfried2018quantum}, and chapter 20 of \cite{Taylor:1972pty}). This is intuitive because the Schrödinger equation governing both bound and unbound states is the same, differing only in energy. As a result, it is natural to expect a connection between them through analytic continuation in energy. The appearance of a residue in this relation is also intuitive: bound-state wavefunctions are normalizable, whereas unbound wavefunctions are not, leading to the expectation that the latter exhibit a simple pole as the bound-state energy is approached.

In this section, we verify that the residues of the scattering amplitude poles in the energy plane correspond precisely to the residues of the transition from the in-state to a bound state, followed by the transition from that bound state to the out-state (\ref{transition main result}). Throughout this process, no arbitrary scale appears at intermediate steps, demonstrating the internal consistency of the Coulomb amplitude and its independence from arbitrary scales.

This calculation may also be relevant given the recent interest in developing a scattering-to-bound map in gravitational-wave physics \cite{Kalin:2019rwq, Adamo_2024,Saketh:2021sri,Cho:2021arx,Gonzo:2023goe}. While much of the recent work on gravitational-wave emission has focused on black-hole scattering \cite{Bern:2019nnu, Damour:2016} (for pedagogical reviews, see \cite{bjerrumbohr2022sagexreviewscatteringamplitudes, kosower2022sagexreviewscatteringamplitudes}), detectors like LIGO primarily measure waveforms from black-hole mergers \cite{Abbott:2016blz}. Thus, scattering-amplitude calculations can be leveraged if a precise relation—known as the scattering-to-bound map—can be established between waveforms emitted during scattering and those emitted during the inspiral phase.  The recent work \cite{Adamo_2024} has taken an important step in this direction by applying the aforementioned relation between bound and unbound states—via analytic continuation and residue extraction—to the problem of gravitational-wave emission.

Consider the relativistic scattering amplitude for a particle on a Coulomb background. For concreteness, we focus on particles rather than antiparticles and assume an attractive potential (\(\alpha < 0\)). For simplicity, we work in the rest frame of the source throughout this section. The amplitude is given by  
\begin{gather}
      \braket{p_2^{\t{out}}|p_1^{\t{in}}}_{\t{SQED}}\defined \mathcal{A}(E,\theta)=-i\gamma\frac{4\pi^2}{|\vec{p}_1|} \delta(E_1-E_2) \frac{\Gamma(1+i\gamma)}{\Gamma(1-i\gamma)}\Bigg(\frac{2|\vec{p}_1|}{|\vec{p}_1-\vec{p}_2|}\Bigg)^{2+2i\gamma},\label{start amp}
\end{gather}
where we consider the scattering amplitude as a function of the complex variable \(E\) and the scattering angle \(\theta\). In general, it is important to specify which two complex variables one is working with \cite{Mizera_2024}, as the analytic structure in the \(E\)-plane depends on this choice. For instance, the analytic properties of \(\mathcal{A}(E,\theta)\) may differ from those of \(\mathcal{A}(E,t)\), where \(t = (p_1 - p_2)^2\). The function \(\Gamma(1+i\gamma)\) exhibits simple poles when its argument is a negative integer or zero,  
\begin{gather}
    i\gamma=-n,\qquad n=1,2,3,\dots
\end{gather}
As is well known \cite{Taylor:1972pty}, this corresponds to the bound-state energy spectrum,
\begin{gather}
    i\alpha \frac{E}{\sqrt{E^2-m^2}}=-n,\quad \implies\quad  
    E_n=\frac{m}{\sqrt{1+\frac{\alpha^2}{n^2}}}.\label{boundenergy}
\end{gather}
Equation (\ref{boundenergy}) has two solutions for \(E\), corresponding to positive- and negative-energy states. To ensure that the bound-state energy smoothly asymptotes to the rest-mass energy of a free particle as the coupling \(\alpha\) tends to zero, one must select the positive-energy solution.
An important subtlety about the positive-energy solution is that it arises from resolving the branch cut in \(\sqrt{E_n^2 - m^2}\), specifically by choosing \(\sqrt{-1} = -i\). This choice leads to the momentum  (despite the absolute-value notation, \( |\vec{p}| \) should be considered a complex variable)
\begin{gather}
    |\vec{p}| = \sqrt{E_n^2 - m^2} = -i\frac{\alpha}{n} E_n.
\end{gather}  
For an attractive potential (\(\alpha < 0\)), this simplifies to  
\begin{gather}
    |\vec{p}| = i\frac{|\alpha|}{n}E_n = -i\eta_n,
\end{gather} 
where \(\eta_n\) is the same parameter that appears in the argument of the bound-state wavefunctions in (\ref{eq:etas}). This corresponds to the well-known fact that bound states appear as poles in the scattering amplitude on the positive imaginary \( |\vec{p}| \)-axis (see, e.g., chapter 2 of \cite{Mizera_2024}). This is also intuitive, as only along this axis do the radial wavefunctions decay to zero at radial infinity. We compute the residue of the amplitude (\ref{start amp}) at an \(E_n\) pole,
\begin{align}
     \underset{{E\rightarrow E_n}}{\t{Res}}   \mathcal{A}(E,\theta)
     &=(-1)^{n-1}4i\pi^2\frac{\alpha E_n^2}{[n!]^2m^2}\delta(E_1-E_2)\Big(\frac{2}{1-\hat{p}_1\.\hat{p}_2}\Big)^{1-n},\label{res}
\end{align}
where we computed the residue of the Gamma function as
\begin{align}
    \underset{{E\rightarrow E_n}}{\t{Res}}\Gamma(1+i\gamma)&=\frac{(-1)^{n}}{(n-1)!}\Big(\frac{\DD}{\DD E}i\alpha \frac{E}{\sqrt{E^2-m^2}}\at{E=E_n}\Big)^{-1}\\
    &=\frac{(-1)^{n-1}}{(n)!}\frac{\alpha^2}{m^2n^2}E_n^3.
\end{align}
At this point, we note that the IR scale selected by the Coulomb wavefunctions, the \( |\vec{p}| \) in the numerator of (\ref{start amp}), takes on a specific value at this residue. Had we instead used an arbitrary scale \(\mu\), the residue would differ by a factor of \((-i\mu / \eta_n)^{-n}\).

We now relate (\ref{res}) to the transition amplitudes between continuum and bound states. The continuum-state wavefunctions exhibit a simple pole at the bound-state energy due to their Gamma-function normalization factors. Using (\ref{partialwavescatter}) for the in-scattering wavefunction in terms of spherical harmonics, we find
\begin{align}
   \underset{E\rightarrow E_n}{\t{Res}}  f^{+}_{\t{in}}(x,p_1,u_s^r)=\sum_{l=0}^{n-1}P(n,l)\sum_{m=-l}^lB_{nlm}(\vec{r})\, Y_{lm}^{\star}(\hat{p}_1)\label{confirm theorem}\\
   P(n,l)=4\pi (i)^{n+l}(-1)\frac{1}{\sqrt{2(n-l-1)!(l+n)!}}\frac{\sqrt{|\alpha|}}{m n}E_n,\label{PN}
\end{align}
which states that the residue of the continuum wavefunctions is proportional to the bound-state wavefunctions, with a proportionality factor \( P(n,l) Y_{lm}^{\star}(\hat{p}_1) \), which is consistent with the aforementioned general theorem from scattering theory \cite{Mizera_2024, Newton:1982qc, gottfried2018quantum, Taylor:1972pty}. The sum over \( l \) reflects the \( n \)-fold degeneracy of the $n^{\t{th}}$ energy level \( E_n \) of hydrogen. This degeneracy is broken when including the \( A^2 |\phi|^2 \) interaction.

In the following, we will also need the residue of the complex conjugate of the out-state. The order of operations is important: we first take the complex conjugate and then evaluate the residue at the bound-state energy,
\begin{align}
   \underset{E\rightarrow E_n}{\t{Res}}  \Big[f^{+,\star}_{\t{out}}(x,p_2,u_s^r)\Big]=\sum_{l=0}^{n-1}P(n,l)\sum_{m=-l}^lB_{nlm}^{\star}(\vec{r})Y_{lm}(\hat{p}_2).\label{conjugate residue}
\end{align}
Here, \( P(n,l) \) is the same as that defined previously in (\ref{PN}). It is straightforward to compute the transition amplitude for these analytically continued in/out wavefunctions to all bound states degenerate with the energy level \(E_n\) (i.e., the modes with \( l = 0,1,\dots,n-1 \) and their associated \( m = -l, \dots, l \)). Since the bound states are orthonormal with respect to the SQED inner product (\ref{boundortho}), we find
\begin{align}
    &\sum_{l=0}^{n-1}\sum_{m=-l}^{m=l}\Big(\underset{E\rightarrow E_n}{\t{Res}}\braket{p_2^{\t{out}}|B_{nlm}}_{\t{SQED}}\Big)\Big(\underset{E\rightarrow E_n}{\t{Res}}\braket{B_{nlm}|p_1^{\t{in}}}_{\t{SQED}}\Big)\label{residue first}\\
    &=8\pi^2(-1)^{n+1}\frac{\alpha}{m^2n^2}E_n^2\sum_{l=0}^{n-1}\frac{(-1)^l}{(n-l-1)!(l+n)!}\sum_{m=-l}^lY_{lm}(\hat{p_2})Y^{\star}_{lm}(\hat{p_1})\label{first line PN}\\
    &=2\pi(-1)^{n+1}\frac{\alpha}{m^2n^2}E_n^2\sum_{l=0}^{n-1}\frac{(-1)^l}{(n-l-1)!(l+n)!}(2l+1)P_l(\hat{p}_1\.\hat{p}_2)\label{second line is easy}\\
    &=2^{2-n}\frac{1}{\Gamma^2(n)}\pi(-1)^{n+1}\frac{\alpha}{m^2n^2}E_n^2(1-\hat{p}_1\.\hat{p}_2)^{n-1}.\label{resres}
\end{align}
To compute $\underset{E\rightarrow E_n}{\t{Res}}\braket{p_2^{\t{out}}|B_{nlm}}_{\t{SQED}}$ in (\ref{residue first}), we first extract the residue of the continuum wavefunction at the bound-state energy $E_n$, as given in (\ref{conjugate residue}), and then evaluate the SQED inner product. Equation (\ref{first line PN}) follows from the orthonormality of the bound-state wavefunctions, along with the proportionality factors \( P(n,l) \) given in (\ref{PN}). Equation (\ref{second line is easy}) is simply the sum over spherical harmonics, which yields the corresponding Legendre polynomial. In the final equality, we used the identity
\begin{gather}
    (1-x)^{n-1}=2^{n-1}\Gamma^2(n)\sum_{l=0}^{n-1}\frac{(-1)^l}{\Gamma(n-l)\Gamma(1+l+n)}(2l+1)P_{l}(x),\qquad n \in \mathbb{Z}, \quad n \geq 1
\end{gather}
which can be verified by projecting both sides onto the $j^{\t{th}}$ Legendre polynomial by integrating with \( \int_{-1}^{1} \DD x \, P_j(x) \). The reader may wonder why $\braket{p^{\t{in}}|B_{nlm}}_{\t{SQED}} \neq 0$, given that we previously stated in (\ref{boundscatterortho}) that continuum wavefunctions are orthogonal to bound states. The resolution is that this orthogonality holds only for continuum states with energies above the threshold, whereas in this case, we are analytically continuing them to energies below the threshold.

Comparing (\ref{res}) to (\ref{resres}), we conclude that
\begin{gather}
    \boxed{ \underset{E\rightarrow E_n}{\t{Res}}\mathcal{A}(E,\theta)=2\pi i\sum_{l=0}^{n-1}\sum_{m=-l}^{m=l}\Big(\underset{E\rightarrow E_n}{\t{Res}}\braket{p_2^{\t{out}}|B_{nlm}}\Big)\Big(\underset{E\rightarrow E_n}{\t{Res}}\braket{B_{nlm}|p_1^{\t{in}}}\Big).}\label{transition main result}
\end{gather}
That is, the residue of the amplitude at the bound-state pole factorizes into a product of scattering-to-bound-state transitions, with all bound states degenerate at the given energy level contributing as intermediate states.

We emphasize that no arbitrary infrared scale enters at any stage of our derivation. In contrast, had we used amplitudes regulated by an arbitrary IR scale, the resulting relation would inherit that dependence. 
By comparison, methods such as the Dollard evolution operator, dimensional regularization, or introducing a photon mass yield Coulomb amplitudes that depend explicitly on a reference scale (see, e.g., \cite{Kabat_1992}). In these cases, the amplitude takes the form
\begin{gather}
      \mathcal{A}^{\t{reg.}}(E_1,\theta)=-i\gamma\frac{4\pi^2}{|\vec{p}_1|} \delta(E_1-E_2) \frac{\Gamma(1+i\gamma)}{\Gamma(1-i\gamma)}\Bigg(\frac{2\mu}{|\vec{p}_1-\vec{p}_2|}\Bigg)^{2+2i\gamma}.\label{arb mu}
\end{gather}
Here, the residue of the amplitude at the poles of the Gamma function is proportional to 
\( \mu^{-n} \), introducing an explicit dependence on an arbitrary scale.

More generally, in any scattering amplitude involving a regulator $\mu$, different choices of $\mu$—as functions of the external momenta—can affect important analytic properties of the amplitude. These include whether the Steinmann relations hold, the locations of branch points, the Regge behavior, and the factorization structure at bound-state poles.

\section*{Acknowledgments}
I am grateful to Tim Adamo, Anton Ilderton, Hofie Hannesdottir, and Riccardo Gonzo for their feedback on drafts of the manuscript. I also thank Andrew McLeod and Einan Gardi for valuable discussions. I thank George Sterman and Ozan Erdo\u{g}an for correspondence. This work is supported by the enhanced research expenses
\verb|RF\ERE\221030| associated with the Royal Society grant \verb|URF\R1\221233|. OpenAI Codex was used to assist with proofreading, bibliography checking, LaTeX formatting, and verification of internal cross-references. All suggestions were reviewed by the author, who takes full responsibility for the manuscript.

\appendix

\section{Completeness relations}\label{app: completeness relations and identities}
\subsection{Completeness relation for spherically symmetric wavefunctions}\label{app:spherecomplete}
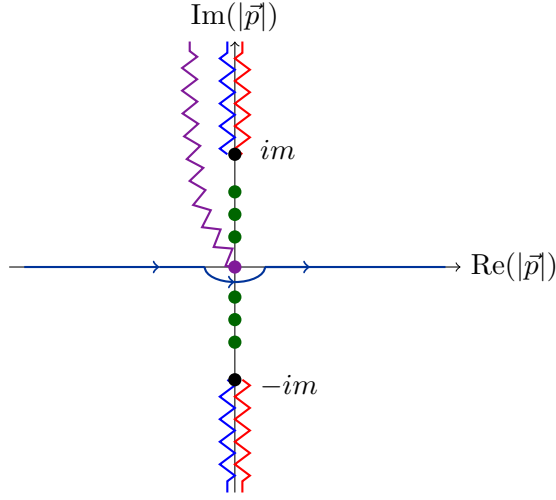
\begin{figure}
    \centering
    \begin{tikzpicture}
    \definecolor{darkgreen}{rgb}{0.0, 0.4, 0.0} 
    \definecolor{navyblue}{rgb}{0.0, 0.2, 0.6}  
    \definecolor{purplebranch}{rgb}{0.5, 0.1, 0.6}  

    \draw[->] (-3,0) -- (3,0) node[right] {Re($|\vec{p}|$)};
    \draw[->] (0,-3) -- (0,3) node[above] {Im($|\vec{p}|$)};
    
    \draw[thick,red,decorate,decoration={zigzag,amplitude=0.1cm,segment length=0.3cm}] (0.1,1.5) -- (0.1,3);
    \draw[thick,red,decorate,decoration={zigzag,amplitude=0.1cm,segment length=0.3cm}] (0.1,-1.5) -- (0.1,-3);

    \draw[thick,blue,decorate,decoration={zigzag,amplitude=0.1cm,segment length=0.3cm}] (-0.1,1.5) -- (-0.1,3);
    \draw[thick,blue,decorate,decoration={zigzag,amplitude=0.1cm,segment length=0.3cm}] (-0.1,-1.5) -- (-0.1,-3);
 
    \draw[thick,purplebranch,decorate,decoration={zigzag,amplitude=0.1cm,segment length=0.3cm}]
        (0,0) to[out=120, in=-90] (-0.5,1.2) to[out=100, in=-90] (-0.6,3);

    \fill[black] (0,1.5) circle (2.5pt);
    \fill[black] (0,-1.5) circle (2.5pt);
    \fill[purplebranch] (0,0) circle (2.5pt); 

    \fill[darkgreen] (0,1.0) circle (2.5pt);
    \fill[darkgreen] (0,0.7) circle (2.5pt);
    \fill[darkgreen] (0,0.4) circle (2.5pt);

    \fill[darkgreen] (0,-1.0) circle (2.5pt);
    \fill[darkgreen] (0,-0.7) circle (2.5pt);
    \fill[darkgreen] (0,-0.4) circle (2.5pt);

    \draw[thick, navyblue,->] (-2.8,0) -- (-1,0);
    \draw[thick, navyblue] (-1,0) -- (-0.4,0);
     \draw[thick, navyblue,->] (0.4,0) -- (1,0);
    \draw[thick, navyblue, postaction={decorate},
          decoration={markings, mark=at position 0.5 with {\arrow{>}}}] 
          (-0.4,0) arc (180:360:0.4 and 0.2); 
    \draw[thick, navyblue] (0.4,0) -- (2.8,0);

    \node[above right] at (0.2,1.3) {$i m$}; 
    \node[below right] at (0.2,-1.3) {$-i m$}; 
\end{tikzpicture}
    \caption{Analytic structure of the integrand (\ref{eq:to analyze analytic}) in the complex $|\vec{p}|$-plane. The purple branch cut is due to the confluent hypergeometric function of the second kind $\Psi(a,b,z)$ exhibiting a branch cut along the negative real $z$ line. The dark blue curve is the integration contour, which avoids passing through this branch cut. The blue (antiparticle) and red (particle) branch cuts at $|\vec{p}|=\pm i m$ arise due to $\gamma$'s dependence on the mass-shell energy $E=\sqrt{|\vec{p}|^2+m^2}$. However, as we are summing over particle and antiparticle wavefunctions $\rho=\pm$, the particle and antiparticle branch cuts cancel one another. The green dots represent the bound-state poles of the Gamma function, of which there are infinitely many approaching the origin.}
    \label{fig:analytic structure of completeness integrand}
\end{figure}

In this appendix, we will prove the completeness relation (\ref{Completespherical}) for the spherically symmetric relativistic Coulomb wavefunctions. The proof of completeness for the non-relativistic spherically symmetric Coulomb wavefunctions can be found in Mukunda's paper \cite{Makunda} (see Appendix C of \cite{completeness2} and the discussion around equation 3.43 in the textbook \cite{Michel:2021jkx}, which expands on Mukunda's proof with further rigor), as well as an alternative approach presented in \cite{Mukhamedzhanov_2008}. We will generalize Mukunda's proof to the relativistic setting. An interesting novelty in the relativistic case is that one must include negative-energy (antiparticle) wavefunctions to obtain a completeness relation. Consequently, bound states always contribute to the completeness relation, unlike in the non-relativistic case, where bound states contribute only for attractive potentials, i.e., when the charges have opposite signs, \( e_1 e_s < 0 \). However, with the inclusion of antiparticles, if the potential is repulsive for particles, then antiparticles will bind to it.\\
Following the strategy of \cite{Makunda}, we define the following integral, which resembles, and will turn out to be, the radial continuum contribution of a completeness relation\footnote{The relativistic and non-relativistic Coulomb wavefunctions possess a singularity at \( |\vec{p}| = 0 \) due to their dependence on \( \gamma \propto \frac{1}{|\vec{p}|} \). An analysis of the integrand (\ref{eq:integrand}) for non-relativistic wavefunctions indicates that the integrand vanishes sufficiently fast as \( |\vec{p}| \to 0 \), ensuring a zero contribution to the integral \cite{Makunda,completeness2,Michel:2021jkx}. For this reason, some authors \cite{completeness2} prefer to introduce a lower bound on the integration, which they later take to zero, though this does not affect the final conclusions. The same analysis and conclusion apply to the relativistic wavefunctions, as they exhibit the same \( |\vec{p}| \to 0 \) behavior.}
\begin{gather}
    J_l(r,r')=\sum_{\rho=\pm}\int_{0}^{\infty}\DD |\vec{p}|\,\,|\vec{p}|^2 R^{\star}_{l}(|\vec{p}|r,\rho \gamma)R_{l}(|\vec{p}|r',\rho \gamma)\label{eq:integrand}
\end{gather}
where $R_l$ denotes the relativistic unbound/continuum radial wavefunctions (\ref{eq:unboundradialsphericallysymetric}). There is no need to complex conjugate as the radial wavefunctions are purely real\footnote{$R^{\star}_{l}(|\vec{p}|r,\rho \gamma)=R_{l}(|\vec{p}|r,\rho \gamma)$ follows from the identity
\begin{equation}
{}_1F_1(a, b; -z) = e^{-z} {}_1F_1(b-a, b; z).    
\end{equation}
}. Using the definitions of the continuum radial wavefunctions (\ref{eq:unboundradialsphericallysymetric})
\begin{align*}
    &J_l(r,r')=\frac{(4rr')^l}{[(2l+1 !)]^2}\sum_{\rho=\pm}\int_{0}^{\infty}\DD |\vec{p}|\,\,|\vec{p}|^{2l+2}\Gamma(1+l+i\gamma)\Gamma(1+l-i\gamma)\\
    &e^{-\rho \pi\gamma}e^{-i|\vec{p}|r}{}_1F_1\Big(1+l-i\rho \gamma,2l+2,2i|\vec{p}|r\Big)e^{-i|\vec{p}|r'}{}_1F_1\Big(1+l-i\rho \gamma,2l+2,2i|\vec{p}|r'\Big)\label{eq:startint}.\numberthis
\end{align*}
If it were not for the $e^{-\rho \pi \gamma}$ factor in the second line, then the integrand would be a completely even function of $|\vec{p}|$. To obtain an integrand symmetric under $|\vec{p}|\to-|\vec{p}|$, we next express the confluent hypergeometric function of the first kind in terms of the confluent hypergeometric function of the second kind (A.4 of \cite{Makunda})
\begin{equation}
    \frac{{}_1F_1(a, b; z)}{\Gamma(b)} = 
\exp(\pm i\pi a) \frac{\Psi(a, b; z)}{\Gamma(b - a)} 
+ \exp\big[z \pm i\pi(a - b)\big] 
\frac{\Psi(b - a, b; -z)}{\Gamma(a)}.
\end{equation}
This identity requires that \( z \) is not real and negative; that \( b \neq 0, -1, -2, \dots \); that \( -z = z \exp(\mp i \pi) \); and that the \( +(-) \) sign goes with \( \operatorname{Im} z > 0 \, (< 0) \). We will apply this identity to one of the Coulomb wavefunctions in the integrand (in our case $\operatorname{Im} z>0$):
\begin{align*}
&e^{-i|\vec{p}|r}{}_1F_1\big(1+l-i\rho\gamma, 2l+2; 2i|\vec{p}|r\big) = (-1)^{l+1} \Gamma(2l+2) \, e^{ \pi\rho \gamma}\\
&\times \bigg[e^{-i|\vec{p}|r}
\frac{\Psi\big(1+l-i\rho\gamma, 2l+2; 2i|\vec{p}|r\big)}{\Gamma(l+1 + i\rho \gamma)}
+ e^{i|\vec{p}|r} \frac{\Psi\big(1+l+i\rho\gamma, 2l+2; -2i|\vec{p}|r\big)}{\Gamma(1+l-i\rho\gamma)}
\bigg].\numberthis\label{transformconfluent}
\end{align*}
Applying (\ref{transformconfluent}) to (\ref{eq:startint}) (note that the $e^{\pi\rho\gamma}$ in (\ref{transformconfluent}) cancels the unwanted factor in (\ref{eq:startint})) gives
\begin{align*}
    J_l(r,r')=&(-1)^{l+1} \frac{(4rr')^l}{(2l +1)!}\sum_{\rho=\pm}\int_{0}^{\infty}\DD |\vec{p}|\,\,|\vec{p}|^{2l+2}\Gamma(1+l+i\gamma)\Gamma(1+l-i\gamma)\\
    &\times\bigg[e^{-i|\vec{p}|r}
\frac{\Psi\big(1+l-i\rho\gamma, 2l+2; 2i|\vec{p}|r\big)}{\Gamma(l+1 + i\rho\gamma)}
+ e^{i|\vec{p}|r} \frac{\Psi\big(1+l+i\rho\gamma, 2l+2; -2i|\vec{p}|r\big)}{\Gamma(1+l-i\rho\gamma)}
\bigg]\\
    &\times e^{-i|\vec{p}|r'}{}_1F_1\Big(1+l-i\rho\gamma,2l+2,2i|\vec{p}|r'\Big).\numberthis
\end{align*}
The integrand is now a completely even function of $|\vec{p}|$, so we can write the integral as
\begin{align*}
    &J_l(r,r')=(-1)^{l+1} \frac{(4rr')^l}{(2l+1) !}\sum_{\rho=\pm}\int_{-\infty}^{\infty}\DD |\vec{p}|\,\,|\vec{p}|^{2l+2}\Gamma(1+l-i\rho\gamma)\\
    &e^{-i|\vec{p}|(r+r')}
{}_1F_1\Big(1+l-i\rho\gamma,2l+2,2i|\vec{p}|r'\Big)\Psi\big(1+l-i\rho\gamma, 2l+2; 2i|\vec{p}|r\big)\numberthis\label{eq:to analyze analytic}.
\end{align*}
 We now analyze the analytic structure of the integrand (\ref{eq:to analyze analytic}) in the complex $|\vec{p}|$-plane; see Figure \ref{fig:analytic structure of completeness integrand}. We note that one of the interesting differences between the non-relativistic and relativistic integrands is the presence of the branch cuts at $|\vec{p}|=\pm i m$ in the complex $|\vec{p}|$-plane due to the mass-shell energy $E=\sqrt{m^2+|\vec{p}|^2}$, which features in the Coulomb wavefunctions via their dependence on $\gamma=\alpha \frac{E}{|\vec{p}|}$. However, once we sum over antiparticle and particle contributions, these branch cuts cancel one another. We note that $\Psi(a,b,2i|\vec{p}|r)$ has a branch cut along the positive imaginary $|\vec{p}|$-axis. The integration contour in (\ref{eq:to analyze analytic}) is defined to pass below this branch cut when circling the origin. Thus, a more rigorous starting point at (\ref{eq:integrand}) would be to use a contour with a half-arc at the $|\vec{p}|=0$ origin instead \cite{Mukhamedzhanov_2008,completeness2}. However, the final completeness relation is not affected by this, as the integrand (\ref{eq:to analyze analytic}) vanishes sufficiently fast for the half-arc in Figure \ref{fig:analytic structure of completeness integrand} to give a zero contribution to the integral \cite{Makunda,completeness2,Michel:2021jkx}. The remaining singularities are the poles of the Gamma function, which will correspond to bound states, as we will see shortly.
 We deform the integration contour into the lower half-plane (LHP), as in Figure \ref{fig:deform contour}, and pick up two terms: (1) an arc at infinity $I_l(r,r')$ and (2) a bound-state contribution $B_l(r,r')$ from the residues at the poles of the $\Gamma$ function,
\begin{figure}
    \begin{center}
\begin{tikzpicture}[scale=0.8]
    \definecolor{darkgreen}{rgb}{0.0, 0.4, 0.0}
    \definecolor{navyblue}{rgb}{0.0, 0.2, 0.6}
    \definecolor{purplebranch}{rgb}{0.5, 0.1, 0.6}

    \def\dotspacing{0.8}

    \begin{scope}
        \draw[->] (-3,0) -- (3,0);
        \draw[->] (0,-3) -- (0,3) node[above] {Im($|\vec{p}|$)};

        \draw[thick, purplebranch, decorate, decoration={zigzag, amplitude=0.1cm, segment length=0.3cm}]
            (0,0) to[out=120, in=-90] (-0.5,1.2) to[out=100, in=-90] (-0.6,3);

        \fill[purplebranch] (0,0) circle (2.5pt);

        \foreach \y in {\dotspacing,2*\dotspacing,3*\dotspacing,-\dotspacing,-2*\dotspacing,-3*\dotspacing} {
            \fill[darkgreen] (0,\y) circle (2.5pt);
        }

        \draw[very thick, navyblue,->] (-2.8,0) -- (-1,0);
        \draw[very thick, navyblue] (-1,0) -- (-0.4,0);
        \draw[very thick, navyblue,->] (0.4,0) -- (1,0);
        \draw[very thick, navyblue, postaction={decorate},
              decoration={markings, mark=at position 0.5 with {\arrow{>}}}]
              (-0.4,0) arc (180:360:0.4 and 0.2);
        \draw[very thick, navyblue] (0.4,0) -- (2.8,0);
    \end{scope}

    \begin{scope}[xshift=8cm] 
        \draw[->] (-3,0) -- (3,0);
        \draw[->] (0,-3) -- (0,3) node[above] {Im($|\vec{p}|$)};
        \node[right] at (3,0) {Re($|\vec{p}|$)};

        \draw[thick, purplebranch, decorate, decoration={zigzag, amplitude=0.1cm, segment length=0.3cm}]
            (0,0) to[out=120, in=-90] (-0.5,1.2) to[out=100, in=-90] (-0.6,3);

        \fill[purplebranch] (0,0) circle (2.5pt);

        \foreach \y in {\dotspacing,2*\dotspacing,3*\dotspacing,-\dotspacing,-2*\dotspacing,-3*\dotspacing} {
            \fill[darkgreen] (0,\y) circle (2.5pt);
        }

        \foreach \y in {-\dotspacing,-2*\dotspacing,-3*\dotspacing} {
            \draw[very thick, navyblue, postaction={decorate},
                  decoration={markings, mark=at position 0.25 with {\arrow{>}}, 
                                      mark=at position 0.75 with {\arrow{>}}}]
                  (0.3,\y) arc (360:0:0.3);
        }

        \draw[very thick, navyblue, postaction={decorate},
              decoration={markings, mark=at position 0.25 with {\arrow{>}}, 
                                  mark=at position 0.75 with {\arrow{>}}}] 
              (-2.8,0) arc (180:360:2.8 and 2.9);

    \end{scope}

    \node at (4,0) {\LARGE $=$};

\end{tikzpicture}
\end{center}
    \caption{Deforming the integration contour of (\ref{eq:to analyze analytic}).}
    \label{fig:deform contour}
\end{figure}
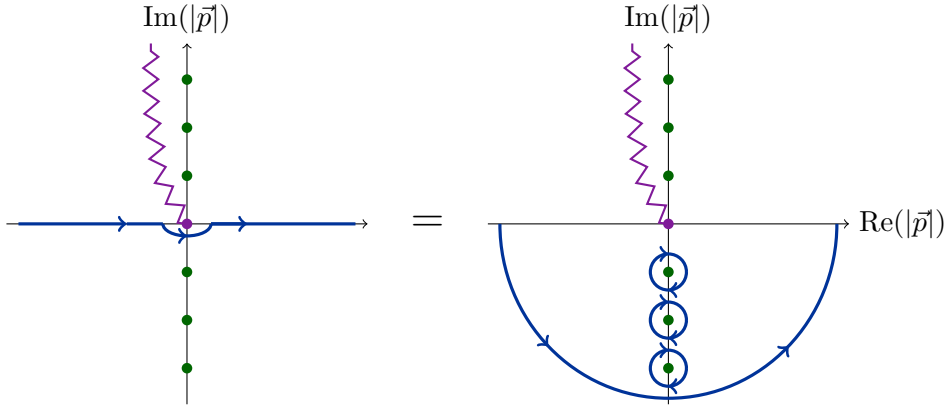
\begin{gather}
    J_{l}(r,r')=I_{l}(r,r')+B_{l}(r,r').\label{simple J}
\end{gather}
\subsection*{Arc at infinity}
We begin with the arc at infinity. The contour for the arc at infinity is
\begin{gather}
    |\vec{p}|= K e^{i\theta},\qquad \theta \in [0,-\pi]
\end{gather}
where we will take the limit $K\rightarrow \infty$.
We note that the relativistic and non-relativistic cases exhibit different behavior as $|\vec{p}|\rightarrow \infty$. For the relativistic case we have
\begin{gather}
  \lim_{|\vec{p}|\rightarrow\infty}  \gamma=\lim_{|\vec{p}|\rightarrow\infty}\alpha \frac{E}{|\vec{p}|}= \alpha
\end{gather}
versus $\gamma=\alpha \frac{m}{|\vec{p}|}\rightarrow 0$ for the non-relativistic case. We note the leading terms as
\begin{align*}
 \lim_{|\vec{p}|^2\rightarrow \infty,\, \operatorname{Im}|\vec{p}|<0} \,&{}_1F_1\Big(1+l-i\gamma,2l+2,2i|\vec{p}|r'\Big)\\
 &=\Gamma(2+2l)\Bigg(\frac{e^{2i|\vec{p}|r}(2i|\vec{p}|r)^{-1-l-i\alpha}}{\Gamma(1+l-i\alpha)}-\frac{(-2i|\vec{p}|r)^{-1-l+i\alpha}}{\Gamma(1+l+i\alpha)}\Bigg)\numberthis\label{eq:expandp}\\
    \lim_{|\vec{p}|^2\rightarrow \infty,\, \operatorname{Im}|\vec{p}|<0}   & \Psi\Big(1+l-i\gamma,2l+2,2i|\vec{p}|r'\Big)=(2i|\vec{p}|r')^{-1-l+i\alpha}\numberthis
 \end{align*}
Simplifying terms we have
\begin{align}
    I_{l}(r,r')
    &=\frac{1}{4 rr'}\sum_{\rho=\pm}\Big(\frac{r'}{r}\Big)^{i\rho \alpha}\int_{C_\infty}\DD |\vec{p}|\,\,e^{i|\vec{p}|(r-r')}-(-1)^{-1-l}r^{2i\rho\alpha}\frac{\Gamma(1+l-i\rho\alpha)}{\Gamma(1+l+i\rho\alpha)}e^{-i|\vec{p}|(r+r')}\label{eqn: large p}
\end{align}
where $C_{\infty}$ is the counter-clockwise half-circle in the LHP. We parameterize this contour as
\begin{gather}
    |\vec{p}|= K e^{i\theta},\qquad \theta \in [0,-\pi]\\
    \DD |\vec{p}|= iK e^{i\theta}\DD\theta
\end{gather}
so we obtain for the first term in (\ref{eqn: large p})
\begin{align}
   \lim_{K\rightarrow \infty} \int_{C_{\infty}}\DD |\vec{p}|\,\,e^{i|\vec{p}|(r-r')}&=\lim_{K\rightarrow \infty}iK\int_{-\pi}^{0}\DD\theta e^{i\theta}\exp\Big(i(r-r')K e^{i\theta}\Big)\\
    &=\lim_{K\rightarrow \infty}2\frac{\sin \Big(K(r-r')\Big)}{r-r'}\\
    &= 2\pi \delta(r-r').\label{distribution}
\end{align}
The distributional equality (\ref{distribution}) is explained at Eq. 2.10 of \cite{Makunda}. The second term in (\ref{eqn: large p}) will give $\delta(r+r')$ and hence is zero for positive $r,r'>0$. On the support of $\delta(r-r')$ we have that $\Big(\frac{r'}{r}\Big)^{i\alpha}=1$ and hence the contribution from the arc at infinity gives
\begin{gather}
    I_{l}(r,r')= \frac{\pi}{r^2}\delta(r-r').
\end{gather}
\subsubsection*{Bound state contributions}
We now address the clockwise $(-2\pi i)$ residues arising from the poles of the Gamma function $\Gamma(1+l-i\rho\gamma)$ in the lower half of the $|\vec{p}|$-plane. These are situated at
\begin{gather}
    i\rho\alpha \frac{\sqrt{|\vec{p}|^2+m^2}}{|\vec{p}|}=n\qquad n=l+1,l+2,l+3,\dots\\
    |\vec{p}|_{n}=i\rho \frac{m\alpha}{\sqrt{n^2+\alpha^2}}=i \eta_{n}
\end{gather}
where $\eta_n$ is the same $\eta_n$ appearing in the variables used to describe the bound states (\ref{eq:etas}). As we are only picking up the residues in the LHP, we see that if $\alpha>0$ (repulsive), then we pick up the antiparticle $\rho<0$ bound states, and vice versa, in agreement with the expectation that particles bind to attractive potentials and antiparticles to repulsive potentials. This is an interesting distinction between the relativistic and non-relativistic cases: in the former, bound states always contribute due to the necessity of antiparticles in relativistic theories. To compute the residue of the Gamma function, we use
\begin{gather}
    \underset{x\rightarrow n}{\text{Res}}\,\,\Gamma(n)=\frac{(-1)^n}{(-n)!},\qquad n \in 0,-1,-2,\dots
\end{gather}
Accounting for a Jacobian factor, we find that the residues at the poles in the LHP are
\begin{gather}
    \t{Res}_{|\vec{p}|\rightarrow |\vec{p}|_{n}}\Gamma(1+l-i\rho \gamma)=i\frac{(-1)^{n-l}}{(n-l-1)!}\frac{m n |\alpha| }{(n^2+\alpha^2)^{\frac{3}{2}}}
\end{gather}
The bound-state contribution is then the sum of the residues at the poles of the Gamma function in the LHP,
\begin{align*}
    &B_{l}(r,r')= -i\sum_{n=l+1}^{\infty}\frac{(4rr')^l}{[(2l+1)!]^2}(-2\pi i)( \eta_{n})^{2l+2}\frac{(l+n)!}{(n-l-1)!}\frac{m n |\alpha| }{(n^2+\alpha^2)^{\frac{3}{2}}}  \\
    &e^{-\eta_{n}(r+r')}
{}_1F_1\Big(1+l+n,2l+2,2 \eta_{n} r'\Big){}_1F_1\big(1+l+n, 2l+2; 2\eta_{n}r\big)\numberthis \label{B}
\end{align*}
where we used \cite{Makunda}
\begin{align*}
    &\Psi(l+1-n, 2l+2; 2r|\eta_n|)\\ 
&= (-1)^{n-l-1} \frac{(l+n)!}{(2l+1)!} 
{}_1F_1(l+1-n, 2l+2; 2r|\eta_n|),\qquad l \in 0,1,2,\dots\numberthis
\end{align*}
as well as
\begin{gather}
    {}_1F_1(a,b,-z)=e^{-z}{}_1F_1(b-a,b,z).
\end{gather}
We recognize the terms in (\ref{B}) as the bound-state wavefunctions (\ref{eq:boundstateradial}),
\begin{gather}
    B_l(r,r')=-\pi \sum_{n=l+1}^{\infty} E_n R_{nl}(r)R_{nl}(r') 
\end{gather}
Rewriting (\ref{simple J}), $J=B+I$, as $J-B=I$, we conclude the following completeness relation for the radial wavefunctions:

\begin{gather}
 \pi \sum_{n=l+1}^{\infty} E_n R_{nl}(r)R_{nl}(r')   + \sum_{\rho=\pm}\int_{0}^{\infty}\DD |\vec{p}|\,\,|\vec{p}|^2 R^{\star}_{l}(|\vec{p}|r,\rho \gamma)R_{l}(|\vec{p}|r',\rho \gamma)=\frac{\pi}{r^2}\delta(r-r')
\end{gather}
Using this completeness relation, we can derive a completeness relation for the bound-state wavefunctions $B_{nlm}(\vec{x},t)$ (\ref{eq:fullbound}) and continuum wavefunctions (\ref{eq:fullcont}), which include the spherical-harmonic factors, if we use
\begin{gather}
\sum_{l,m} Y_{lm}(\theta, \phi) Y^*_{lm}(\theta', \phi') = \delta(\cos\theta - \cos\theta') \delta(\phi - \phi')\\
\delta^3(\vec{x} - \vec{x}') = \frac{\delta(r - r')}{r^2} \delta(\cos\theta - \cos\theta') \delta(\phi - \phi').
\end{gather}
From these relations, we arrive at the following equal-time completeness relation:
\begin{mdframed}
    \begin{align*}
   &\sum_{l=0}^{\infty}\sum_{m=-l}^l\Bigg( \pi \sum_{n=l+1}^{\infty} E_n B_{nlm}(\vec{x},t)B^{\star}_{nlm}(\vec{x}\,',t)+\sum_{\rho=\pm}\int_{0}^{\infty}\DD |\vec{p}|\,\,|\vec{p}|^2 f^{\rho}_{lm}(\vec{x},t,|\vec{p}|)f^{\star \rho}_{lm}(\vec{x}\,',t,|\vec{p}|)\Bigg)\\
   &=\pi\delta^3(\vec{x}-\vec{x}\,')\numberthis\label{Completesphericalappendix}
\end{align*}
\end{mdframed}
The equal-time condition is necessary to cancel the $e^{\pm i\rho E t}$ factors.

\subsection{Completeness of scattering solutions}\label{app:scattercomplete}
Using the completeness of the spherically symmetric wavefunctions (\ref{Completesphericalappendix}), the completeness relation for the scattering solutions follows straightforwardly. Consider the quantity
\begin{gather}
    Q(\vec{x},\vec{x}')=\sum_{\rho=\pm}\int \DD^3\vec{p}\, f_{\t{in}}^{\rho}(\vec{p},\vec{x})f_{\t{in}}^{\star\rho}(\vec{p},\vec{x}\,')\label{Q}
\end{gather}
where $f_{\t{in}}^{\rho}(\vec{p},\vec{x}\,')$ are the in-state scattering solutions (\ref{restframeinstate}).
Recall that we can expand these wavefunctions in terms of the spherically symmetric wavefunctions (\ref{partialwavescatter})
    \begin{gather}
    f^{\rho}_{\t{in}}(x,p,u_s^r)=4\pi\sum_{l,m}i^l\sqrt{\frac{\Gamma(1+l+i\rho\gamma)}{\Gamma(1+l-i\rho\gamma)}}f^{\rho}_{lm}(x,|\vec{p}|)Y^{\star}_{lm}(\hat{p}).\label{expand}
\end{gather}
Using this expansion in (\ref{Q}) and then performing the angular $\hat{p}$ integral using
\begin{equation}
\int_{\theta=0}^{\pi} \int_{\varphi=0}^{2\pi} Y_{\ell}^{m} Y_{\ell'}^{m'*} \, d\Omega = \delta_{\ell\ell'} \delta_{mm'} ,
\end{equation}
and then using that $i^l\sqrt{\frac{\Gamma(1+l+i\rho\gamma)}{\Gamma(1+l-i\rho\gamma)}}$ is a pure phase for real kinematics gives
\begin{gather}
    Q=16\pi^2\sum_{\rho=\pm}\int_0^{\infty}\DD|\vec{p}|\,|\vec{p}|^2f^{\rho}_{lm}(x,|\vec{p}|)f^{\star,\rho}_{lm}(\vec{x}\,',|\vec{p}|)\label{halfway}
\end{gather}
 Comparing with (\ref{Completesphericalappendix}), we conclude the completeness relation

 \begin{mdframed}
    \begin{align*}
   &\sum_{l=0}^{\infty}\sum_{m=-l}^l\Bigg(  \sum_{n=l+1}^{\infty} E_n B_{nlm}(\vec{x},t)B^{\star}_{nlm}(\vec{x}\,',t)+\frac{1}{2}\sum_{\rho=\pm}\int \frac{\DD^3\vec{p}}{(2\pi)^3}\, f_{\t{in}}^{\rho}(\vec{p},\vec{x},t)f_{\t{in}}^{\star\rho}(\vec{p},\vec{x}\,',t)\Bigg)\\
   &=\delta^3(\vec{x}-\vec{x}\,').\numberthis\label{Completescatter}
\end{align*}
\end{mdframed}
Again, the equal-time condition is necessary to cancel the $e^{\pm i\rho E t}$ factors present in the wavefunctions.
Similarly, using out-state scattering solutions yields an identical completeness relation, as the only difference for such solutions is the phase of the partial-wave coefficients in the expansion (\ref{expand}), which cancels when taking the modulus at (\ref{halfway}).
\subsection{Further identities}\label{app: further identities}
Hostler \cite{hostler1964coulomb} derives the completeness relation for the spherically symmetric Coulomb wavefunctions using a circular argument, assuming completeness to recover its explicit form along with other useful identities. Since we have independently established completeness in this appendix, we can use Hostler’s approach to (1) cross-check our result and (2) derive additional identities. While Hostler’s argument applies to spherical wavefunctions, we are interested in scattering wavefunctions; adapting his method to this case is straightforward.

Given that scattering states (\ref{restframeinstate}) and bound states (\ref{eq:fullbound}) form a complete basis for functions in \( L^2(\mathbb{R}^3) \) (\ref{Completescatter}), any solution to the semi-free EQM (\ref{EQMrestframe}) can be expressed as a linear combination of these wavefunctions (here we work in the rest frame of the source):
\begin{gather}
    \phi(\vec{x},t)=\int\frac{\DD^3\vec{p}}{(2\pi)^3 2E_p}\Bigg(a^{+}_{\t{in}}(\vec{p})f^{+}_{\t{in}}(x,p)-a^{-}_{\t{in}}(\vec{p})f^{-}_{\t{in}}(x,p)\Bigg)-\t{sign}[\alpha]\sum_{n,l,m}b_{nlm}B_{nlm}(x),
\end{gather}
where the coefficients can be extracted via the SQED inner product:
\begin{gather}
    a^{\pm}_{\t{in}}(\vec{p})=\braket{f^{\pm}_{\t{in}}(\vec{p})|\phi}_{\t{SQED}},\quad b_{nlm}=\braket{B_{nlm}|\phi}_{\t{SQED}}\label{expansions}
\end{gather}
which follows from the normalization of the in-state wavefunctions (\ref{normofinout}) and bound-state wavefunctions (\ref{boundortho}). Let us consider the expansion (\ref{expansions}) at time $t=0$ and use the explicit expressions for the expansion coefficients.
\begin{align*}
&\phi(\vec{x},0)=\Bigg[\t{sign}[\alpha]\sum_{n,l,m}B_{nlm}(x)\int\DD^3\vec{y}B_{nlm}^{\star}(y)\Bigg(-i\,\dot{\phi}(y)+\Big(-\,\t{sign}[\alpha]E_n+ \frac{2\alpha}{|\vec{y}|}\Big)\phi(y)\Bigg)\\
&+\int\frac{\DD^3\vec{p}}{(2\pi)^3 2E_p}f^{+}_{\t{in}}(x,p)\int\DD^3\vec{y}f^{+,\star}_{\t{in}}(y,p)\Bigg(i\dot{\phi}(y)-\phi(y)\Big(E_p+\frac{2\alpha}{|\vec{y}|} \Big)\Bigg)\\
&+\int\frac{\DD^3\vec{p}}{(2\pi)^3 2E_p}f^{-}_{\t{in}}(x,p)\int\DD^3\vec{y}f^{-,\star}_{\t{in}}(y,p)\Bigg(-i\dot{\phi}(y)+\phi(y)\Big(-E_p+\frac{2\alpha}{|\vec{y}|} \Big)\Bigg)\Bigg]\Bigg\vert_{x^0=y^0=0}\numberthis\label{explicitexpansion}
\end{align*}
Now, because the EQM (\ref{EQMrestframe}) are second order in time, we can choose $\phi(\vec{x},t=0)$ and $\dot{\phi}(\vec{x},t=0)$ independently of one another. In particular, if we choose $ \phi(\vec{x},t=0)=0$ in (\ref{explicitexpansion}), then the coefficients of the remaining $\dot{\phi}(y)$ terms must vanish for each point in $\vec{y}$-space. Consequently, we obtain the relation
\begin{align*}
\Bigg[\sum_{\rho=\pm}\rho\int\frac{\DD^3\vec{p}}{(2\pi)^3 2E_p}f^{\rho}_{\t{in}}(x,p)f^{\rho,\star}_{\t{in}}(y,p)-\,\t{sign}[\alpha]\sum_{n,l,m}B_{nlm}(x)B_{nlm}^{\star}(y)\Bigg]\Bigg\vert_{x^0=y^0=0}=0\numberthis\label{equal time vanishing}
\end{align*}
In the main text, when we examine causality, we will need the covariant form of this expression. The essence of the argument leading to (\ref{equal time vanishing}) is that, for a solution to the equations of motion (\ref{eq:freemotion}), we can independently specify on an arbitrary Cauchy surface $\Sigma$, where all points are spacelike separated, the field configuration $\phi\vert_{\Sigma}$ and its conjugate momentum $n\.\d \phi\vert_{\Sigma}$, where $n^{\mu}$ is the normal vector to the Cauchy surface. The covariant wavefunctions form a complete basis for solutions to the EQM in an arbitrary reference frame, and hence we can expand the field on this arbitrary Cauchy slice using these wavefunctions and use the SQED inner product again to extract the coefficient, much as in (\ref{equal time vanishing}) but with $\dot{\phi}(y)$ replaced by $n\.\d\phi(y)$ and the spatial integral instead taken over the Cauchy surface $\int\DD^3\Sigma_{\mu}$. By examining the coefficients of $n\.\d \phi(y)$, we again obtain a vanishing relation for each pair of points on the Cauchy surface. As all points on a Cauchy surface have to be spacelike separated, we arrive at the covariant form of the constraint:
\begin{align*}
\theta(-(x-y)^2)\Bigg[\sum_{\rho=\pm}\rho\int\frac{\DD^3\vec{p}}{(2\pi)^3 2E_p}&f^{\rho}_{\t{in}}(x,p,u_s)f^{\rho,\star}_{\t{in}}(y,p,u_s)\\
&-\,\t{sign}[\alpha]\sum_{n,l,m}B_{nlm}(x,u_s)B_{nlm}^{\star}(y,u_s)\Bigg]=0\numberthis\label{equal time vanishing covariant}
\end{align*}
where we must use the wavefunctions in their covariant form, without assuming that we are working in the rest frame of the source.

To derive the next relation, we examine the $\frac{\alpha}{|\vec{y}|}$ terms in (\ref{explicitexpansion}). They have the same coefficients as in (\ref{equal time vanishing}), and thus the $\frac{\alpha}{r}$ terms cancel one another in (\ref{explicitexpansion}). Therefore, all that remains in (\ref{explicitexpansion}) are the terms with the energy coefficients $E_n$ and $E_p$. In order for the equality (\ref{explicitexpansion}) to hold with only these terms remaining, it must be the case that (note that $\t{sign}[\alpha]^2=+1$)
    \begin{align*}
   &\Bigg[\sum_{l=0}^{\infty}\sum_{m=-l}^l\Bigg(  \sum_{n=l+1}^{\infty} E_n B_{nlm}(x)B^{\star}_{nlm}(y)+\frac{1}{2}\sum_{\rho=\pm}\int \frac{\DD^3\vec{p}}{(2\pi)^3}\, f_{\t{in}}^{\rho}(x,p)f_{\t{in}}^{\star\rho}(y,p)\Bigg)\Bigg]\Bigg\vert_{x^0=y^0=0}\\
   &=\delta^3(\vec{x}-\vec{y})\label{completeappendix3}\numberthis
\end{align*}
which is again the completeness relation (\ref{Completescatter}), thus providing a cross-check on all the factors appearing in the completeness relation.
Considering instead the time derivative of (\ref{explicitexpansion}) gives
\begin{align*}
&\Bigg[\t{sign}[\alpha]\sum_{n,l,m}E^2_nB_{nlm}(x)B_{nlm}^{\star}(y)-\frac{1}{2}\sum_{\rho=\pm}\rho\int\frac{\DD^3\vec{p}}{(2\pi)^3 }E_p\,f^{+}_{\t{in}}(x,p)f^{+,\star}_{\t{in}}(y,p)\Bigg]\Bigg\vert_{x^0=y^0=0}   \\
&=\frac{2\alpha}{|\vec{y}|}\delta^3(\vec{x}-\vec{y}).\numberthis
\end{align*}

\section{Time-ordered Green's function in a Coulomb background}\label{app:greens}
In this appendix, we verify that the time-ordered Green's function for the Coulomb field in the rest frame of the source (omitting the $A^2\phi$ interaction term)

\begin{gather}
  G\big(x,y|A(u_s)\big)=  \braket{A(u_s)|\t{T} \phi(x)\phi^{\star}(y)|A(u_s)}\label{Greensapp}\\
       \Bigg( \Box + m^2 + 2i\frac{\alpha}{r} \partial_t \Bigg) G_F(x,y, u_s^r)=-i\delta^4(x-y)\label{tosolve}
\end{gather}
is given by 
\begin{align*}
   G_F(x,y, u_s^r)=\Bigg(& \sum_{l=0}^{\infty}\sum_{m=-l}^l \sum_{n=l+1}^{\infty} \theta\Big(-\t{sign}[\alpha](x^0-y^0)\Big) B_{nlm}(x)B^{\star}_{nlm}(y)\\
   &+\frac{1}{2}\sum_{\rho=\pm}\int \frac{\DD^3\vec{p}}{(2\pi)^3E_p}\theta(\rho(x^0-y^0)) f_{\t{in}}^{\rho}(x,\vec{p})f_{\t{in}}^{\star\rho}(y,\vec{p})\Bigg).\numberthis\label{Greensexplicit}
\end{align*}
Before proceeding with the derivation, we note two features to help make sense of the structure of Eq.~(\ref{Greensexplicit}).
For zero coupling, the bound states \( B_{nlm} \) vanish, and the scattering Coulomb wavefunctions reduce to on-shell plane waves, so that Eq.~(\ref{Greensexplicit}) reduces to the standard free Feynman propagator. 

The presence of the \(\theta\big(-\t{sign}[\alpha](x^0 - y^0)\big)\) can be understood as follows. The Green's function (\ref{Greensexplicit}) describes the propagation of a particle from \( y \) to \( x \) when \( y^0 < x^0 \), or the propagation of an antiparticle from \( x \) to \( y \) when \( x^0 < y^0 \). If \(\alpha < 0\), only particle bound states exist (as opposed to antiparticle bound states), meaning these bound states can propagate only when \( y^0 < x^0 \). This explains the presence of the \(\theta\)-function multiplying the bound-state wavefunctions. Note that \(\phi^{\star}\) creates particles when acting on the vacuum.
\\
We now verify that Eq.~(\ref{Greensexplicit}) satisfies Eq.~(\ref{tosolve}). If the time-ordering \(\theta\)-functions were absent in Eq.~(\ref{Greensexplicit}), the differential operator in Eq.~(\ref{tosolve}) would annihilate the expression entirely, as both the bound states \( B_{nlm} \) and the scattering wavefunctions \( f_{\text{in}}^{\rho} \) satisfy the corresponding equations of motion. Therefore, the only contributions come from the differential operator acting on the \(\theta\)-functions.
We start with the \( \frac{\alpha}{r} \partial_{x^0} \) term acting on the \(\theta\)-functions. Using the identity \( \partial_{x^0} \theta(\rho(x^0 - y^0)) = \rho \delta(x^0 - y^0) \), we obtain
    \begin{align*}
   \delta(x^0-y^0)\Bigg(& \sum_{l=0}^{\infty}\sum_{m=-l}^l \sum_{n=l+1}^{\infty} -\t{sign}[\alpha] B_{nlm}(x)B^{\star}_{nlm}(y)\\
   &+\frac{1}{2}\sum_{\rho=\pm}\rho\int \frac{\DD^3\vec{p}}{(2\pi)^3E_p} f_{\t{in}}^{\rho}(x,\vec{p})f_{\t{in}}^{\star\rho}(y,\vec{p})\Bigg)=0.\numberthis
\end{align*}
This vanishes due to the equal-time relation in (\ref{equal time vanishing}). Thus, the only non-vanishing contribution from the differential operator in (\ref{tosolve}) comes from the \( \partial_{x^0}^2 \) term acting on both the time-ordering \(\theta\)-functions and the wavefunctions. There are two such terms, for instance, when acting on the $\theta$-functions multiplying the scattering wavefunctions:
\begin{align}
   &\Bigg(\d^2_{x^0} \theta(\rho(x^0-y^0)) \Bigg)f_{\t{in}}^{\rho}(x,\vec{p})+2\Bigg(\d_{x^0} \theta(\rho(x^0-y^0)) \Bigg)\d_{x^0}f_{\t{in}}^{\rho}(x,\vec{p})=-iE_p\delta(x^0-y^0)f_{\t{in}}^{\rho}(x,\vec{p})
\end{align}
where we used the distributional identity $\delta'(x)=-\delta(x)\d_x$. Similarly, for the bound-state wavefunctions, there are two such terms:
\begin{align*}
   &\Bigg(\d^2_{x^0} \theta(-\t{sign}[\alpha](x^0-y^0))\Bigg) B_{nlm}(x)+2\Bigg(\d_{x^0} \theta(-\t{sign}[\alpha](x^0-y^0))\Bigg)\d_{x^0} B_{nlm}(x) \\
   &=-iE_n B_{nlm}(x)\delta(x^0-y^0)\numberthis
\end{align*}
Combining these results, we obtain:
\begin{align*}
         & \Bigg( \Box + m^2 + 2i\frac{\alpha}{r} \partial_t \Bigg) G_F(x,y, u_s^r)\\
         &=-i\delta(x^0-y^0)
         \Bigg( \sum_{l=0}^{\infty}\sum_{m=-l}^l \sum_{n=l+1}^{\infty} E_nB_{nlm}(x)B^{\star}_{nlm}(y)
         +\frac{1}{2}\sum_{\rho=\pm}\int \frac{\DD^3\vec{p}}{(2\pi)^3}f_{\t{in}}^{\rho}(x,\vec{p})f_{\t{in}}^{\star\rho}(y,\vec{p})\Bigg)\numberthis.
\end{align*}
The right-hand side matches the equal-time completeness relation in Eq. (\ref{completeappendix3}), thus verifying that (\ref{Greensexplicit}) satisfies Eq.~(\ref{tosolve}).

\section{Late-time dynamics of plane-wave creation and annihilation operators}\label{app:latetimedynamics}
Without making any assumptions, we can expand a position-space correlator at any time in terms of an on-shell plane-wave basis with time-dependent coefficients
\begin{gather}
    \phi(x)=\int\frac{\DD^3\vec{p}}{(2\pi)^32E_p}[a_{\vec{p}}(t)e^{-ip\. x}+a^{\star}_{\vec{p}}(t)e^{ip\. x}].\label{planewaveexp}
\end{gather}
We can extract these coefficients at any time using the KG inner product
\begin{gather}
    a_{\vec{p}}(t)=\frac{\dot{a}_{\vec{p}}(t)}{2i E_p}+i\int\DD^3\vec{x}\, e^{ip\.x}\overset{\leftrightarrow}{\d_0}\phi(x).
\end{gather}
Taking the time derivative of this expression gives
\begin{align}
       \dot{a}_{\vec{p}}(t)&=\frac{\ddot{a}_{\vec{p}}(t)}{2i E_p}+i\int\DD^3\vec{x}\, e^{ip\.x}\overset{\leftrightarrow}{\d}_{00}\phi(x)\\
       &=\frac{\ddot{a}_{\vec{p}}(t)}{2i E_p}+i\int\DD^3\vec{x}\, e^{ip\.x}(E_p^2+\vec{\nabla}^2-m^2-2ie_1A^{\mu}\d_{\mu})\phi(x)\label{Schwingerdyson}\\
       &=\frac{\ddot{a}_{\vec{p}}(t)}{2i E_p}+i\int\DD^3\vec{x}\, e^{ip\.x}(E_p^2+\overset{\leftarrow}{\nabla}^2-m^2-2ie_1A^{\mu}\d_{\mu})\phi(x)\label{IBP}\\
       &=\frac{\ddot{a}_{\vec{p}}(t)}{2i E_p}+2e_1\int\DD^3\vec{x}\, e^{ip\.x}A^{\mu}\d_{\mu}\phi(x)\label{dynamics}
\end{align}
In (\ref{Schwingerdyson}) we used that quantum fields obey the classical equations of motion (\ref{eq:freemotion}) up to contact terms with other fields occurring in the correlator\footnote{We assume that these contact terms do not contribute, as we expect that, for wave packets of particles at late times, none of the particle trajectories intersect for generic kinematics.} according to the Schwinger-Dyson equations. In (\ref{IBP}), we assume that integration by parts can be applied to the Laplacian and then use the on-shell condition of the plane wave. Incidentally, these are also standard steps in textbook derivations of the LSZ reduction formula. Since we are interested in the late-time dynamics, we assume that \( \ddot{a}_{\vec{p}}(t) \) vanishes sufficiently fast to be neglected. The consistency of this assumption can be verified using our final result (\ref{firstederive}). Using the plane-wave expansion (\ref{planewaveexp}) in (\ref{dynamics}), and working in the rest frame of the source where \( A^{\mu} \partial_{\mu} = \frac{e_s}{4\pi} \frac{1}{|\vec{x}|} \partial_0 \), we obtain
\begin{align}
    \lim_{t\rightarrow \pm \infty}\dot{a}_{\vec{p}}(t)&=\lim_{t\rightarrow \pm \infty}2\alpha \int\DD^3\vec{x}\, e^{ip\.x}\frac{1}{|\vec{x}|}\dot{\phi}(x)\\
    &=\lim_{t\rightarrow \pm \infty}\frac{\alpha}{\pi^2} \int\DD^3\vec{x}\int\DD^3\vec{q}\, e^{ip\.x}\frac{e^{i\vec{q}\.\vec{x}}}{|\vec{q}|^2}\d_0\int\frac{\DD^3\vec{k}}{(2\pi)^32E_k}[a_{\vec{k}}(t)e^{-ik\. x}+a^{\star}_{\vec{k}}(t)e^{ik\. x}]\label{FT}\\
    &=\lim_{t\rightarrow \pm \infty}-\frac{i\alpha}{2\pi^2} \int\DD^3\vec{q}\, e^{it(E_p-E_{p-q})}\frac{1}{|\vec{q}|^2}a_{\vec{p}-\vec{q}}(t)\label{manysteps}
\end{align}
where we inserted the Fourier transform of the \( \frac{1}{|\vec{x}|} \) potential in (\ref{FT}). In passing to (\ref{manysteps}), we omitted the terms where the time derivative acts on the creation operators, as our final result (\ref{firstederive}) will show that these scale as \( \dot{a}_p \sim \frac{1}{t} \), and the remaining integrals introduce an additional factor of \( \frac{1}{t} \). Thus, these terms can be neglected when analyzing the late-time behavior. The \( a^{\star}_{\vec{k}} \) operators do not contribute because the phase factor \( e^{it(E_p + E_{p+q})} \) has no stationary phase. The phase in (\ref{manysteps}) reads
\begin{align}
    E_p-E_{p-q}&=\sqrt{m^2+|\vec{p}|^2}-\sqrt{m^2+|\vec{p}-\vec{q}|^2}\\
    &=\frac{\vec{p}\.\vec{q}}{E_p}+\mathcal{O}\left(\frac{|\vec{q}|^2}{E_p}\right)\label{smallq}
\end{align}
where we took the small $|\vec{q}|$ limit in going to (\ref{smallq}) to derive the stationary-phase condition. Thus, we see that at large times $t$, the region where $\frac{\vec{p}\.\vec{q}}{E_p}\sim \frac{1}{t}$ provides the dominant contribution to the dynamics. We can therefore approximate the integral (\ref{manysteps}) by introducing a $\frac{1}{t}$ cutoff on the $|\vec{q}|$ radial integral
\begin{align}
    \lim_{t\rightarrow \pm \infty}\dot{a}_{\vec{p}}(t)
    &=\lim_{t\rightarrow \pm \infty}-\frac{i\alpha}{2\pi^2} a_{\vec{p}}(t)\int_0^{\frac{1}{t}}\DD|\vec{q}|\int\DD^2\Omega_{\hat{q}}\, e^{it|\vec{q}|\frac{\vec{p}\.\hat{q}}{E_p}}\\
    &=\lim_{t\rightarrow \pm \infty}-\frac{\alpha}{2\pi^2  t}E_p a_{\vec{p}}(t)\int\DD^2\Omega_{\hat{q}}\, \frac{1}{\vec{p}\.\hat{q}}(e^{i\frac{\vec{p}\.\hat{q}}{E_p}}-1)\\
    &=\lim_{t\rightarrow \pm \infty}-\frac{2i\gamma}{\pi}\frac{1}{t} a_{\vec{p}}(t)\t{Si}\left(\frac{|\vec{p}|}{E_p}\right)
\end{align}
where for small \( |\vec{q}| \), we approximated \( a_{\vec{p}-\vec{q}}(t) \approx a_{\vec{p}}(t) \). However, this approximation is not strictly necessary to reach our final conclusion that the asymptotic dynamics is not free; it simply allows for an illuminating, solvable differential equation later on. Here $\gamma=\alpha\frac{E_p}{|\vec{p}|}$ and $\t{Si}$ is the sine integral function. The important point is the scaling
\begin{gather}
     \lim_{t\rightarrow \pm \infty}\dot{a}_{\vec{p}}(t)=\frac{i\delta}{t}a_{\vec{p}}(t)\implies\label{firstederive}\\
     a_{\vec{p}}(t)=a_{\vec{p}}(t_0)\left(\frac{t}{t_0}\right)^{i\delta}
\end{gather}
where $\delta$ is a constant in time. Thus, we see that $a_{\vec{p}}(t)$ does not tend to a constant at late times, and hence the asymptotic dynamics is not free. This should be contrasted with the dynamics of a short-range potential $\frac{1}{|\vec{x}|^2}$, where the same argument would lead to
\begin{gather}
     \lim_{t\rightarrow \pm \infty}\dot{a}_{\vec{p}}(t)=\frac{i\delta}{t^2}a_{\vec{p}}(t)\implies\\
     a_{\vec{p}}(t)=C(t_0)e^{-i\frac{\delta}{t}},\label{eq:latetimeshortpotential}
\end{gather}
which tends to a constant at $t\rightarrow \pm \infty$ and hence the asymptotic dynamics is free for such potentials.

\bibliography{coulomb.bib}
\bibliographystyle{JHEP}
\end{document}